\documentclass[acmsmall]{acmart}
\usepackage{multirow}
\usepackage[inline]{enumitem}
\usepackage{csquotes}
\usepackage{caption}
\usepackage{longtable}
\usepackage{pgf}
\usepackage{tablefootnote}
\usepackage{xspace}
\usepackage{subcaption}
\usepackage{tikz}
\usepackage{todonotes}

\setcopyright{acmcopyright}
\acmJournal{TOIS}
\acmYear{2021} \acmVolume{39} \acmNumber{2} \acmArticle{20} \acmMonth{1} \acmPrice{15.00}
\acmDOI{10.1145/3434185}

\author{Maurizio Ferrari Dacrema}
\email{maurizio.ferrari@polimi.it}
\orcid{0000-0001-7103-2788}
\author{Simone Boglio}
\email{simone.boglio@mail.polimi.it}
\author{Paolo Cremonesi}
\email{paolo.cremonesi@polimi.it}
\orcid{0000-0002-1253-8081}
\affiliation{%
  \institution{Politecnico di Milano}
  \country{Italy}
  \city{Milano}
}

\author{Dietmar Jannach}
\affiliation{%
 \institution{University of Klagenfurt}
 \city{Klagenfurt}
 \country{Austria}
}
\email{dietmar.jannach@aau.at}
\orcid{0000-0002-4698-8507}

\newcommand{\palpha}{\ensuremath{\text{P}^3\!\alpha}\xspace}
\newcommand{\palphabold}{\ensuremath{\text{\textbf{P}}^\textbf{3}\!{\boldsymbol \alpha}}\xspace}
\newcommand{\pbeta}{\ensuremath{\text{RP}^3\!\beta}\xspace}
\newcommand{\pbetabold}{\ensuremath{\text{\textbf{RP}}^\textbf{3}\!{\boldsymbol \beta}}\xspace}

\newcommand{\iALS}{iALS\xspace}
\newcommand{\EASER}{EASE$^R$\xspace}
\newcommand{\pbetanotyetpublished}{We report \pbeta\cite{paudel2017Rp3beta} for completeness although the DL algorithm we evaluate here predates its publication.}
\newcommand\blue[1]{\textcolor{black}{{#1}}}

\begin{document}

\title{A Troubling Analysis of Reproducibility and Progress in Recommender Systems Research}
\renewcommand{\shortauthors}{M. Ferrari Dacrema et al.}
% Have the word "phantom progress" in one of the titles.

\begin{abstract}
%\djx{Should reconsider the title. Our main criticism is not on deep learning (as some think), but on methodology.}
The design of algorithms that generate personalized ranked item lists is a central topic of research in the field of recommender systems.
In the past few years, in particular, approaches based on deep learning (neural) techniques have become dominant in the literature. For all of them, substantial progress over the state-of-the-art is claimed. However, indications exist of certain problems in today's research practice, e.g., with respect to the choice and optimization of the baselines used for comparison, raising questions about the published claims.
%In order to obtain a better understanding of the actual progress, we have tried to reproduce recent results in the area of neural recommendation approaches based on collaborative filtering.
In order to obtain a better understanding of the actual progress, we have compared recent results in the area of neural recommendation approaches based on collaborative filtering against a consistent set of existing simple baselines.
The worrying outcome of the analysis of these recent works---all were published at prestigious scientific conferences between 2015 and 2018---is that 11 out of the 12 reproducible neural approaches can be outperformed by conceptually simple methods, e.g., based on the nearest-neighbor heuristic or linear models.
None of the computationally complex neural methods was actually consistently better than already existing learning-based techniques, e.g., using matrix factorization or linear models. In our analysis, we discuss common issues in today's research practice, which, despite the many papers that are published on the topic, have apparently led the field to a certain level of stagnation.\footnote{This paper significantly extends or own previous work presented in \cite{Ferraridacremaetal2019,DBLP:conf/ijcai/DacremaCJ20}.}
\end{abstract}


\maketitle

\section{Introduction}
Personalized recommendations are a common feature of many modern online services, e.g., on e-commerce, media streaming, and social media sites. In many cases, these recommendations are generated using \emph{collaborative filtering} (CF) techniques \cite{BerkovskyCF2019}. Such techniques leverage the preference or activity profiles of a large user community to predict which are the most relevant items for the individual customer. Early technical approaches to build collaborative filtering recommender systems (RS) were based on nearest neighbor techniques and date back to the 1990s \cite{Resnick:1994:GOA:192844.192905,Linden2003}. In the following years numerous machine learning approaches were proposed for the \emph{top-n} recommendation task, which is to rank the available items according to their assumed relevance for the user. In the most recent years, recommendation methods based on deep learning (DL) or ``neural'' technology have become particularly popular.
While these methods are often computationally complex, their solid success in other application areas---such as language or image processing---led to deep learning techniques nowadays dominating the recommender systems research landscape.

However, there are a number of indications that using increasingly deeper learning methods is not as beneficial as one would expect. For example, in two recent papers, \cite{Lin:2019:NHC:3308774.3308781} and \cite{Yang:2019:CEH:3331184.3331340}, the authors report that for a certain information retrieval task recent neural methods are actually not better than long-existing non-neural ones. In the domain of time series prediction, Makridakis et al. \cite{Makridakis2018} compared various statistical methods with traditional and recent neural methods and observed that statistical methods were mostly favorable over even the most recent machine learning techniques. In the area of recommender systems, the empirical analyses in \cite{Ludewig2018} and \cite{LudewigMauro2019} showed that sometimes almost trivial methods can outperform the latest neural methods for the task of session-based recommendations. These findings are, however, not tied to recent deep learning approaches alone.
The observation that the reported improvements in the literature for certain retrieval tasks ``don't add up'' was put forward back in 2009 \cite{Armstrong:2009:IDA:1645953.1646031}. Most often, the reason for such \emph{phantom progress} lies in the choice and poor optimization of the baselines used in the experiments. Lin \cite{lin2019recantation}, while recognizing the progress made by transformer models for certain tasks of document retrieval, observed it is not rare to find articles comparing neural ranking algorithms only against weak baselines, raising questions regarding the empirical rigor in the field.  For the domain of rating prediction for recommender systems, Rendle et al. in \cite{DBLP:journals/corr/abs-1905-01395} found that many algorithms that were published between 2015 and 2019 actually did not outperform longer-existing methods, when these were properly tuned.

In their paper, Rendle et al. attribute the problem not only to poorly-optimized baselines but also to the sometimes missing standardized benchmarks. In some sense, this finding is surprising since the research community over the last decades has indeed developed certain standards regarding the evaluation of \emph{top-n} recommendation algorithms. Typically, the input to an algorithm is a matrix of user-item interactions and the task is to compute relevance scores for the missing entries or just to rank the items. A number of public datasets exists that can be used for reproducible evaluation, as well as various well-established evaluation metrics. Finally, the technical details of newly proposed algorithms are often laid out in great detail in scientific papers and sometimes authors even share the code they used in their experiments.

This methodological approach for benchmarking algorithms seems very solid at first sight and suitable to determine if an algorithm is favorable over another in a specific combination of
\begin{enumerate*}[label=\textit{(\roman*)}]
\item performance measure, \item evaluation procedure and \item dataset,
\end{enumerate*} at least when we assume that both algorithms are properly optimized. However, the claims made in many research papers are much more general. As can be observed in other fields of computer science, many papers claim a significant improvement over the ``state-of-the-art'', but do not explicitly state such claim to be only supported under very specific experimental conditions. In today's recommender systems research scholarship the researcher has ample freedom in selecting the specific experimental conditions, i.e., which metrics, which protocol, which datasets, and which baselines to use. %Furthermore, since research in applied machine learning is often driven by the ``hunt'' for the best model \cite{DBLP:journals/corr/abs-1206-4656} and not by theoretical or practical considerations, usually authors do not have to justify their choices.

Given these observations regarding potential methodological issues in our field, we were wondering to what extent recent deep learning methods actually help to achieve progress in creating \emph{top-n} recommendation lists based on user-item rating matrices. To that purpose, we  conducted an extensive set of experiments in which we tried to reproduce the results reported in papers that were recently published in top-tier scientific conferences. We scanned the proceedings of several conference series and identified 26 relevant papers. We could reproduce 12 (46\%) of them with reasonable effort. After fine-tuning % both the proposed models \emph{and}
a number of established baselines algorithms,
it turned out that in 8 of the reproducible cases, simple and long-known approaches (e.g., based on nearest neighbors) outperformed the latest neural methods.
% \djx{Previous version said that we tuned the proposed models; commented this statement out, please check.}
Furthermore, when including established linear models based on machine learning (e.g., based on matrix factorization), \emph{we found that only 1 out of 12 recent neural methods was clearly better than the non-neural  baselines}. And even in this case, this was true for only one of the considered datasets.  Overall, these results indicate certain signs of stagnation in the context of applying machine learning methods for \emph{top-n} recommendation tasks, despite many papers published every year claiming substantial progress. %Furthermore, our observations strongly suggest that we should rethink our current research practice. %Improvements that ``don't add up'' were reported in the related field of Information Retrieval almost ten years ago \cite{Armstrong:2009:IDA:1645953.1646031}, and it seems that we nowadays observe such a ``phantom progress'' also in the field of recommender systems.

The rest of the paper is organized as follows. Next, in Section \ref{sec:selection-of-works}, we describe how we selected relevant works considered in our study. Section \ref{sec:methodology} provides details about our research methodology and Section \ref{sec:results} lists our results in detail. The potential reasons for the observed phenomena are finally discussed in Section \ref{sec:discussion}.

\section{Identifying Relevant and Reproducible Works}
\label{sec:selection-of-works}
In order to make sure our research is neither focusing on a few hand-picked examples nor singling out any individual researcher, we followed a systematic approach to identify recent papers that are relevant for our analysis. To obtain a suitable sample for our research, we considered papers which fulfilled the following constraints.

\begin{enumerate}
\item \textbf{Topic.} The paper proposed a new neural collaborative filtering recommendation method for \emph{top-n} recommendation tasks. Hybrid techniques with a collaborative filtering component were considered as well. We limited ourselves to works that focus on the traditional item ranking task. Papers that dealt with other tasks, e.g., session-based recommendation or group recommendation, were not considered to be in the scope of our study. Furthermore, to be considered, a paper had to report at least one ranking or classification accuracy metric. Papers that focused solely on rating or relevance prediction tasks, e.g., using the RMSE, were not taken into account.
\item \textbf{Publication date.} The paper was published between 2015 and 2018 in one of the following conferences: SIGIR, KDD, TheWebConf (WWW), IJCAI, WSDM, ACM RecSys. All of the mentioned conferences are typical outlets for recommender systems research in computer science. Furthermore, all of them, except ACM RecSys, are rated with A$^*$ in the Australian CORE ranking system. ACM RecSys, in contrast, is entirely focused on recommender systems, which is why we included it in the study as well. We identified relevant papers by scanning the proceedings of these conference series in a manual process.
\item \textbf{Reproducible experimental setup.} The experimental setting was reproducible, with reasonable effort, based on source code published by the authors. This constitutes a precondition for the reproducibility of the published results.\footnote{In theory, research papers should contain all relevant information that are needed to implement the proposed algorithm. In reality, however, certain details are sometimes omitted in length-restricted conference papers. For that reason, and to ensure that the results that we report here are reliable, we followed a conservative approach and limited ourselves to the papers where the original authors themselves provided an implementation of their method.}
\begin{itemize}
\item In particular, the source code should include the implementation of the model and its  data structures, a train loop and a way of computing the recommendations, see also the discussions by Collberg and Proebsting \cite{collberg2016repeatability}. For some of the relevant papers we found that some source code was published but was missing major parts, e.g., it consisted only of a skeleton of the algorithm, or did not work. For those papers that were deemed relevant but for which the source code was not publicly available or was not runnable, we contacted all authors of the papers by e-mail. When there was no positive response after 30 days, we considered the source code of the paper to be not available.%  We considered these papers as being non-reproducible as well.
\item At least one of the datasets that were used in the original paper for evaluation was publicly available. In some cases, the authors also provided the data splits for training and testing that were used in their experiments. If the data splits were not available, the experimental setup of the papers was considered as reproducible only if they contained sufficient information  about data pre-processing and splitting.\footnote{In case we encountered problems with the provided code, the data, or the reproduction of the results, we also contacted the authors for assistance.}
\end{itemize}
%\item \textbf{Reproducible results.} For the papers having a reproducible experimental setup, the numerical results reported in the paper could be closely reproduced. This definition of reproducibility is sometimes referred to as \emph{replicability}.\footnote{SIGIR guidelines: \url{https://sigir.org/wp-content/uploads/2018/07/p004.pdf}, see also ACM Artifact Review and Badging: \url{https://www.acm.org/publications/policies/artifact-review-badging}.} \pc{For the shake of completeness, we included in our analysis also the papers with non-reproducible numerical results, as one of our goal is the comparison with strong baselines. }
\end{enumerate}

Following this approach, we identified \textbf{26} relevant papers.
Of these, \textbf{12} were considered having a reproducible experimental setup, according to our classification scheme.
%and \textbf{9} also had fully reproducible numerical results.
Table \ref{tab:reproducibility-stats} summarizes which works were considered relevant and which ones had a reproducible setup.\footnote{%Among the papers with a reproducible experimental setup we can distinguish between those where the numerical results reported in the paper can be closely reproduced.
Among the papers with a reproducible experimental setup, there were some works where the numerical results could not be reproduced exactly, namely \cite{Zhang2018CoupledCF,Zhang2018Neurec,Xue2017DeepMF}. The reproducibility of the numerical results is however not a criterion for the inclusion or exclusion of the algorithm in our analysis. %The works having a reproducible experimental setup but not reproducible numerical results are .
Our approach therefore combines aspects of reproducibility and replicability according to the terminology in the SIGIR guidelines: \url{https://sigir.org/wp-content/uploads/2018/07/p004.pdf}, see also ACM Artifact Review and Badging: \url{https://www.acm.org/publications/policies/artifact-review-badging}.}

% TODO: Can be just written in the text in case of space problems
\begin{table}[h!t]
\centering
\caption{Statistics of relevant and reproducible works on deep learning algorithms for \emph{top-n} recommendation per conference series from 2015 to 2018. %The works having a reproducible experimental setup but not reproducible numerical results are \cite{Zhang2018CoupledCF,Zhang2018Neurec,Xue2017DeepMF}.
} %\mf{new, last sentence}}
\small
\label{tab:reproducibility-stats}
\begin{tabular}{lll|l}
  Conference  & Rep. Setup ratio  & Reproducible Setup & Non-Reproducible Setup\\ \midrule
  KDD  & 3/4 (75\%)
        &  \cite{hu2018leveragingmetapathcontext}, \cite{li2017collaborativevariationalautoencoder}, \cite{wang2015collaborativedeeplearning}
        & \cite{tay2018multipointercoattention}\\
  IJCAI & 5/7 (71\%)
        & \cite{He:2018:OPN:3304889.3304969}, \cite{Zhang2018Neurec}, \cite{Zhang2018CoupledCF}, \cite{Cheng2018DELFAD}, \cite{Xue2017DeepMF}
        & \cite{Nguyen:2018:NNP:3304415.3304640}, \cite{Xuinproceedings}\\
  WWW  & 2/4 (50\%)
        & \cite{he2017neural}, \cite{liang2018variationalautoencodersforCF}
        & \cite{tay2018latentrelationalmetric}, \cite{elkahky2015multiviewdeeplearningcrossdomain} \\
  SIGIR & 1/3 (30\%)
        & \cite{ebesu2018collaborative}
        &  \cite{manotumruksa2018contextualattention}, \cite{chen2017attentivecomponentlevelattention}\\
  RecSys  & 1/7 (14\%)
        & \cite{Zheng:2018:SCF:3240323.3240343}
        & \cite{Sun:2018:RKG:3240323.3240361}, \cite{Bharadhwaj:2018:RRG:3240323.3240383}, \cite{Sachdeva:2018:ANA:3240323.3240397}, \cite{Tuan:2017:CNS:3109859.3109900}, \cite{Kim:2016:CMF:2959100.2959165},  \cite{Vasile:2016:MPE:2959100.2959160}\\
  WSDM & 0/1 (0\%)
        &
        & \cite{wu2016collaborativeCDAE} \\ %- \mf{shall we keept it? This paper was a baseline in others}\\
  % DJ: We could if it were reproducible.
  % TODO: I think we could mention it in the discussion. Did we ask the authors for the code? \mf{The conference was treated as all others, proceedings checked and email to the authors, no reply}
  \midrule
  Total & 12/26 (46\%) &  \\
  \bottomrule
%   \multicolumn{3}{p{0.5\columnwidth}}{\emph{Non-reproducible:} KDD: \cite{tay2018multipointercoattention}, IJCAI: \cite{Nguyen:2018:NNP:3304415.3304640}, \cite{Xuinproceedings}, WWW: \cite{tay2018latentrelationalmetric}, \cite{elkahky2015multiviewdeeplearningcrossdomain}, SIGIR: \cite{manotumruksa2018contextualattention}, \cite{chen2017attentivecomponentlevelattention}, RecSys: \cite{Sun:2018:RKG:3240323.3240361}, \cite{Bharadhwaj:2018:RRG:3240323.3240383}, \cite{Sachdeva:2018:ANA:3240323.3240397}, \cite{Tuan:2017:CNS:3109859.3109900}, \cite{Kim:2016:CMF:2959100.2959165},  \cite{Vasile:2016:MPE:2959100.2959160}, WSDM: \cite{wu2016collaborativeCDAE}}  \\
\end{tabular}
\end{table}

Our first contribution in this work is therefore an analysis of the reproducibility---at least when using our specific practical definition---of research works published on neural collaborative filtering. We generally found that the share of papers that can be reproduced based on the provided source code by the authors is still relatively low. When looking at the statistics over the years, we can observe a certain trend towards authors sharing their source code more often. One possible reason is that reproducibility in general is considered a positive point in the reviewing process, e.g., at KDD.
%\mf{mention which conferences have reproducibility questions in the review process? KDD does IJCAI maybe not}

%XXX Comment from Paolo XXX
%We should point out that \palpha and \pbeta are almost unknown algorithms very seldom used as baselines, despite they are **very simple** %and **effective** and published in **relevant** venues.
%XXX
% DJ: Done in the description of baselines; should also mention this in the final discussion.

\section{Evaluation Methodology}
\label{sec:methodology}
The core of our study was to re-run the experiments reported in the original papers following the original experimental methodology, including additional baseline methods which were systematically fine-tuned just like the newly proposed methods.

To ensure the reproducibility of this study, we share all the data, the source code used for pre-processing, hyperparameter optimization, algorithms, and the evaluation as well as all hyperparameter values and results online.\footnote{\url{https://github.com/MaurizioFD/RecSys2019_DeepLearning_Evaluation}}

\subsection{Measurement Approach}
Our analysis of the relevant papers shows that researchers use all sorts of datasets, evaluation protocols, and metrics in their experiments, see also Section \ref{subsec:methodological-issues}. % \textbf{X}. \mf{table available in Long Paper} % TODO Provide some summary statistics at the end or an overview.
To make our analyses and comparisons as fair as possible, we decided to run our evaluations in exactly the same way as the authors of the originally proposed method did, i.e., using their datasets, their protocol, and their performance metrics. To obtain a broader understanding of the model performance, we also included additional baselines% and measurements \mf{which additional measurements? We have the computational time and the evaluation with all metrics, the latter however I don't think we are using}
.\footnote{An alternative would have been to integrate all methods in one unified framework for evaluation, as done in \cite{Ludewig2018}, and evaluate them on a set of representative datasets. This approach would allow a direct comparison of neural approaches as in \cite{LudewigMauro2019}, which was however not the goal of our work.} %\mf{Another consideration. If they almost all can be outperformed by simple baselines, to what end run a thorough comparison among non-competitive DL algorithms?}}
% TODO (just a comment, DJ): I think such a comparison can make sense to see how the algorithms compare to each other. Also, when doing such a comparison, we could provide a new benchmarking environment for these types of algorithms. where everything is done in a sound manner.

In order to ensure all algorithms are evaluated under the same conditions, we re-factored the original code so that we could include it in our evaluation framework along with the additional baselines.
The core algorithm implementations remained unaltered.
%We made evaluations for those datasets that were also used in the original papers, at least for those that were publicly available or shared by the authors. Whenever available, we also used the data splits that were provided by the authors. When the data splits were not provided, we created the data splits based on the information in the papers.
We evaluated the algorithms using the datasets reported in the original papers, provided that they were either publicly available or shared by the authors. We also used the original train/test split whenever the authors shared it, otherwise we created the data split ourselves following the information provided in the original paper.

Extensive hyperparameter optimization was performed for all examined baselines. For the investigated neural methods, in all but one case we relied on the optimal hyperparameters reported in the original papers. This is appropriate as we used the same datasets and evaluation procedures in the experiments. The only exception is the SpectralCF algorithm (Section \ref{sec:spectralcf}), for which we performed a new hyperparameter optimization ourselves due to an issue with the provided dataset splits, as will be discussed later.
%\mf{We optimized the number of epochs in all of them}
%In some cases, however, we had to optimize the number of training \emph{epochs} also for the neural approaches.
Since the number of epochs and the stopping criteria are usually not described in the original papers, for all machine learning models we select the number of epochs via \emph{early-stopping}, see also Section \ref{sec:earlystopping}.
%Since usually neither the number of training epochs nor indications on the criteria used to stop the training are reported in the original papers and that in a number of implementations available on Github this parameter was inappropriately selected on the test set, we apply \emph{early-stopping} (Section \ref{sec:earlystopping}) on all machine learning models. %Early stopping is a widely used technique to select the optimal number of train epochs based on a validation set and is easily available in many libraries like Keras.\footnote{Early stopping available in Keras \url{https://keras.io/callbacks/\#earlystopping}} The idea of early stopping is to periodically evaluate the model on the validation data while the model is being trained, and stop the training when for a certain number of validation steps the model did not improve over the best solution found so far. Due to how time consuming evaluation steps can be we run 5 train epochs before each validation step. We choose to early-stop the training if for 5 consecutive validation steps the recommendation quality of the model is worse than the best one found so far.

The optimal hyperparameters were selected via a Bayesian search \cite{antenucci2018artist,NIPS2014_5324,Freno:2015:ORM:2783258.2788579}, using the Scikit-Optimize\footnote{\url{https://scikit-optimize.github.io/}} implementation. We considered 50 cases for each algorithm during this search. The first 15 of them were used as initial random points. Once the optimal hyperparameters were determined, including the number of epochs, the final model was fitted on the union of train and validation data using those optimal hyperparameters. The considered hyperparameter ranges and distributions are listed in Appendix \ref{sec:hyperparameter_range}.

%\footnote{In theory, it is not necessary to optimize the hyperparameters for the originally proposed methods, as optimal parameters are usually provided in the papers and optimized for the given evaluation setup.}
%\mf{For the DL algorithms we did optimize only the number of epochs and used the hyperparameters reported in the original paper.
%The only exception is SpectralCF for which we could not use the original hyperparameters due to the data split being wrong.}.

\subsection{Baselines}
Over the last 25 years, a multitude of algorithms of different types were proposed. In order to obtain a picture that is as broad as possible, we selected algorithms of different families for inclusion in our measurements. An overview of all used baselines is given in Table \ref{tab:overview-of-baselines} and the respective hyperparameter ranges are reported in Appendix \ref{sec:hyperparameter_range}.

Our analysis shows that a large variety of other baselines is used in the works we investigated. This variety of baselines (together with the variety of datasets) represents one of the underlying problems that we identified, because it limits the comparability of results across papers. In our work, we therefore benchmark the proposed models against a consistent set of known algorithms of different families.
%The question if the authors of the original works could have further optimized their own baselines is, however, not in the scope of our present work.

\begin{table}[h!t]
\caption{Overview of Baseline Methods}
{\def\arraystretch{1.3}\tabcolsep=5pt
\label{tab:overview-of-baselines}
\begin{tabular}{p{3cm}lp{7.5cm}c}
\toprule
\emph{Family} & \emph{Method} & \emph{Description} \\ \midrule
\multirow{1}{3cm}{Non-personalized}
        & TopPopular & Recommends the most popular items to everyone \cite{Cremonesi:2010:PRA:1864708.1864721}\\ \midrule
\multirow{2}{3cm}{Nearest-Neighbor}
        & UserKNN & User-based k-nearest neighbors \cite{Resnick:1994:GOA:192844.192905} \\ \cline{2-3}
        & ItemKNN & Item-based k-nearest neighbors \cite{sarwar2001item} \\ \midrule
\multirow{2}{3cm}{Graph-based}
         & \palpha & A graph-based method based on random walks \cite{cooper2014P3alpha} \\ \cline{2-3}
         & \pbeta  & An extension of \palpha \cite{paudel2017Rp3beta} \\ \midrule
\multirow{4}{3cm}{Content-Based and Hybrid}
         & ItemKNN-CBF & ItemKNN with content-based similarity \cite{lops2011content}\\ \cline{2-3}
         & ItemKNN-CFCBF & A simple item-based hybrid CBF/CF approach \cite{mobasher2003semantically} \\ \cline{2-3}
         & UserKNN-CBF & UserKNN with content-based similarity \\ \cline{2-3}
         & UserKNN-CFCBF & A simple user-based hybrid CBF/CF approach \\ \midrule
\multirow{4}{3cm}{Non-Neural Machine Learning}
         & \iALS & Matrix factorization for implicit feedback data \cite{hu2008IALS} \\ \cline{2-3}
         & pureSVD  & A basic matrix factorization method \cite{Cremonesi:2010:PRA:1864708.1864721} \\ \cline{2-3}
         & SLIM & A scalable linear model  \cite{ning2011SLIM, levy2013SLIM_ElasticNet}  \\
         \cline{2-3}
         & \EASER  & A recent linear model, similar to auto-encoders \cite{DBLP:conf/www/Steck19} \\
\bottomrule
\end{tabular}
}
\end{table}

%\pc{should we call the "Non-neural machine learning" techniques as "linear machine learning"? The definition of a deep learning technique is "a cascade of multiple non-linear processing layers"}
%\djx{I think nearest-neighbor methods are usually considered to be ``non-linear''. In particular in the context of classification problems the decision boundary is non-linear I would say.}
%\pc{Here I'm referring to SLIM, iALS, EASE-R, which are machine learning methods in which the model is linear (SLIM or EASE-R or linearized (iALS)} \djx{Oh, I see, don't know about iALS exactly. If we call these methods linear, should we provide a very brief explanation as well?}

% DJ: WE do not need the random baseline.
\subsubsection{Popularity-Based Ranking}
Recommending the most popular items to everyone is a common strategy in practice. The method \textbf{TopPopular} implements this non-personalized recommendation approach. The popularity of an item is determined by its number of implicit or explicit ratings in the given dataset.

\subsubsection{Nearest-Neighbor Methods}
Nearest-neighbor techniques were used in the early GroupLens system \cite{Resnick:1994:GOA:192844.192905} and first successful reports of collaborative filtering systems also used nearest-neighbor techniques \cite{Linden2003}. We consider both \emph{user-based} and \emph{item-based} variants, \textbf{UserKNN} and \textbf{ItemKNN}.

Many variants of the basic nearest-neighbor prediction scheme were proposed over the years, see \cite{Breese:1998:EAP:2074094.2074100} for an early performance comparison. In this work, we therefore consider different variations of the nearest-neighbor techniques as well. For both UserKNN and ItemKNN, the following hyperparameters can be set and were optimized in our experiments, their ranges are
reported in Appendix \ref{sec:hyperparameter_range}.%\mf{Style: Instead of describing the range of some hyperparameters in the text, referring every time to the appendix}
\begin{itemize}
\item \emph{Neighborhood Size}: This main parameter determines how many neighbors are considered for prediction. %We tried neighborhood sizes from 5 to 1000.
\item \emph{Similarity Measure}: We made experiments with the Jaccard coefficient \cite{QamarGCL08} as well as Cosine \cite{sarwar2001item}, Asymmetric Cosine \cite{aiolli2013AsymmetricCosine}, Dice-S{\o}rensen \cite{Dice1945} and Tversky \cite{Tversky1977} similarities.
Some of these similarity measures also have their own parameters, as reported in Appendix \ref{sec:hyperparameter_range}, which we optimized as well.
\item \emph{Shrinkage}: As proposed in  \cite{bell2007improved}, we used a parameter (the \emph{shrink term}) to lower the similarity between items that have only few interactions in common. The shrinkage is applied to all similarities.% We tried shrink term values between 0 and 1000.
\item \emph{Feature Weighting}: Using feature weighting for ratings was proposed in \cite{wang2008probabilistic}. In our experiments, we both tested configurations with no weighting and weighting with either the TF-IDF or the BM25 scheme.
\item \emph{Normalization}: This setting determines if we should consider the denominator in the similarity measure as normalization. Only some of the similarity measures % \djx{only some of the ``similarity measures''?}
have this parameter.
\end{itemize}

\subsubsection{Graph-based Methods}
Traditional nearest-neighbor models consider ``direct'' neighborhoods by computing similarities between pairs of objects. Graph-based models can help to overcome this possible limitation relying on a broader interpretation of neighborhoods. In our study, we consider two such graph-based methods called \palpha \cite{cooper2014P3alpha} and \pbeta \cite{paudel2017Rp3beta}. Both methods often lead to good recommendation quality at low computational cost. Interestingly, based on the comparatively limited number of citations they received, these two methods appear to be almost unknown in the community and seldom used as baselines, despite the fact that they are very simple, effective and have been published in top-tier venues.

\begin{itemize}
\item \emph{\palphabold}:
    This method implements a two-steps random walk from users to items and vice-versa, where the probabilities to jump between users and items are computed from the normalized ratings raised to the power of $\alpha$.
    The method is equivalent to a KNN item-based CF algorithm, with the similarity matrix being computed as the dot-product of the probability vectors \cite{cooper2014P3alpha}.
    In addition to what is described in the original algorithm, we normalize each row of the similarity matrix with its $l1$ norm.
    The hyperparameters of the algorithm include the size of the neighborhood and the value for $\alpha$.
\item \emph{\pbetabold}:
    This is an improved version of \palpha proposed in \cite{paudel2017Rp3beta}.
    In \pbeta, each similarity between two items is computed with {\palpha} and divided by the popularity of the items raised to the power of $\beta$.
    Again, we normalize each row of the similarity matrix with its $l1$ norm.
    If $\beta$ is 0, {\pbeta} is equivalent to \palpha.
    The hyperparameters of the algorithm are the size of the neighborhood and the values for $\alpha$ and $\beta$.
\end{itemize}

\subsubsection{Content-based and hybrid Methods}
Some of the neural methods investigated in this paper include side information about items or users.
We have therefore included two simple baselines that make usage of content information.

\begin{itemize}
\item \textbf{ItemKNN-CBF}, \textbf{UserKNN-CBF}:
A neighborhood-based content-based-filtering (CBF) approach, where we compute the item (or user) similarities based on the items' (or user's) content features (attributes) \cite{lops2011content}.
We tested the same set of similarity measures described for the collaborative KNN methods (Jaccard coefficient, Cosine, Asymmetric Cosine, Dice-S{\o}rensen and Tversky similarity).
The hyperparameters are the same as for the ItemKNN and UserKNN methods.

\item \textbf{ItemKNN-CFCBF}, \textbf{UserKNN-CFCBF}:
% TODO: rephrased, please check.
A hybrid algorithm based on item-item (or user-user) similarities and described in \cite{mobasher2003semantically}.
The similarity between items is computed by first concatenating, for each item, the vector of implicit ratings (collaborative features) and the vector of item attributes (content features)
%-- %$\left[ \vec{r}_i \, , w \vec{f}_i \right]$ --
and by later computing the similarity between the concatenated vectors. In case of user-user similarities the algorithm operates in a similar way,
%\djx{Did not know the word ``specularly''. Use another one?}
concatenating the vector of implicit ratings of each user with the user's content feature vector.
The hyperparameters and similarity measures are the same as for ItemKNN, plus a parameter $w$ that controls the relative importance of the content features with respect to the collaborative features.
When $w$ is 0, this algorithm is equivalent to the pure collaborative versions, either ItemKNN or UserKNN.
\end{itemize}

\subsubsection{Non-Neural Machine Learning Approaches}
A wide variety of machine learning models were proposed for \emph{top-n} recommendation tasks in the literature.
In our experiments, we included a number of comparably basic models from the literature as representatives of which methods were often considered the state-of-the-art in pre-neural times.

\begin{itemize}
\item \emph{Matrix Factorization (MF) Techniques}: The application of matrix decomposition methods for collaborative filtering problems was investigated already in the early years of recommender systems \cite{Billsus:1998:LCI:645527.657311}, and became a de-facto standard after the Netflix prize competition (2006-2009). We made experiments with many variants, but will limit our discussion to two main techniques which proved to consistently lead to competitive results among the different MF techniques.
    \begin{itemize}
    \item \textbf{\iALS}: In their seminal work \cite{hu2008IALS}, Hu et al. proposed an \emph{Alternating Least Squares} approach for implicit feedback datasets, which turns implicit feedback signals into confidence values. The authors also proposed a particular optimization method that has the advantage of scaling well on larger datasets.
    A number of hyperparameters can be tuned for the method, including the number of latent factors, the confidence scaling and the regularization factor.
    \item \textbf{PureSVD}: This method corresponds to a basic matrix factorization approach as proposed in \cite{Cremonesi:2010:PRA:1864708.1864721}. To implement PureSVD, we used a standard SVD decomposition method provided in the \small{\texttt{scikit-learn}} \normalsize package for Python.%(\small{\texttt{sklearn.utils.extmath.randomized\_svd}}\normalsize)
    \footnote{\url{https://scikit-learn.org/stable/modules/generated/sklearn.utils.extmath.randomized_svd.html}} The only hyperparameter of this method is the number of latent factors.
    \end{itemize}

\item \emph{Sparse Linear Models (SLIM)}: SLIM was proposed as a well-performing regression-based method for \emph{top-n} recommendation tasks in \cite{ning2011SLIM}. In our work, we use the more scalable variant proposed in \cite{levy2013SLIM_ElasticNet} (\textbf{SLIM ElasticNet}) which learns the item similarity matrix one item at a time (e.g. one column $w$ at a time) by solving a regression problem in such a way that the interactions for the target item $y$ are learned by using all other interactions as training data. To implement \emph{SLIM ElasticNet} we used a standard ElasticNet solver provided in the \small{\texttt{scikit-learn}} \normalsize package for Python.\footnote{\url{https://scikit-learn.org/stable/modules/generated/sklearn.linear_model.ElasticNet.html}}
The hyperparameters of this method include the ratio of \emph{l1} and \emph{l2} regularizations as well as a regularization magnitude coefficient.
\item \emph{\textbf{\EASER}}: In a recent article \cite{DBLP:conf/www/Steck19} the author showed that an ``embarrassingly shallow'' linear model, which shares similarities with an auto-encoder, can produce highly-accurate recommendations that often outperform existing and much more complex techniques. A peculiarity of this model is the existence of a closed-form solution for the training objective which results in very fast training.
The only hyperparameter is the choice of the regularization factor.

This algorithm has been published in 2019 and, as such, the papers covered by our study could not include \EASER as a baseline.
However, we include \EASER to investigate whether shallow auto-encoders are able to provide, on average, more accurate recommendations with respect to complex deep-learning architectures.

\end{itemize}

%\mf{Sometimes the names of baselines are italic sometimes they are not. I can ensure all is uniform but we should choose a format}
%\djx{Let us use the non-italic form.}
% \\emph{([^}]*)}
% \pc{If possible, we should insert a table with the list of hyperparameters and their range (if numerical).} \mf{DONE, Appendix A}

\subsection{Early Stopping Approach}
\label{sec:earlystopping}

Many machine learning models are trained for a \emph{number of epochs} in which the model's performance varies, first increasing and then stabilizing, while usually exhibiting some degree of variance. The number of epochs therefore represents another important parameter to be determined. %, just like any other.
However, it is worth noting that in the articles we have analyzed neither the number of epochs nor the stopping criteria are usually mentioned. The procedure in which this parameter was chosen in the original articles is therefore not clear. Looking at the code shared by the authors we could observe that, in some cases, the number of epochs was inappropriately selected via an evaluation done on the test data, therefore causing information leakage from the test data. In other cases, the reported metric values were inappropriately taken from different epochs. % or the reported metrics were inconsistent as each could be referred to a different epoch.

%In our experiments we find the number of epochs by optimizing the recommendation quality on the validation set.
%Since usually neither the number of training epochs nor indications on the criteria used to stop the training are reported in the original papers and that in a number of implementations available on Github this parameter was inappropriately selected on the test set, we apply \emph{early-stopping} (Section \ref{sec:earlystopping}) on all machine learning models.
%To select the optimal number of epochs we use an \emph{early stopping} approach.
Early stopping is a widely used technique to select the optimal number of train epochs and is available in many libraries like Keras.\footnote{For early stopping in Keras, see \url{https://keras.io/callbacks/\#earlystopping}} The idea of early stopping is to periodically evaluate the model on the validation data, while the model is being trained, and stop the training when for a certain number of validation steps the model quality has not improved over the best solution found so far. Early stopping has the advantage of selecting the number of epochs with a transparent criterion, avoiding arbitrary manual optimization, and often results in shorter training times.

%To that purpose,
To implement early stopping, we use two independent copies of the current model. One is the model that is still being trained, the other is the model frozen at the epoch with the best recommendation quality found so far. %The first model is trained for a given number of epochs, followed by an evaluation step using the validation set.
If the trained model, after further epochs, exhibits better recommendation quality than the best one found so far, the best model is updated. %In case the recommendation quality of the model does not improve over the best one for a certain number of epochs, training is stopped before a pre-specified maximum number of epochs is reached.
%Early stopping is a widely used technique to select the optimal number of train epochs based on a validation set and is easily available in many libraries like Keras.\footnote{Early stopping available in Keras \url{https://keras.io/callbacks/\#earlystopping}} The idea of early stopping is to periodically evaluate the model on the validation data while the model is being trained, and stop the training when for a certain number of validation steps the model did not improve over the best solution found so far. Due to how time consuming evaluation steps can be we run 5 train epochs before each validation step. We choose to early-stop the training if for 5 consecutive validation steps the recommendation quality of the model is worse than the best one found so far.
Since the evaluation step is time consuming, we run five train epochs before each validation step. Moreover, we choose to stop the training if for 5 consecutive validation steps the recommendation quality of the current model is worse than the best one found so far.
%This early stopping criterion is also adopted in the Keras framework\footnote{\url{https://keras.io/callbacks/\#earlystopping}}.
%Note that this rule is not as strict as the one applied in \cite{he2017neuralfactorizationmachines}, where in the publicly available implementation it can be obsevered that the training stops only if for 5 consecutive epochs the recommendation quality on the validation set is monotonically getting worse.

\section{Results -- Analysis of Reproducibility and Progress}
\label{sec:results}
%\mf{Analysis of reproducibility and progress (?)}

In this section, we summarize the main observations of our experiments. For each analyzed paper, we describe the basic idea of the method and the summary of our results. A more detailed analysis is reported in Appendix \ref{sec:detailed_analysis}, where we describe
\begin{enumerate*}[label=\textit{(\roman*)}]
% \item the basic idea of the method;
\item the baseline algorithms and datasets that were used in the original paper;
\item the outcomes reported in the original work;
\item our results after including and optimizing additional baselines.
\end{enumerate*}

%\djx{I think it is good to report that computing was heavy. We should probably be clearer what we mean by ``result'', ``measurement'', and ``experiment''. What do the ``900 results'' refer to? What are the 41000 experiments? In one view, we could say the entire paper is one experiment. Not sure if we should have the detailed information in the footnote. Could go into the main text as well. As a reader, I would immediately look what the 253 days refer to. But maybe it is okay.}
%The experimental evaluation reported in this paper required significant computational effort. We report more than 900 results across the considered algorithms, datasets and experimental conditions. When taking into account the hyperparameter tuning procedure, the total number of experiments raises to 41,000, corresponding to a computation time of 253 days.\footnote{The computation time refers to the total instance time for one AWS instance p3.2xlarge, with 8 vCPU, 30GB RAM, and one Tesla V100-SXM2-16GB GPU. The detailed measurements are available in the online material (Section \ref{sec:methodology})}

\begin{figure}[h!t]
% \includesvg[width=\textwidth]{figure/overview-neural-methods-2.svg}
\includegraphics[width=0.9\linewidth]{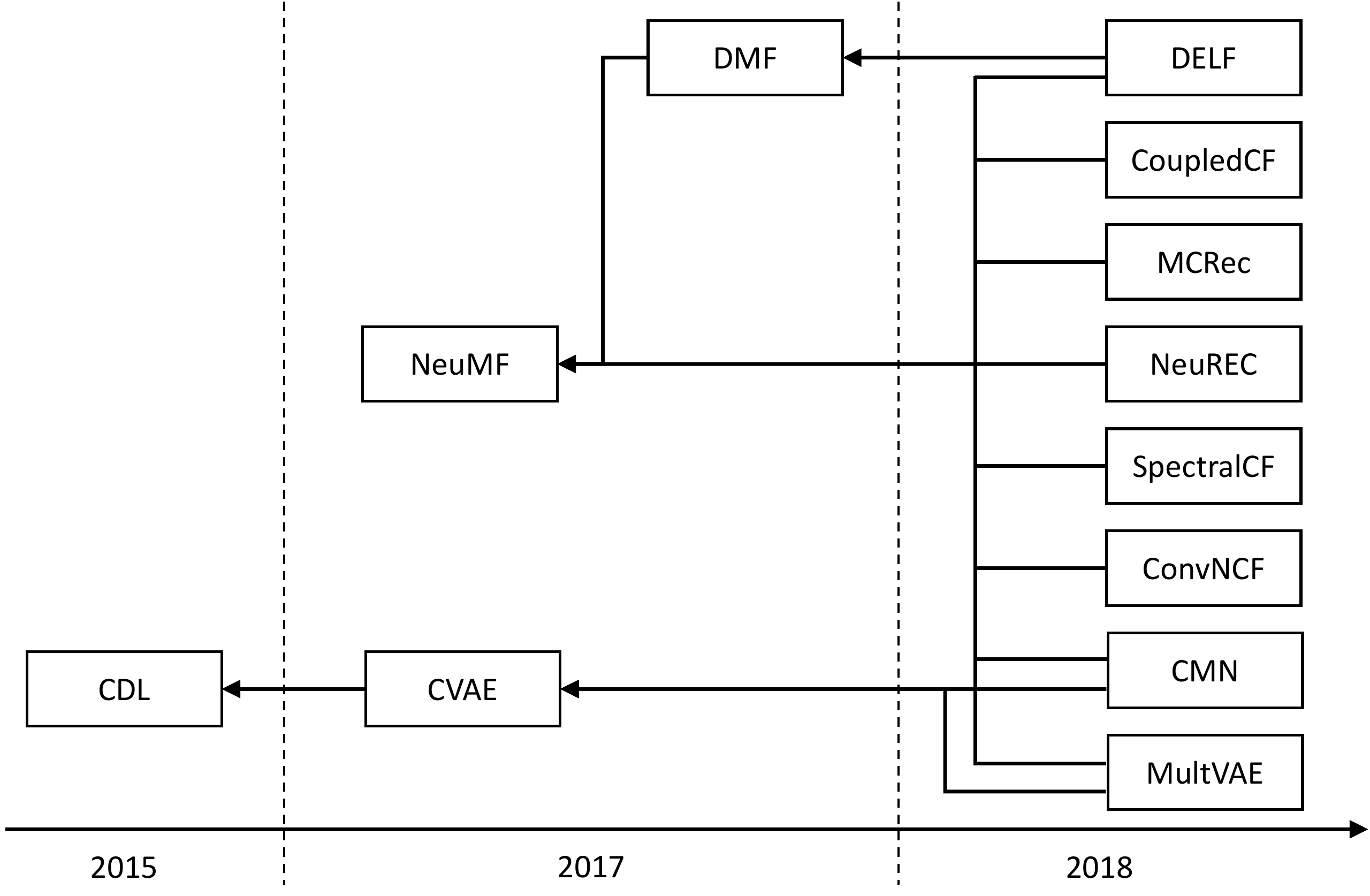}
\caption{Overview of Neural Methods, arrows indicate when a newer method used another one as baseline in the experiments.}
\label{fig:overview-neural-methods}
\end{figure}

The analyzed papers were published between 2015 and 2018. We organize the discussion of the papers mostly by year of publication. The list of considered papers is shown in Table \ref{tab:overview-of-compared-methods}.
% Table \ref{tab:overview-of-results} summarizes the main findings.
An overview of the temporal development and the dependencies between the approaches can be found in Figure \ref{fig:overview-neural-methods}.

 \begin{longtable}[h!t]{p{7cm}l}
 \caption{} \\
 \label{tab:overview-of-compared-methods}
 \small
 %\small
 %\begin{tabular}{p{4cm}lp{7.3cm}}
 % paper title, conference/year, overview of results
 %\toprule
 %\hline
 \textbf{Paper} & \textbf{Conference} \\ \hline
 Collaborative Deep Learning for Recommender Systems (CDL) \cite{wang2015collaborativedeeplearning}\footnote{\url{https://github.com/js05212/CDL}} & KDD '15  \\  \hline
 Collaborative Variational Autoencoder for Recommender Systems \cite{li2017collaborativevariationalautoencoder}\footnote{\url{https://github.com/eelxpeng/CollaborativeVAE}} & KDD '17  \\ \hline
 Neural Collaborative Filtering \cite{he2017neural}\footnote{\url{https://github.com/hexiangnan/neural_collaborative_filtering}} & WWW '17  \\  \hline
 Deep Matrix Factorization Models for Recommender Systems \cite{Xue2017DeepMF}\footnote{The source code was not publicly available but the authors shared it with us upon request.} & IJCAI '17 \\  \hline
 Variational Autoencoders for Collaborative Filtering \cite{liang2018variationalautoencodersforCF}\footnote{\url{https://github.com/dawenl/vae_cf}} & WWW '18 \\ \hline
 NeuRec: On Nonlinear Transformation for Personalized Ranking \cite{Zhang2018Neurec}\footnote{\url{https://github.com/cheungdaven/NeuRec}} & IJCAI '18  \\ \hline
 CoupledCF: Learning Explicit and Implicit User-item Couplings in Recommendation for Deep Collaborative Filtering \cite{Zhang2018CoupledCF}\footnote{\url{https://github.com/zhqgui/CoupledCF}} & IJCAI '18
 \\ \hline
 DELF: A Dual-Embedding based Deep Latent Factor Model for Recommendation \cite{Cheng2018DELFAD}\footnote{The source code was not publicly available but the authors shared it with us upon request.} & IJCAI '18 \\ \hline
 Outer Product-based Neural Collaborative Filtering \cite{He:2018:OPN:3304889.3304969}\footnote{\url{https://github.com/duxy-me/ConvNCF}} & IJCAI '18  \\ \hline
 Leveraging Meta-path based Context for \emph{top-n} Recommendation with a Neural Co-attention Model \cite{hu2018leveragingmetapathcontext}\footnote{\url{https://github.com/librahu/MCRec}} & KDD '18  \\ \hline
 Collaborative Memory Network for Recommendation Systems \cite{ebesu2018collaborative}\footnote{\url{https://github.com/tebesu/CollaborativeMemoryNetwork}} & SIGIR '18\\ \hline
 Spectral Collaborative Filtering \cite{Zheng:2018:SCF:3240323.3240343}\footnote{\url{https://github.com/lzheng21/SpectralCF}} & RecSys '18\\ \hline
 %\bottomrule
 %\end{tabular}
 \end{longtable}

% TODO: make names of algorithms consistent and improve the figure.

The experimental evaluation reported in this paper required significant computational effort. %Considering all baselines and reproducible deep learning algorithms, due to the number of different datasets and pre-processing procedures, we report the recommendation quality of more than 900 models.
%\djx{Shortened a sentence here.}
In total, we report the recommendation quality of more than 900 trained models (counting both baselines and reproducible deep learning algorithms). When taking into account the hyperparameter tuning procedure, 41,000 models were fitted, corresponding to a total computation time of 253 days.\footnote{The computation time refers to the total instance time for one AWS instance p3.2xlarge, with 8 vCPU, 30GB RAM, and one Tesla V100-SXM2-16GB GPU. The detailed measurements are available in the online material.} % (Section \ref{sec:methodology})}

% ---------------------------------------------------------------------------------
\subsection{Collaborative Deep Learning for Recommender Systems (CDL)}
% \label{sec:cdl}
% ---------------------------------------------------------------------------------
\paragraph{Method}
\emph{CDL} is the earliest method in our analysis, published at KDD '15 \cite{wang2015collaborativedeeplearning}\footnote{\url{https://github.com/js05212/CDL}}. CDL is a hybrid method which applies deep learning to jointly learn a deep representation of content information and collaborative information. Technically, it is a probabilistic feed-forward model for joint learning of a stacked denoising autoencoder and collaborative filtering.

\paragraph{Result summary}
According to our experiments, simple hybrid baselines outperform CDL in three out of four datasets. On a dense dataset, CDL is also outperformed by pure collaborative baselines with recommendation lists shorter than 100 items. CDL is only better than our baselines on one small and very sparse dataset with only one interaction per user in the training set.
From a methodological perspective, there is no indication of why comparably long list lengths, up to 300, were used for evaluation in the paper and why no measurements were reported for list lengths below 50, which is commonly the case in the literature.

% ---------------------------------------------------------------------------------
\subsection{Collaborative Variational Autoencoder (CVAE)}
% \label{sec:cvae}
% ---------------------------------------------------------------------------------
\paragraph{Method}
Like CDL, the \emph{CVAE} method \cite{li2017collaborativevariationalautoencoder}\footnote{\url{https://github.com/eelxpeng/CollaborativeVAE}} is a hybrid technique that relies both on content information and collaborative features for recommending. The work was published at KDD '17. Technically, the model learns deep latent representations from content data in an unsupervised manner and also considers implicit relationships between items and users from both content and ratings. Unlike previous works proposing denoising autoencoders,
%\djx{Can we rephrase ``with denoising criteria''? Unclear to me. ``that use/implement denoising criteria''?}
CVAE learns a latent distribution for content in the latent space instead of the observation space through an inference network.

\paragraph{Result summary}
While CVAE, according to the original experiments, outperforms previous neural methods including CDL \cite{wang2015collaborativedeeplearning}, our experiments indicate these neural methods were not necessarily strong baselines. Our own experiments confirm the improvements of CVAE over the previous CDL method. However, similarly to CDL, simple hybrid methods outperformed CVAE in three out of four datasets. CVAE is instead superior in one very sparse dataset that had only one interaction per user in the training set.

% ---------------------------------------------------------------------------------
\subsection{Neural Collaborative Filtering (NCF)}
% \label{sec:ncf}
% ---------------------------------------------------------------------------------
\paragraph{Method}
The \emph{Neural network-based Collaborative Filtering} (\emph{NCF}) \cite{he2017neural}\footnote{\url{https://github.com/hexiangnan/neural_collaborative_filtering}} framework was presented at WWW '17 and rapidly became very influential, being used as a baseline for most later neural recommendation approaches, as shown in Figure \ref{fig:overview-neural-methods}.
The framework generalizes matrix factorization in a way that the commonly used inner product is replaced by a neural architecture, which can learn different types of functions from the data and therefore can also model non-linearities. Different variants are considered in the paper: Generalized Matrix Factorization, Multi-Layer Perceptron, and Neural Matrix Factorization, where the last one, called NeuMF is an ensemble of the other two.  In our evaluation, we only consider Neural Matrix Factorization, because this method led to the best results.

\paragraph{Result summary}
In our experiments we could observe that, for one out of two datasets, nearest-neighbor methods outperform NCF on all measurements. For the other dataset, NFC was better than the nearest neighbor techniques, but not better than non-neural machine learning methods. Regarding methodological issues, we observed in the original source code that the number of training epochs, which should be optimized on the validation set, was erroneously optimized on the test data.\footnote{In addition, Rendle et al. \cite{ncfvsmf2020} confirmed that NCF is not able to consistently outperform simple non-neural baselines and  experimentally showed that learning a dot product with a multilayer perceptron is not trivial.}

% ---------------------------------------------------------------------------------
\subsection{Deep Matrix Factorization (DMF)}
% \label{sec:dmf}
% ---------------------------------------------------------------------------------
\paragraph{Method}
\emph{Deep Matrix Factorization Models} (\emph{DMF}) were proposed at IJCAI '17 \cite{Xue2017DeepMF}\footnote{The source code was not publicly available but the authors shared it with us upon request.}.
% \djx{The footnote says that the source code is not publicly available. What does this mean - did they share the code only with us? If so, we should mention this}
As an input to their model, the authors first build a user-item matrix from explicit ratings and implicit feedback, which is then used by a deep \emph{structure learning} architecture. One key aspect here is that a common low-dimensional space for representing users and items is used.
Furthermore, the authors develop a new loss function based on cross entropy that considers both implicit feedback and explicit ratings.

\paragraph{Result summary}
The proposed method is outperformed on three out of four datasets by our long-known baselines. Only for one very sparse dataset, the proposed method was better than our baseline methods, in particular with respect to the Hit Rate\footnote{The Hit Rate measures if a single hidden true positive was part of the top-n recommendation list, see, e.g., \cite{Zhang2018CoupledCF} for a formal definition. The metric is similar to Recall.}.
Regarding methodological aspects, we observed in the source code that the Hit Rate and NDCG results are reported as the best value obtained by evaluating continuously on the test data during the training phase, regardless of the epoch. In our experiments we report the results associated to the epoch selected via early-stopping on the validation data.

% ---------------------------------------------------------
\subsection{Variational Autoencoders for Collaborative Filtering (Mult-VAE)}
% \label{sec:vae}
% ---------------------------------------------------------
\paragraph{Method}
In \cite{liang2018variationalautoencodersforCF}\footnote{\url{https://github.com/dawenl/vae_cf}}, the authors propose a collaborative filtering method for recommendation based on implicit feedback using variational autoencoders. The method is called \emph{Mult-VAE} and was presented at WWW '18. Technically, the paper introduces a generative model with multinomial likelihood and a different regularization parameter for the learning objective, and uses Bayesian inference for parameter estimation. The authors furthermore show that there is an efficient way to tune the parameter using annealing.

\paragraph{Result summary}
We could reproduce the results reported in the original paper for two out of three datasets and found that the proposed method outperforms all neighborhood-based baselines on all metrics by a large margin. The algorithm also outperforms all non-neural machine-learning baselines on one dataset. For the second dataset, however, the SLIM method was better when the optimization target was the same as the evaluation measure (NDCG).

% ------------------------------------------------------------------------
\subsection{NeuRec: On Nonlinear Transformation for Personalized Ranking}
% \label{sec:neurec}
% ------------------------------------------------------------------------
\paragraph{Method}
\emph{NeuRec} \cite{Zhang2018Neurec}\footnote{\url{https://github.com/cheungdaven/NeuRec}} was presented at IJCAI '18. The work aims at learning user-item relationships from implicit feedback and combines latent factor models with neural networks in order to capture both linear and non-linear dependencies in the data. Technically, the user-item interaction matrix is first mapped into a low-dimensional space with multi-layered networks. Recommendations are then generated by computing the inner product of item and user latent factors. A user-based and an item-based variant are proposed.

\paragraph{Result summary}
Even though the authors published a runnable implementation of their method and provided detailed information on the hyperparameters, we could not obtain the results reported in the original paper.
\blue{We contacted the authors but even with their help the reason for this differences could not be clarified.
In our experiments,} the method is consistently outperformed by many of our baselines on three out of four commonly used datasets. Only on one small dataset and for one individual measure at a short list length the proposed method is slightly better than our baselines.
Regarding methodological aspects, we found again that the number of epochs was erroneously optimized on the test data: based on the provided source code, the best results of NeuRec are reported for different measures at potentially different training epochs. In our experiments we report the results associated to the epoch selected via early-stopping on the validation data.

% ------------------------------------------------------------------------
\subsection{CoupledCF: Learning Explicit and Implicit User-item Couplings}
% \label{sec:coupledCF}
% ------------------------------------------------------------------------
\paragraph{Method}
\emph{CoupledCF} \cite{Zhang2018CoupledCF}\footnote{\url{https://github.com/zhqgui/CoupledCF}} was also presented at IJCAI '18. The approach is based on the observation that users and items in real-world datasets are not independent and identically distributed. The proposed method therefore aims to learn implicit and explicit couplings between users and items and to thereby leverage available side information (e.g., user demographics, item features) more effectively. Technically, a complex architecture is used, involving a CNN for learning the couplings based on the side information and a deep CF model that considers explicit and implicit interactions between users and items.\footnote{\label{fnote:cnn_on_embeddings}The claim that the outer product of embeddings is equivalent to an image and that the CNN allows to model embedding correlations has been later questioned by Ferrari Dacrema et al. \cite{FerrariDacrema2020convolution}.
Similar considerations apply for ConvNCF, discussed in Section \ref{sec:convcf}.
}

\paragraph{Result summary}
For one out of two datasets, the non-deep machine learning models were consistently better than the proposed method. For the second dataset, nearest-neighbor methods were preferable. Regarding methodological aspects, from the available source code we could observe that the number of training epochs was selected using the test data and that the provided data splits exhibits inconsistencies with the splitting procedure described in the paper. Furthermore, as described in the original paper, the default hyperparameters for some baselines are used, meaning they were not properly optimized.

\subsection{DELF: A Dual-Embedding based Deep Latent Factor Model for Recommendation}
% \label{sec:delf}
% ------------------------------------------------------------------------
\paragraph{Method}
The \emph{DELF} model \cite{Cheng2018DELFAD}\footnote{The source code was not publicly available but the authors shared it with us upon request.}, presented at IJCAI '18, was designed for \emph{top-n} recommendation tasks given implicit feedback data.
Inspired by previous work (NSVD) \cite{Paterek2007ImprovingRS}, the authors propose to learn \emph{dual} embeddings to capture certain interactions in the data.
Instead of using only the common user embedding, the authors propose to learn an additional item-based user embedding and vice versa for item embeddings. The embeddings are then combined to model non-linear interactions between users and items within a deep learning architecture. Through this approach the authors  generalize ideas of NSVD and Neural Collaborative Filtering (NCF). Two variants of the approach, \emph{DELF-MLP} and \emph{DELF-EF} were investigated in the original paper.

\paragraph{Result summary}

% \djx{Revised this paragraph. I felt that the main message regarding the outcomes was lost.}
Two datasets, \emph{MovieLens} and \emph{Amazon Music}, were used for the evaluation. On the movies dataset, \emph{DELF} was consistently outperformed by our machine learning baselines. The performance of \emph{DELF} on the music dataset was, however, substantially better than all baselines. An investigation of this phenomenon revealed that when using the \emph{Amazon Music} with a time-based splitting criterion, more than half of the test items never appeared in the training set, which is an uncommon setup for evaluating pure collaborative filtering approaches. The relatively high accuracy of \emph{DELF} therefore stems from its tendency to push cold items---for which it could not have learned anything in the training phase---to the top end of the recommendation lists. An additional experiment in which cold items are not considered in the evaluation---which is a more suitable setup when evaluating pure collaborative filtering methods---shows that the performance of \emph{DELF} again drops below that of the machine learning baselines. Looking at methodological aspects, like in other works discussed here, the authors did not optimize the number of epochs on the validation set but took the best values obtained when testing.

%A problem was identified with one of the dataset which contained a substantial number of cold items, despite the DELF model being a pure collaborative algorithms and, therefore, unable to model cold items meaningfully. Further analysis showed the competitiveness of DELF in that datasets was due to its tendency to frequently recommend cold items. After the removal of this anomaly, the method was consistently outperformed by an established implicit-feedback baseline (iALS). For one of the datasets, several existing baselines were always better than the proposed method. Looking at methodological aspects, like in other works discussed here, the authors did not optimize the number of epochs on the validation set but took the best values using the test.

% ------------------------------------------------------------------------

% ------------------------------------------------------------------------
\subsection{Outer Product-based Neural Collaborative Filtering (ConvNCF)}
 \label{sec:convcf}
% ------------------------------------------------------------------------
\paragraph{Method}
The \emph{ConvNCF} method \cite{He:2018:OPN:3304889.3304969}\footnote{\url{https://github.com/duxy-me/ConvNCF}} was presented at IJCAI '18. Its main idea is to explicitly model the pairwise correlations between the dimensions of the embedding using an outer product. With this technique, the authors aim to create an \emph{interaction map}, which is more expressive than existing methods that use simple concatenations of embeddings or element-wise products.\textsuperscript{\ref{fnote:cnn_on_embeddings}}

\paragraph{Result summary}
Two datasets were used for the evaluation, traditional nearest-neighbor baselines are consistently better than the proposed method on one dataset and better except for one measurement on the other.
Regarding methodological aspects, based on the provided
source code the number of epochs, as in other papers, was determined on the test data. Furthermore, the authors decided to set the embedding size to the constant value of 64 for all baselines. However, the embedding size is a hyperparameter to be tuned for each dataset and for each embedding-based algorithm, as different models with different objective functions or training procedures will likely require different values for it.

% ------------------------------------------------------------------------
\subsection{Leveraging Meta-path based Context (MCRec)}
% \label{sec:mcrec}
% ------------------------------------------------------------------------
\paragraph{Method}
The \emph{MCRec} \cite{hu2018leveragingmetapathcontext}\footnote{\url{https://github.com/librahu/MCRec}} method was published at KDD '18. It is a hybrid method that uses side information about the recommendable items in the recommendation process. The side information is represented as a network structure, and meta-paths are relation sequences that connect objects in this graph. Technically, the authors use a priority-based sampling technique to select more informative paths instances and a novel co-attention mechanism to improve the representations of meta-path based context, users and items.

\paragraph{Result summary}
The authors provided an implementation which had the meta-paths hard-coded, and we therefore could reliably reproduce the results only for the small MovieLens dataset. For this dataset, however, it turned out that the traditional item-based nearest-neighbor technique was better than MCRec on all performance measures.% (Precision, Recall, and NDCG).

In the context of this work, additional problems were identified based on the provided source code.
For example, the accuracy metrics reported by the source code correspond to the maximum values that are obtained across different epochs when evaluating on the test data.
Furthermore, in the original article, the hyperparameters of the examined baselines were said to be taken from the original papers and not optimized for the datasets used in the evaluation.
Finally, the NDCG metric was implemented in an uncommon way.

% ------------------------------------------------------------------------
\subsection{Collaborative Memory Network for Recommendation System (CMN)}
% \label{sec:cmn}

% ------------------------------------------------------------------------
\paragraph{Method}
Collaborative Memory Networks (CMN) \cite{ebesu2018collaborative}\footnote{\url{https://github.com/tebesu/CollaborativeMemoryNetwork}} was presented at SIGIR '18, and it represents a collaborative-filtering approach based on memory networks and neural attention mechanisms. The underlying idea of the approach is to combine latent factor with neighborhood (memory-based) approaches in a non-linear fashion.

\paragraph{Result summary}
For two datasets, out of three, the recent CMN method was outperformed by most of the simpler baselines. On the third dataset, CMN was much better than our personalized baselines, but this dataset has such a skewed distribution that recommending the most popular items to everyone was by far the most effective method in this evaluation. The relatively good performance of CMN in this setting is therefore attributed to the higher popularity bias of CMN.

% ------------------------------------------------------------------------
\subsection{Spectral Collaborative Filtering (SpectralCF)}
% \label{sec:spectralcf}
% ------------------------------------------------------------------------
\paragraph{Method}
\emph{SpectralCF}, presented at RecSys '18 \cite{Zheng:2018:SCF:3240323.3240343}\footnote{\url{https://github.com/lzheng21/SpectralCF}} is a graph-based approach. Users and items are represented as a bipartite graph.
The novelty of this method is a convolution approach which operates on the \emph{spectral domain}.
The method considers both proximity and connectivity information in the graph, which is assumed to be particularly helpful for cold-start problems.

\paragraph{Result summary}
In our initial experiments, we found that the algorithm was competitive with our baselines only for one of three datasets. Specifically, it was the dataset for which the authors shared the train-test split. An investigation revealed the distribution of the data in the provided test set was very different from what we would likely obtain by applying a random sampling procedure. After creating the train-test splits by our own, we found that SpectralCF does not work as expected and consistently exhibits lower performance when compared with personalized and even non-personalized baselines.

\begin{figure}[h!t]
\begin{subfigure}{.48\textwidth}
  \centering
%   \includegraphics[width=.8\linewidth]{image1}
    % \resizebox{\linewidth}{!}{%
    % \input{figure/SpectralCF_movielens1m_original_popularity_plot.pdf}
    % }
    \includegraphics[width=\linewidth]{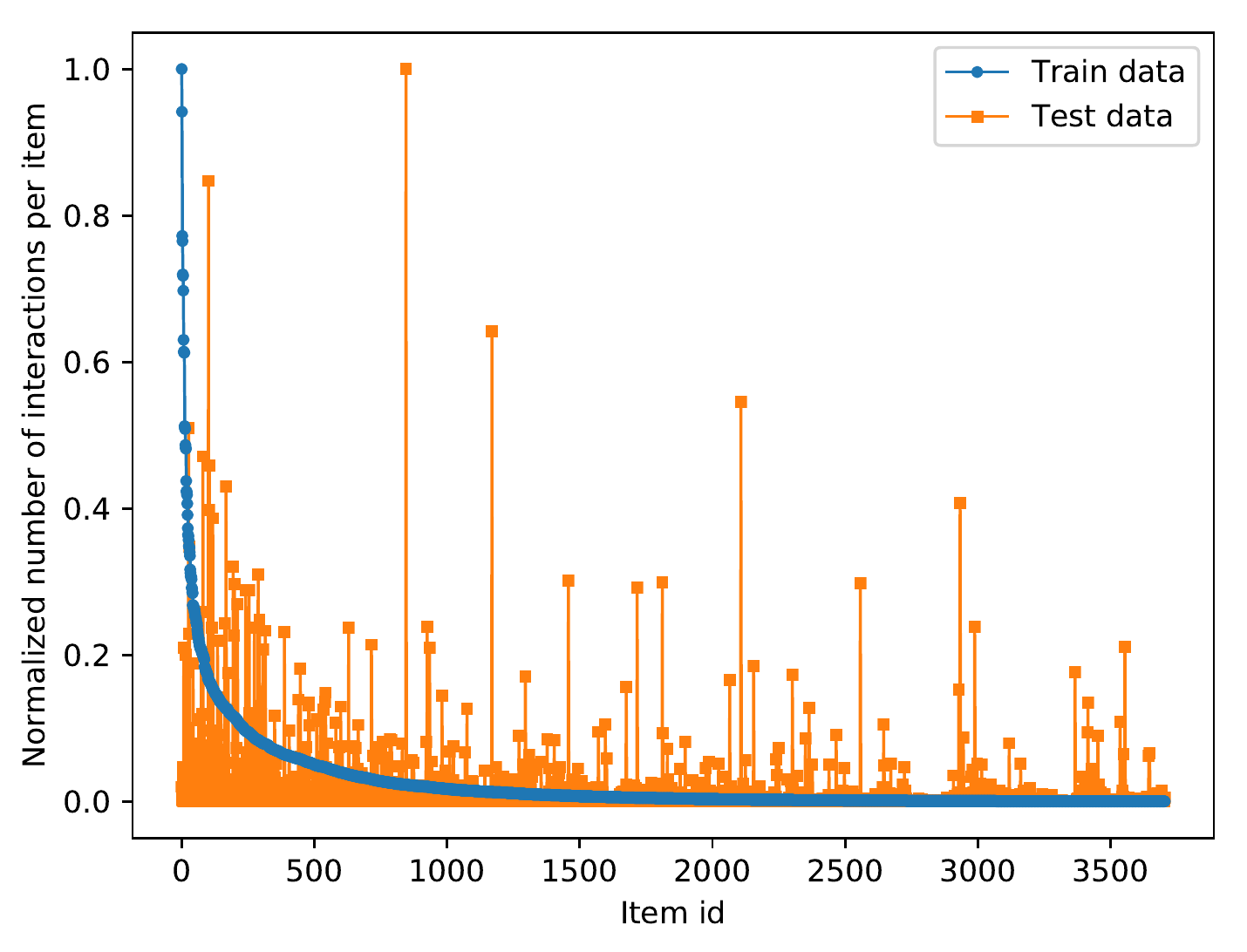}
    \caption{Normalized popularity distributions of the train and test splits provided by the original authors.}
    \label{fig:spectralcf-observed-distribution}
\end{subfigure}%
\hfill
\begin{subfigure}{.48\textwidth}
  \centering
\includegraphics[width=\linewidth]{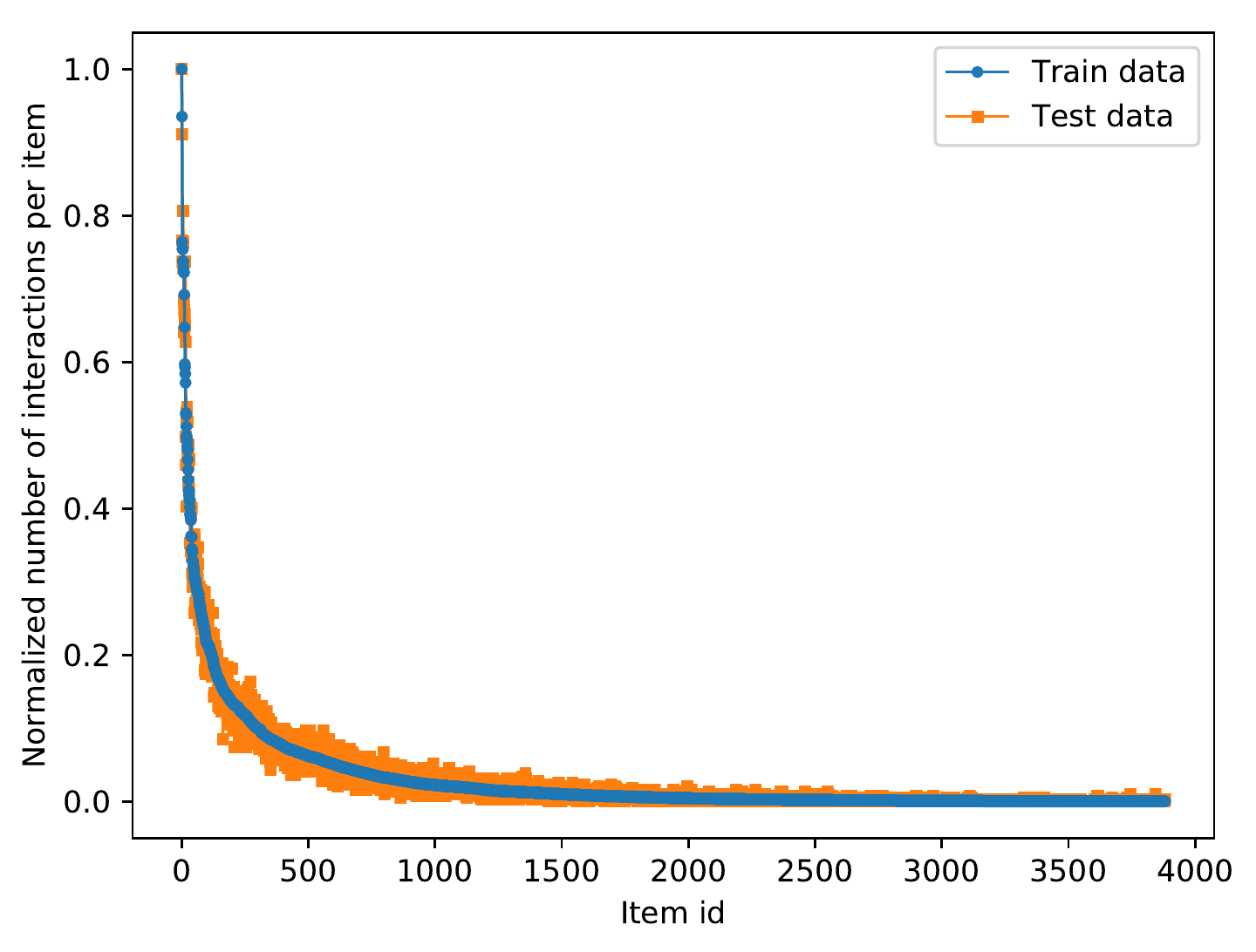}
    \caption{Normalized popularity distributions of the train and test splits generated by us.}
    \label{fig:spectralcf-expected-distribution}
\end{subfigure}
\label{fig:spectralcf-comparison-distribution}
\caption{Normalized popularity distributions of the train and test splits for SpectralCF, the value 1 corresponds to the most popular item in that split. For a random split, as can be seen in Figure \ref{fig:spectralcf-expected-distribution}, the normalized values of both splits are, on average, similar. In the split provided by the original authors, however, as can be seen in Figure \ref{fig:spectralcf-observed-distribution}, train and test data have quite different distributions.
}
\end{figure}

To illustrate the data splitting problem, we compared the popularity distribution of the items in the training and the test split in the provided data, see Figure \ref{fig:spectralcf-expected-distribution}. The figure plots the normalized popularity (i.e., the popularity of an item over the popularity of the most popular item) for both training and test items.
Items are ranked based on their popularity in the training set, with the most popular training items being on the left.
In case of a true random split, the normalized popularity values of the items in the training and the test split should be relatively close.
However, the figure shows the split provided by the authors has some major deviations.
Figure \ref{fig:spectralcf-expected-distribution} shows the popularity distributions of our random split, which are almost identical between training and test sets.
%a major issue regarding the evaluation was spotted since the provided train-test split have a very different popularity distribution from what would be the likely result of a random sampling procedure. After correction, the proposed method does not work as expected, leading to non-competitive performance against both personalized and non-personalized baselines.

\section{Discussion}
\label{sec:discussion}
Our work indicates that rather limited progress was made through deep learning approaches applied to the classic \emph{top-n} recommendation problem, at least when considering the papers that were analyzed in the present work and which were published at top-level scientific conferences.
While many papers claiming to make advances over the state of the art were published, these mostly seemed to amount to phantom progress.
In this section, we will review the possible causes for the apparent stagnation in this field. We hope that an increased awareness of these causes can help ensuring progress in the future and lead to the development of even more powerful top-n recommendation approaches.
We will structure our discussions along three dimensions: reproducibility, methodology, and fundamental considerations. Please note that almost all discussed issues are not tied to deep learning techniques, and our paper should therefore not be understood as a criticism of deep learning in general or in recommender systems research.

\subsection{Reproducibility}
The reproducibility statistics for the conferences we analyzed in this study are reported in Table \ref{tab:reproducibility-stats}. Using our specific criteria, we found that 12 of 26 papers could be reproduced based on the provided code and publicly available data sets. In recent years, we could observe an increased trend for researchers to share the source code of their algorithms and experiments, probably due to the increased importance of reproducibility as a criterion in the academic peer-reviewing process. Nonetheless, we found that for more than half of the papers the source code was either not provided or it was lacking important details to the extent that a reliable reproduction of the results was made difficult or impossible.

The reasons for not sharing the code, not even in private with other researchers, sometimes remain obscure.
Technically, a scientific paper should contain all necessary information for others to reproduce the work. Nonetheless, providing the artifacts (code, data sets) that were used in the experiments is important, as
%there are sometimes details which are not reported in the papers
details may sometimes be missing from the papers, e.g., for space reasons,
%but can have a significant effect on the observed results.
that could have a significant impact on the observed results.
% TODO: maybe add a reference or provide an example.
Generally, however, establishing reproducibility in the context of algorithmic proposals, as discussed in our paper, is relatively easy compared to other disciplines. Nowadays, applied research in deep learning is mostly based on public libraries and freely accessible software tools. Virtualization technology can also be leveraged effectively to make the entire software environment used for the experiments available to the community. Furthermore, while many algorithms have some non-deterministic elements, the overall performance is usually quite stable over several runs, which should allow other researchers to obtain at least comparable results when using the artifacts of the original authors.

Sometimes, researchers argue that they cannot provide the code because of the rules of the organization employing them.
% This, however, seems contradictory.
% Once a method is proposed and published in a scientific paper, others could in principle use it as-is, without restrictions.
% The competitive gain of organizations therefore seems limited.
%As such, the intellectual property of organizations can be better protected by other means (e.g., patents).
In some cases, the data used in the experiments is also not made available to the community, which means that no one will ever have the chance to verify, and possibly contest or falsify, the claimed results under the exact same conditions. This puts  academic researchers in a certain dilemma. The scientific community is highly interested in both real-world problems and solutions that work in practice. From a scientific perspective, however, reproducibility and verifiability (or falsifiability) are the main guiding principles of scientific research.

In the context of our study, we contacted the authors of papers where the artifacts were not publicly available. While some of the authors responded to our inquiries and shared their code with us, we found that in the great majority of cases (10 out of 14 for non reproducible papers) we received no reply.
%\pc{ add some statistics .} \mf{done}
We can only speculate about the reasons for these phenomena.
Clearly, providing code for others can lead to a substantial amount of extra work for the researcher. Furthermore, if reviewers do not put much emphasis on this, there might be not enough incentives for the researcher to go the extra mile \cite{collberg2016repeatability,stodden2010scientific}. %, even though this  % even though citations might be more ..
At the same time, publishing all artifacts can make certain assumptions or limitations of a method more obvious, e.g., when the scalability of a method is limited. In general, however, it should be in the interest of the community, and hence of researchers themselves, that others can reproduce, confirm or falsify their results, as this is a cornerstone of scientific research in the first place. For a discussion of the role of reproducibility as a scientific principle in other areas of computer science, see, e.g., \cite{freire_et_al:DR:2016:5817}.

%\mf{from Dietmar: One thing I think we should add somewhere as a footnote, is about our communication with the authors of NeuMF. I am thinking that we can mention that we were in contact with them, and that they reported that they could achieved slightly improved results on one dataset when using a different architecture, hyperparameter ranges, and optimization procedure. We can then say that we in general do not claim that improved results are not possible for the investigated methods, but finding configurations that are better than those reported in the published methods is not in the goal of our work.  }
%\todo[inline]{DJ: Could add guidelines here: refer to \url{https://ai.facebook.com/blog/new-code-completeness-checklist-and-reproducibility-updates}}

% also cover scalability as a subtopic
\subsection{Methodological Issues}
\label{subsec:methodological-issues}
% this paper about time-series things. choice of baselines
Throughout the paper, we reported a number of methodological issues that contributed to the limited progress we have observed in the analyzed articles. Here, we summarize our main observations.

\begin{itemize}
\item \emph{Choice of baselines:} Algorithms for \emph{top-n} recommendation problems have been investigated for decades, and now it is less than clear what represents the state-of-the-art. Determining the state-of-the-art can be difficult or even impossible due to the fact that there exist no general ``best'' algorithm. In fact, performance rankings of algorithms depend on a variety of factors, including the dataset characteristics or the evaluation procedure and measure. Another observation is that often only complex machine learning methods are considered as baselines. Comparably simple methods like \palpha and \pbeta are not widely known in the community, even though they were published at top-level venues and often lead to strong results.
\item \emph{Propagation of weak baselines:} Nowadays, methods like NCF are often considered as competitive state-of-the-art baselines, even though our analysis showed that they are often not better than relatively simple and well-known techniques.
\item \emph{Lack of proper tuning of baselines:} This is probably the most striking observation of our analysis and is not specifically tied to deep learning approaches \cite{DBLP:journals/corr/abs-1905-01395} or to recommendation problems \cite{Lin:2019:NHC:3308774.3308781,lin2019recantation}. Researchers apparently invest significant efforts in optimizing their own new method but do not pay the same attention to their baselines. Sometimes, authors simply pick the hyperparameter settings reported to be optimal from a previous paper, even though those may refer to a different dataset or experimental procedure.
Probably, this behavior might be the result of a \emph{confirmation bias}, i.e., the tendency to search for results that affirm rather than refute prior research hypotheses.
% than of researchers deliberately not tuning the baselines to the optimum. % TODO: expand or remove?
\end{itemize}

Regarding the choice of the baselines, we found that in some cases researchers reuse the experimental design that was used in previous studies, i.e., they use the same datasets, evaluation protocol and metrics, to demonstrate progress.
This is in principle very meaningful as it helps the comparison of results.
However, this also means that the experimental setup is not questioned anymore, because it has been used by many researchers before. Some papers in our present study are based on two quite small CiteULike datasets that were used in some early deep learning papers. These datasets were reused by other authors later on, without questioning if these quite small datasets were representative for a larger set of real-world recommendation problems or whether they were useful at all to analyze certain phenomena.
In our analysis, we even found that the Epinions dataset had such characteristics that the non-personalized recommendation of the most popular items was favorable over any personalized technique.

In general, the reuse of experimental designs is still the exception rather than the norm. In our analysis, we found that researchers use a large variety of evaluation protocols, datasets, and performance measures in their experiments. This is expected in articles having different goals and aimed at different scenarios. However, in most cases, there is no particular argumentation on why a certain metric is used or why particular datasets from a certain domain serve as a basis for the research. To illustrate these phenomena, we show in Table \ref{tab:paper-statistics-dataset} which datasets were used for evaluation in the reproducible papers.

\begin{table*}[h!t]
    \caption{Datasets used by reproducible papers.
    %The only dataset we did not use in our experiments because we could not ensure we were reproducing the original experimental setup is MSD (see Section \ref{sec:vae}).
    }
    \label{tab:paper-statistics-dataset}
    \footnotesize
%\resizebox{\linewidth}{!}{%
    \begin{tabular}{ll|ll}
    \textbf{Dataset}  	&   \textbf{Paper} &   \textbf{Dataset}  	&   \textbf{Paper} \\
    \toprule
    Amazon Movie    &  \cite{Xue2017DeepMF}
        & Amazon Music    &  \cite{Xue2017DeepMF}  \cite{Cheng2018DELFAD}    \\
    CiteULike-a    & \cite{ebesu2018collaborative} \cite{li2017collaborativevariationalautoencoder}  \cite{wang2015collaborativedeeplearning}
        & CiteULike-t     &  \cite{li2017collaborativevariationalautoencoder}  \cite{wang2015collaborativedeeplearning}\\
    Epinions    &  \cite{ebesu2018collaborative}
        & FilmTrust    &  \cite{Zhang2018Neurec}     \\
    Frappe    &  \cite{Zhang2018Neurec}
        & Gowalla    &  \cite{He:2018:OPN:3304889.3304969}     \\
    LastFM    &   \cite{Zheng:2018:SCF:3240323.3240343} \cite{hu2018leveragingmetapathcontext}
        & Movielens Hetrec    &  \cite{Zhang2018Neurec}     \\
    Movielens100K    & \cite{hu2018leveragingmetapathcontext}   \cite{Xue2017DeepMF}
        & MovieLens1M    &  \cite{Zheng:2018:SCF:3240323.3240343} \cite{he2017neural} \cite{Zhang2018Neurec} \cite{Zhang2018CoupledCF}  \cite{Cheng2018DELFAD} \\
    MovieLens20M    &  \cite{liang2018variationalautoencodersforCF}
        & MSD    &   \cite{liang2018variationalautoencodersforCF}    \\
    Netflix Prize    &  \cite{wang2015collaborativedeeplearning}  \cite{liang2018variationalautoencodersforCF}
        & Pinterest    & \cite{ebesu2018collaborative}  \cite{he2017neural}    \\
    Tafeng    &  \cite{Zhang2018CoupledCF}
        & Yelp    &  \cite{Zheng:2018:SCF:3240323.3240343} \cite{hu2018leveragingmetapathcontext} \cite{He:2018:OPN:3304889.3304969}    \\
    \bottomrule
  	\end{tabular}
  	%}
    \quad
\end{table*}

Besides the fact that 12 reproducible articles used 18 different datasets, the authors also relied on quite a number of different data splitting procedures, including \emph{leave-one-out}, \emph{leave-last-out}, \emph{80/20 training/test split}, \emph{hold-out of users} or \emph{retain only 1 or 10 interactions per user}. When evaluating, researchers somewhat arbitrarily used 50, 100, or 1000 negative samples per positive sample for ranking. Moreover, the metrics--including the Hit Rate, Precision, Recall, MAP or NDCG---were measured on a variety of cut-off thresholds between 1 and 300.
In addition, several pre-processing strategies were applied, for example, retaining only users or items with an arbitrary minimum number of interactions.
In most cases, no justification is provided for why a particular choice of pre-processing, metrics and cut-off was selected and which relevant scenario aims to represent.
%% NEW PAOLO
\blue{As a result---as recently pointed out by \cite{canamares2020offline}---each of these choices can lead to comparisons that are vulnerable to misinterpretation, and which may lead to different or even opposite outcomes, depending on the exact combination of the used settings.}

One possible way to obtain a better understanding regarding what represents the state-of-the-art could be the adoption of standardized evaluation setups, where all regarded algorithms are evaluated in the same experimental conditions. Such an approach was for example followed in \cite{Ludewig2018} for session-based recommendation problems, where the same set of measurements was applied for various datasets. In pre-neural times, such comparisons were often enabled by the use of recommender systems libraries like MyMediaLite\footnote{\url{http://www.mymedialite.net/index.html}} or LibRec\footnote{\url{https://www.librec.net/}}, which also included evaluation code. In contrast, the works analyzed in our study mostly implemented the evaluation routines from scratch, leading to certain problems as discussed in the paper.

Regarding the issue of baselines often not being properly tuned, one problematic factor with deep learning methods is that most of them are computationally complex, and some of them have a large number of hyperparameters to tune. Furthermore, it is usually also possible to vary the structure of the network in terms of the layers and nodes in each layer. An exhaustive search through the hyperparameter space for a large number of network architectures is therefore often not possible in reasonable time.
%\djx{Removed the sentence about hyperparameter optimization for the neural methods here. please check the previous sentence.}

%In our work, for example, we optimized the hyperparameters via Bayesian search and examined 50 cases for each dataset per algorithm. We did, however, not explore alternative network structures but used the configuration described in the original papers. \mf{I am not clear what is meant here. We never (except SpectralCF) tuned hyperparameters for DL algorithms}

Regarding absolute running times, let us consider a few examples.\footnote{All reported measurements were made on the same hardware: one AWS instance p3.2xlarge, with 8 vCPU, 30GB RAM, and one Tesla V100-SXM2-16GB GPU. All deep-learning algorithms (with the exception of CDL) were trained using the GPU. All of our baselines were trained only on the CPUs. The detailed measurements are available in the online material (Section \ref{sec:methodology})}
\begin{itemize}

\item The often-cited NCF (NeuMF) method (Section \ref{sec:ncf}) needs four hours to train on one of our machines for the popular MovieLens1M dataset.
Training the best-performing SLIM method on the same machine and dataset requires only 4 minutes.
Interestingly, the more recent \EASER method has better accuracy than NeuMF, but only needs about 12 seconds for training on the same machine.
Training NeuMF on the slightly larger Pinterest dataset (1.5M interactions) took more than two days, which is again much higher when compared to the training times of SLIM (12 minutes) or \EASER (2 minutes).

\item Similar observations can be made for the early CDL method (Section \ref{sec:cdl}). On the larger CiteULike-a dataset with about 55k interactions, CDL needs almost two hours for training, the \EASER method 2 minutes, and our well-performing ItemKNN CFCBF hybrid less than 10 seconds to pre-compute the neighborhoods.

\item The Mult-VAE method (Section \ref{sec:cvae}), which significantly outperformed our baselines on one dataset, seems to be also favorable over both baselines and other neural methods in terms of training times. For the relatively large MovieLens20M dataset, training needs about 21 minutes on our machine. This is much faster than the SLIM algorithm, which needs almost two hours and was competitive with Mult-VAE on the other dataset. \EASER, which is on par or better than SLIM, is again favorable here, requiring only about 3 minutes for training.

\item The algorithm with the longest training time on our machine is DMF. Its best variant, based on binary cross entropy, requires almost 5 days of training on the MovieLens1M dataset, while the simple iALS baseline, which outperforms DMF on all measures, requires only \emph{4 minutes} of training.

\end{itemize}

Limited scalability is not a methodological issue per se, and we do not argue that scalability necessarily has to be a main focus of academic researchers, in particular when it comes to new technical ideas and proposals. Once the ideas are assumed to be viable and effective, questions related to the practical deployment can be addressed in subsequent research, e.g., by using more shallow models that implement the same concepts.

However, today's often enormous costs for tuning the baselines may lure researchers into taking hyperparameter settings from previous papers---even though they were determined for different evaluation setups---and reuse them in their own experiments.
% \djx{Added this here to discuss cross-validation}
Another possible consequence of the sometimes high computation costs can be that researchers often do not apply cross-validation procedures. Instead, they rather base their conclusions on one single train-test split of the data.
%In particular when time-based splitting is applied, the observed gains might therefore at least partially be the result of specifics of the temporal developments that are represented in the data.
For this reason, in many research works the statistical significance of the observed outcomes is not reported.

% TODO?
%\textcolor{blue}{[Maybe add a table with summary statistics and more details about CPU/GPU in a side box / table. Perhaps there is also a better place for these details. Maybe we move it (back) to reproducibility issues. ]}

% other topics. wrong research practice.
% TODO? Should we summarize somewhat minor technical issues?
Besides the general issues reported so far, we observed a number of further more technical issues. These include uncommon implementations of ranking measures, non-randomized data splits, reporting best results across different epochs, and determining the best number of epochs on the test set.
%Nowadays, a number of evaluation frameworks for \emph{top-n} recommendation tasks exist, including, e.g., MyMediaLite\footnote{\url{http://www.mymedialite.net/}} or LibRec\footnote{\url{https://www.librec.net/}}. None of the papers in our analysis relied on such a framework in which common evaluation metrics and procedures would have been pre-implemented.
% TODO: Check if this is true.

% there was a method were they reported precision but measured recall? or was it a session-based algorithm? sr-gnn

\subsection{Fundamental Issues}
% Probably the most important factors that contribute to the apparent stagnation are today's incentive mechanisms in the academic system and our established research practices, see also \cite{troubling-trends-1807.03341}.
Among the factors that may contribute to the apparent stagnation observed both in this study and in other fields \cite{Makridakis2018,Armstrong:2009:IDA:1645953.1646031} are today's incentive mechanisms in the academic system and our established research practices, see also \cite{troubling-trends-1807.03341}. Regarding the incentive system, researchers are more and more evaluated based on the citations their works receive. This might lead to the phenomenon that researchers develop a tendency to investigate problems that are popular and (easily) ``publishable'' \cite{DBLP:journals/corr/abs-1206-4656}.
% With the ongoing ``neural hype''
With the current strong interest in neural algorithms \cite{Lin:2019:NHC:3308774.3308781}, papers that do not propose or use deep learning approaches might get criticized by reviewers for not using state-of-the-art technology. It therefore might appear much easier to publish a neural approach than other works, even though the true value of these complex models is not always fully clear. Furthermore, with the thinness of the reviewer pool, the ``mathiness'' of many papers \cite{troubling-trends-1807.03341} and the sheer amount of machine learning papers submitted every year, it becomes more and more difficult to identify those works that truly move the field forward.

One particularly seductive aspect in the context of algorithmic works on \emph{top-n} recommendation is that there is an implicit general agreement about how research in this area should be done. Unlike in other types of recommender systems research, e.g., research that involves user studies, the experimental offline evaluation design approach is generally pre-determined and is not questioned. Generally, to have a paper accepted at one of the prestigious conferences considered in this paper, one has to (at least) propose a new technical approach that outperforms some state-of-the-art methods on one evaluation protocol and on at least a couple of established metrics and publicly available datasets. Which dataset is used often seems arbitrary, and today's papers in most cases do not motivate their work based on domain-specific considerations or an underlying theory.\footnote{See also \cite{DataSetSelection2002} for a satirical discussion of this general problem in machine learning.} Researchers are also flexible in terms of their evaluation method. As discussed above, various protocol variants and accuracy measures are used today, and in most papers the selection of measures is not explained. Finally, as discussed throughout the paper, it is difficult to know what represents the state-of-the-art in a certain area or for a specific subproblem.

All these degrees of freedom make it very tempting for researchers to focus on accuracy improvements for \emph{top-n} recommendation, where it is in some sense easy to ``prove'' progress and where no special research design has to be developed and argued for. However, the established research model, as discussed, allows for a lot of arbitrariness.
For example, a novel algorithm may be compared against a baseline using a number of accuracy metrics at different cut-off lengths on multiple datasets.
The probability of finding some combinations of measures and datasets by which the novel algorithm seems to outperform the baseline increases with the number of cases examined \cite{gunawardana2015evaluating}.

%\pc{I'm not sure if it makes sense to mention the online competitions. In the research community there is the feeling that many competitions are such that overfitting is better than innovation. I have replaced the Kaggle part with the above two sentences. I moved down here the part on Kaggle, just to see how it was.}\djx{I rephrased it a bit (something was also missing), and added criticism about these competitions. We could remove everything as well.}

Online machine learning competitions as hosted, e.g., on Kaggle\footnote{\url{https:///www.kaggle.com}}, represent the other extreme. On Kaggle and also in specific recommender systems challenges, the dataset and the evaluation method are pre-defined by the organization running the competition. Furthermore, participants do not see the test set and usually do not measure the recommendation quality by themselves, which avoids, as we have discovered in our study, a potential source of mistakes. One typical criticism of such online competitions is that they result in a ``leaderboard chasing'' culture, which can lead to limited innovation.

All in all, current research practice can easily create an illusion of progress in terms of accuracy improvements. But even when this progress is real, a further obstacle lies in us not even knowing with certainty whether these accuracy improvements will lead to better recommendations in practice, neither in terms of business value nor in terms of the quality perception by users. In fact, a number of research works indicate that algorithms with higher offline accuracy do not necessarily lead to better perceived recommendation quality \cite{Rossetti:2016:COO:2959100.2959176,Cremonesi:2012:IPP:2209310.2209314,Garcin:2014:OOE:2645710.2645745,Maksai:2015:POP:2792838.2800184,DBLP:conf/ercimdl/BeelL15}.
For this reason, recommender systems research is different from other areas like image recognition or natural language processing (e.g., automated translation), where accuracy improvements can often directly lead to better systems.

This problem of focusing solely on accuracy optimization has actually been well known in the recommender systems community for many years \cite{McNee:2006:AEA:1125451.1125659}, and the ``\emph{hyper-focus on abstract metrics}'' is also common to other areas of applied machine learning research \cite{DBLP:journals/corr/abs-1206-4656}. As shown in comparative evaluations like \cite{Ludewig2018}, however, there usually \emph{is no best model}, and the ranking of algorithms depends on many factors like dataset characteristics or evaluation approach.
While there is no doubt that being better able to predict the relevance of items for individual users is something useful, considering only abstract accuracy measures appears to be a strong oversimplification of the problem.

% Offline evaluation of recommender systems: all pain and no gain
% https://de.slideshare.net/MarkLevy/pain-27121536?next_slideshow=1

\subsection{Reproducibility and Progress: Guidelines \& Best Practices}
% \mf{draft, reorganizing with bullet lists as subsection of discussion}

The analyses in this paper point to several common issues that may lead to limited reproducibility and progress.
Based on these observations, we provide a set of guidelines and best practices for algorithms research in recommender systems. The guidelines relate both to methodological aspects, i.e., how to ensure that progress is made, and publication aspects, i.e., which information should be reported in research papers and which materials should be provided publicly.
\blue{Moreover, we provide recommendations for journal editors and conference chairs to improve reproducibility and progress in the discipline.}

\subsubsection{Methodology}
\blue{Generally, researchers should select a wide, diverse and optimized set of baselines in order to reduce the chance of reporting progress that is only virtual.}

% \djx{Revised the discussions a bit, trying to shorten.}
\begin{description}
    \item[Include baselines algorithms from different families.]
    In our results there was no individual algorithm, or even no single family of algorithms (e.g., neighborhood-based, machine-learning, non-personalized), which was superior to the others in all experimental configurations.
    Therefore, researchers should strive to include well-optimized baseline algorithms of different families in their experiments and not limit themselves to a certain class of algorithms, e.g., deep learning baselines.

    %\item[Choice of baselines:] As our results have highlighted, there is no single category of algorithms (i.e., neighborhood-based, machine-learning, non-personalized) that is superior to the others in all circumstances. At least one of each should be always reported, to ensure that the new proposed model is not compared only to a single category of baselines which, on that dataset and evaluation protocol, may not be the best performing one. \mf{CHECK. Based on the results of our analysis, we suggest to add ItemKNN and \pbeta, as well as SLIM ElasticNet and \iALS.} Reporting baselines which are incomplete is also bad practice. For example, the shrinkage (see Section \ref{sec:methodology}) is almost never used by the analyzed articles despite being known to improve, sometimes substantially, the recommendation quality of ItemKNN and UserKNN methods.

    \item[Systematically optimize all algorithms.] All algorithms in the comparison must be optimized for each tested dataset and for the given evaluation protocol and metrics. A comparison between optimized and non-optimized algorithms is meaningless. It cannot inform us if a new method advances the state of the art. The optimization must include all relevant parameters and all baselines, even the simplest, including, for example, the \emph{shrink term} for neighborhood-based approaches, which is known to sometimes lead to substantial performance gains. Also the number of training epochs for learning approaches is a parameter to be tuned on validation data.

    %\item[Optimization of all algorithms:] All algorithms should be optimized for the specific dataset and evaluation protocol on the validation data. Using the test data at any stage other than the final evaluation would result in information leakage, which is always bad practice. The use of \emph{default} hyper-parameters and the optimization on only one of the reported datasets are bad experimental practices as well. Usually, any algorithm, even the latest state-of-the-art one, which has not been optimized for the given experimental scenario will lead to non-competitive accuracy and therefore does not constitute a strong baseline. The optimization of the hyper-parameters includes the selection of the number of epochs, like any other, should be optimized on the validation data.

    \item[Carefully choose the hyperparameter optimization strategy.]
    %\djx{This topic was not raised in the paper before. Maybe we have something else to say as well.} \mf{we did not raise that point before, but the only papers that do report information on the hyperparameters do so by listing the values they try, so it is likely a grid search.}
    %\pc{My suggestion is to remove this point, as the reviewers didn't ask for it.} \mf{It is an important aspect, I would keep it}
    Another issue we observed for several papers but have not highlighted so far is that authors often do not report the details of the applied hyperparameter optimization strategy. The chosen strategy can, however, have a significant impact on the results. The commonly used Grid Search approach has certain limitations, in particular that the results may be strongly affected by how the value ranges of continuous valued hyperparameters are discretized. Since even Random Search has shown to be often favorable over Grid search \cite{prel_bergstra2012random}, the use of Random search or Bayesian strategies is recommended.

    %\item[Optimization strategy:] The hyperparameter optimization strategy itself can have a significant impact on the results. The very commonly used Grid search has several important limitations, in particular its results are strongly affected by how the value ranges of continuous valued hyperparameters are discretized, which is a rather arbitrary choice that can be severely constrained by the high computational cost this method has. Furthermore, it has been known for several years that even a Random search is often preferable and more efficient than a Grid search \cite{prel_bergstra2012random}. We encourage to use either Random or Bayesian strategies.

\end{description}

\subsubsection{Reproducibility}
\blue{As pointed out by Cockburn et al.~in \cite{cockburn2020threats}, the pressure to publish academic works encourages researchers to consciously or unconsciously adopt not well justified methodologies---such as excluding certain users or items from the datasets---to make their findings look stronger.
While such decisions can be entirely appropriate if properly motivated, engaging in selective reporting of results is not: flexible data analysis and selective reporting can dramatically increase the risk of only virtual progress.
For example, if a researcher collects 10 accuracy metrics and only reports the significant ones (with significance level at 0.05), then the probability of reporting a progress that is only virtual jumps from 5\% to 40\%.}

%\paragraph{What to report in papers.}
While in principle a paper should contain all necessary information to reproduce the results, this is often conflicting with the space limitations most publication outlets enforce.
% \paragraph{What to Report in Papers}
The following is a list of details that should be reported as a minimum in a paper and it is based on \emph{The Machine Learning Reproducibility Checklist}\footnote{\url{https://www.cs.mcgill.ca/~jpineau/ReproducibilityChecklist-v2.0.pdf}} adopted, e.g., by NeurIPS '18 and NeurIPS '19:
%As a set of minimal required details to report in papers we can refer to \emph{The Machine Learning Reproducibility Checklist}\footnote{\url{https://www.cs.mcgill.ca/~jpineau/ReproducibilityChecklist-v2.0.pdf}} adopted, e.g., by NeurIPS '18 and NeurIPS '19:

\begin{description}
    %\item[Assumptions:]
    \item[State underlying assumptions.]
    The assumptions the model relies upon (e.g., data properties) should be clearly stated and, when necessary, verified to be true for the evaluation setup at hand. For example, the claim that interaction maps derived from the outer product of embeddings are equivalent to images and that CNNs allow to model the embedding correlations (CoupledCF \cite{Zhang2018CoupledCF} and ConvNCF \cite{He:2018:OPN:3304889.3304969}) was not directly verified in the original articles and was later questioned by Ferrari Dacrema et al. \cite{FerrariDacrema2020convolution}. %\djx{Do we have something specific to say regarding the papers we analyzed.} \mf{It comes to mind only the problem with the CNN based ones.}

    %\item[pre-processing:]
    \item[Be clear about data pre-processing.]
    All pre-processing steps should be described and the removal of data should be clearly motivated.
    % In some of the articles we analyze the pre-processing step reduced the size of the dataset of more than 80\% while the article claims to be evaluating the newly proposed model on \emph{real world data}.
    In some of the articles we analyzed (DMF \cite{Xue2017DeepMF} and NeuRec \cite{Zhang2018Neurec}), pre-processing reduced the size of the dataset by more than 80\%. Drastic subsampling of the dataset may entirely change its properties. %, and should be considered bad experimental practice if the need for such pre-processing is not compelling. \mf{maybe too much?}

    \item[Report hyperparameter optimization details.]
    The article should either explicitly mention or provide a reference to the range and distribution of each hyperparameter, and should report the method that was used to select them. This includes the number of epochs the model was trained for and early-stopping criteria. %\djx{Could remove last sentence.} \mf{I would keep it, it is the convergence criteria and in terms of results is rather important.}
\end{description}

%\paragraph{Providing Material for Reproducibility}
Given that small details in the implementation of complex models can matter a lot, what published in a paper alone is often not sufficient to ensure reproducibility of results.
Therefore, it is important for authors to also provide a publicly available implementation.
The following guidelines are partially based on those adopted by NeurIPS\footnote{\url{https://github.com/paperswithcode/releasing-research-code}}. In essence, it should be made as easy as possible for other researchers to redo the experiments.

%which can be used by other researchers is very important.
%Several guidelines regarding what to provide are available. We can refer again to those of NeurIPS\footnote{\url{https://github.com/paperswithcode/releasing-research-code}}:
% \footnote{https://medium.com/paperswithcode/ml-code-completeness-checklist-e9127b168501}

\begin{description}
    \item[Publish the source code of all models, including the baselines.] We often observed in our analysis that researchers only published the core training and prediction routine of the newly proposed model. %, but not the code for the baselines and for the pre-processing and splitting of the datasets.
    We could not find any article providing the source code for the baselines, when those were not already publicly available as part of other projects, and only few published the source code for both the pre-processing and splitting of the datasets (Mult-VAE \cite{liang2018variationalautoencodersforCF} and NeuRec \cite{Zhang2018Neurec}).
    \item[Provide the datasets.] The used datasets, or pointers to them, should be published. If possible, the train and test splits should be published as well.
    \item[Make the reproduction easy for others.]
    Authors should specify all software requirements and dependencies of their software, including the specific version. Provide installation scripts and machine-readable dependency descriptions.
    It is also desirable to provide scripts that execute the entire experimental pipeline, from pre-processing, to training, evaluation and collection of metrics.
    \item[Use persistent repositories.] Institutional repositories or DOIs should be preferred to personal webpages since the latter can have a much shorter lifespan making previously published links unavailable \cite{prel_vandewalle2009reproducible}.\footnote{For example, at the time of our analysis we found that the provided links to two datasets (Tafeng and Frappe) were not active anymore. We could only use these datasets because other researchers had uploaded them independently as part of other publications.}

    %\item[Evaluation script:] A way to evaluate the model computing the recommendations and the metrics reported in the paper.
    %\item[Dependencies:] The complete list of dependencies and the specific version, preferably in a format which makes installing them easy using available tools like conda or pip.
    %\item[Training script:] A script to train the model. Ideally, the script should only require the original dataset and apply all pre-processing operations.
    %\item[Experimental pipeline:] The full experimental pipeline, from data reading to model evaluation, allowing to reproduce the published results.
    %\item[Persistent repository:] Institutional repositories or DOIs should be preferred to personal webpages since the latter can have a much shorter lifespan making previously published links unavailable \cite{prel_vandewalle2009reproducible}.
\end{description}

\subsubsection{Mitigation}

In the spirit of mitigating the current issues in phantom progress, our recommendation to journal editors and conference chairs is to promote submissions replicating previous studies.
Replications of studies contribute to elevate confidence and robustness in findings beyond what is possible from a single study.

However, the main obstacle to reproducibility studies is a publication bias in which reproducibility papers are accepted for publication at a much lower rate than those providing novel results \cite{cockburn2020threats}.
Therefore, we encourage editors and chairs to educate reviewers on the research value of replicating previous studies.

\section{Conclusion}
Our work reveals that despite the large number of new methods that are published on the topic of \emph{top-n} recommendation every year, the progress in this field may actually seem limited, at least when considering the papers analyzed in this work. Our analysis shows that sometimes even relatively simple methods lead to performance levels that are similar or even better than the most recent complex neural methods. In other cases, the computationally expensive neural methods did not outperform well-established matrix factorization approaches or linear models. A number of issues contribute to the observed phenomena. We not only identified different methodological problems but also found that the choice of the baselines and often the lack of a proper optimization of these baselines represent key issues that hamper our progress. These phenomena however are not specific to the domain of recommender systems or to neural approaches.
In the context of the use of deep learning techniques for the important \emph{top-n} recommendation task, our analyses in some ways indicate that the power of deep learning has not been leveraged yet to the full extent as it was done in other applications areas. This, as a result, leads to a number of avenues for future research on neural methods for recommender systems.
%\djx{Do we need to connect the next paragraph better?}

Increasing the reproducibility of published research was identified as one of possible strategies to mitigate some of the observed problems. In the information retrieval research community, questions of replicability and reproducibility have recently received more attention, probably thanks in part to the surprising results from \cite{Lin:2019:NHC:3308774.3308781,Yang:2019:CEH:3331184.3331340}, and \cite{Armstrong:2009:IDA:1645953.1646031}.
An increased awareness and corresponding initiatives are also advisable for the recommender systems community. However, even with better reproducibility, fundamental problems of algorithms-based recommender systems research remain. The reason is that, unlike in some IR tasks, better retrieval or prediction performance does not necessarily lead to recommendations that are better in terms of the users' quality perception or in terms of the business value for providers.

In the end, these problems lead to a certain stagnation of the research field. A large number of researchers hunt for the ``best'' model, even though there are many indications that no such model exists, as the performance ranking of algorithms depends on many factors.
Overall, a paradigmatic shift is therefore required in terms of how we evaluate recommender systems. In the future, it is, however, not only important to address the methodological issues observed in this paper. We also should reconsider how much we can actually learn from offline experiments and computational accuracy measures alone. This, for example, calls for multi-method evaluation approaches that
\begin{enumerate*}[label=\textit{(\roman*)}]
  \item consider the human in the loop and
  \item for research works that are more based on theoretical considerations than on network architecture engineering.
\end{enumerate*}

%Unfortunately, as discussed in \cite{lostinmath2018} for the domain of theoretical physics\footnote{See also \url{http://backreaction.blogspot.com/2018/11/the-present-phase-of-stagnation-in.html}}, there are no strong signs of a crisis, where more and more researchers critically reflect on what has been claimed and what has been actually achieved in the past decades. The ongoing wave of machine learning research, in contrast, seems to add to the stagnation.

% \begin{comment}
% \begin{itemize}
%   \item We found that simple techniques outperform complex models.
%   \item Even though in our baselines we did NOT do anything particular to find an algorithm that works specifically well on some dataset.
%   \item We discussed potential reasons and problems of today's research practice
%   \item Overall we found signs of stagnation, but we need a crisis, see \url{http://backreaction.blogspot.com/2018/11/the-present-phase-of-stagnation-in.html}
%   \item What to do: Increase reproducibility, focus on more interesting questions, approach the problem in a more comprehensive way.
%   \item Refer to IR reproducibility workshop initiative mentioned by Jimmy. But recsys is different from IR, higher accuracy not necessarily better.
% \end{itemize}
% \end{comment}

\clearpage
\appendix
\section{Detailed analysis for reproducible articles}
\label{sec:detailed_analysis}
In this part of the appendix we discuss the analyzed papers in more detail and provide selected numerical results.

% ---------------------------------------------------------------------------------
\subsection{Collaborative Deep Learning for Recommender Systems (CDL)}
\label{sec:cdl}
% ---------------------------------------------------------------------------------

\paragraph{Datasets} The evaluation in the original paper is based on three datasets. Two of them are data samples that were previously collected from the \emph{CiteULike} platform and which were already used in earlier research. One dataset is based on rating data from the Netflix Prize competition, augmented with content information by the authors. In our evaluation, we considered only the CiteULike datasets, because the content information used in combination with the Netflix dataset is not publicly available. Two versions of the CiteULike dataset were considered, a dense version \emph{CiteULike-a} and a sparser one \emph{CiteULike-t}. Both datasets are relatively small (135k and 205k interactions, respectively).
For each of these datasets, experiments were made in two sparsity configurations.
These configurations are described by a parameter $P$, which defines how many interactions per user are left in the training set (with the rest going to the test set).
For parameter $P$, values 1 and 10 were reported, which correspond to 5.5k and 55.5k training interactions, respectively. Note that with $P=1$ there is only one training interaction per user in the training dataset.

\paragraph{Evaluation} Several baseline techniques were explored, among them a number of hybrid matrix factorization approaches, a content-based deep learning technique designed for music recommendation, as well as Collaborative Topic Regression% \cite{wang2011collaborativetopicmodeling}
, a method combining Latent Dirichlet Allocation on the content and collaborative filtering on the interactions.
For evaluation purposes, $P$ interactions for each user were randomly sampled to be part of the training set as mentioned above. The average results of five evaluation runs are reported.
The authors report Recall for comparably long list lengths (50 to 300), and Mean Average Precision for list length 300.
%In the paper on CTR, which used the same datasets, list lengths between 10 and 200 were evaluated.
%\pc{When we refer to baselines used in other papers, sometimes we include a citation for the baseline, sometimes we do not include it. We should be consistent. We have three options: (a) no citations at all, (b) citations for all baselines, (c) citations only for non trivial baselines. I suggest option (c). It is important anyway to be consistent.} \mf{Revised text applying option C, citation provided only for DL baselines we analyze}

\paragraph{Results and Discussion} The authors found their method to outperform all baselines on all measures. We could reproduce their results based on the provided code and dataset.\footnote{In the source code the authors provided it is reported that the original evaluation contained an error such that the absolute values of the evaluation metrics was higher than the correct one, although the relative performance ordering of the algorithms remained unaltered. Once this error is fixed we can reproduce their results.}
To optimize the baselines in our own evaluation, we used 20\% of the training set as a validation set.\footnote{Information about the validation set size was not provided in the original paper. In the evaluation scenario where \emph{P=1}, due to the presence of only 1 training instance per user, any sampling would result in cold users. Therefore, in this scenario, the validation data is also contained in the train data. In the evaluation scenario where \emph{P=10}, training and validation data are disjoint.}
After optimization, our results show that CDL---in three out of four configurations (CiteULike-a with \emph{P=10} and CiteUlike-t with \emph{P=1} and \emph{P=10})---was consistently outperformed by our simple hybrid technique (ItemKNN CFCBF) and, in many cases, also by the pure content-based method (ItemKNN CBF).
Only when removing all but one user interaction from the CiteULike-a dataset (with \emph{P=1}) CDL was, by a large extent, better than any of our baselines.
In particular, in the settings where \emph{P=1}, pure collaborative filtering techniques were, as expected, not competitive.

Table \ref{tab:CDL-results-a10} shows exemplary results for the CiteULike-a dataset (\emph{P=10}), with about 55k interactions in the training dataset.
Detailed results for all datasets can be found in the online appendix (Section \ref{sec:methodology}).
In the table, we highlight in bold those entries where a baseline outperformed CDL.
We can observe that, for shorter and much more typical list lengths, even the simplest collaborative filtering approaches outperform CDL. The \iALS method based on matrix factorization for implicit feedback data was better with respect to CDL in all measurements and cutoff lengths.
Finally, the best results were achieved with the pure content-based method that uses only item features to recommend similar items (ItenKNN CBF).
% TODO: Could mention: Looking at the detailed results fore the KNN methods (omitted here), we found that the choice of the similarity measure can matter.

\begin{table}[h!t]
    \caption{Selected Results for the CDL Method on the CiteULike-a dataset with \emph{P=10}. We highlight in bold those entries where a baseline outperforms CDL.}
    \label{tab:CDL-results-a10}
    \footnotesize
    % \begin{minipage}{\textwidth}
    % \resizebox{.8\linewidth}{!}{%
    \begin{tabular}{lcccccc}
    \toprule
% 	& \multicolumn{6}{c}{a 10}  	\\
		& REC@50 	& REC@100 	& REC@150 	& REC@200 	& REC@250 	& REC@300 	\\
		\midrule
    TopPopular	&0.0040	&0.0078	&0.0103	&0.0204	&0.0230	&0.0258	\\
    \midrule
    UserKNN CF jaccard	&\textbf{0.0806}	&\textbf{0.1207}	&0.1480	&0.1705	&0.1887	&0.2034	\\
    ItemKNN CF cosine	&\textbf{0.0989}	&\textbf{0.1441}	&\textbf{0.1752}	&\textbf{0.1982}	&0.2156	&0.2300	\\
    \midrule
    \palpha	&\textbf{0.0907}	&\textbf{0.1341}	&\textbf{0.1636}	&0.1865	&0.2055	&0.2206	\\
    \pbeta \footnotemark	&\textbf{0.0963}	&\textbf{0.1408}	&\textbf{0.1692}	&0.1908	&0.2090	&0.2239	\\
    \midrule
    \EASER 	& \textbf{0.0839} &           \textbf{0.1253} &           0.1546 &           0.1797 &           0.1988 &           0.2128 	\\
    SLIM 	&\textbf{0.0876}	&\textbf{0.1308}	&\textbf{0.1583}	&0.1821	&0.2005	&0.2165	\\
    PureSVD	& \textbf{0.0715}	&0.1079	&0.1313	&0.1491	&0.1636	&0.1759	\\
    \iALS	& \textbf{0.0779}	&\textbf{0.1388}	&\textbf{0.1834}	&\textbf{0.2186}	&\textbf{0.2472}	&\textbf{0.2706}	\\
    \midrule
    ItemKNN CBF cosine	&\textbf{0.1989}	&\textbf{0.2835}	&\textbf{0.3402}	&\textbf{0.3844}	&\textbf{0.4193}	&\textbf{0.4492}	\\
    ItemKNN CFCBF cosine    &\textbf{0.1858}	&\textbf{0.2816}	&\textbf{0.3445}	&\textbf{0.3930}	&\textbf{0.4335}	&\textbf{0.4642}	\\
    \bottomrule
    CDL	&0.0580	&0.1108	&0.1546	&0.1946	&0.2314	&0.2640	\\
    \bottomrule
  	\end{tabular}
%   	}
%   	\end{minipage}
\end{table}

\footnotetext{\pbetanotyetpublished}

%\pc{In this and other tables, we should remove the name of the similarity from the KNN method, as this is the value of an hyperparamter- Eg.  -ItemKNN CF cosine- should be -ItemKNN-.}

%From a methodological perspective, the following observations can be made. First, there is no indication of why comparably long list lengths were used for evaluation in the paper and why no measurements were reported for list lengths below 50, which is commonly the case in the literature.

From a methodological perspective, there is no indication of why comparably long list lengths were used for evaluation in the paper and why no measurements were reported for list lengths below 50, which is commonly the case in the literature.

\subsection{Collaborative Variational Autoencoder (CVAE)}
\label{sec:cvae}
% ---------------------------------------------------------------------------------

\paragraph{Datasets and Evaluation} The CVAE method is evaluated in the same way as the CDL approach, i.e., two datasets from \emph{CiteULIke} are used and different sparsity configurations are evaluated. Likewise, the authors of CVAE measure Recall at different (long) list lengths. As an additional baseline, the authors include the CDL \cite{wang2015collaborativedeeplearning} method described in the previous section. The hyperparameters for all baseline methods were optimized using a validation set which is, however, not described.

\paragraph{Results and Evaluation}
We could reproduce the results for CVAE.\footnote{In the source code the authors provided it is reported that the original evaluation contained an error such that the absolute values of the evaluation metrics was higher than the correct one, although the relative performance ordering of the algorithms remained unaltered. Once this error is fixed we can reproduce their results.} Table \ref{tab:CVAE-results-a10} shows the results of our experiments for the CiteULike-a dataset with \emph{P=10}, again using the same evaluation measures and protocol as used in the original paper.

\begin{table}[h!t]
    \caption{Experimental results for the CVAE method for the CiteULike-a with \emph{P=10}. We highlight in bold those entries where a baseline outperforms CVAE.}
    \label{tab:CVAE-results-a10}
    \footnotesize
    % \begin{minipage}{\textwidth}
    % \resizebox{\linewidth}{!}{%
    \begin{tabular}{lcccccc}
    \toprule
		& REC@50 	& REC@100 	& REC@150 	& REC@200 	& REC@250 	& REC@300 	\\
		\midrule
    TopPopular	&0.0040	&0.0078	&0.0103	&0.0204	&0.0230	&0.0258	\\
    \midrule
    UserKNN CF jaccard	&\textbf{0.0806}	&0.1207	&0.1480	&0.1705	&0.1887	&0.2034	\\
    ItemKNN CF cosine	&\textbf{0.0989}	&0.1441	&0.1752	&0.1982	&0.2156	&0.2300	\\
    \midrule
    \palpha	&\textbf{0.0907}	&0.1341	&0.1636	&0.1865	&0.2055	&0.2206	\\
    \pbeta \footnotemark	&\textbf{0.0963}	&0.1408	&0.1692	&0.1908	&0.2090	&0.2239	\\
    \midrule
    \EASER  &  \textbf{0.0839} &0.1253 &0.1546 &0.1797 &0.1988 &0.2128 \\ 
    SLIM 	&\textbf{0.0876}	&0.1308	&0.1583	&0.1821	&0.2005	&0.2165	\\
    PureSVD	& 0.0715	&0.1079	&0.1313	&0.1491	&0.1636	&0.1759	\\
    \iALS	&0.0779	&0.1388	&0.1834	&0.2186	&0.2472	&0.2706	\\
    \midrule
    ItemKNN CBF cosine	&\textbf{0.1989}	&\textbf{0.2835}	&\textbf{0.3402}	&\textbf{0.3844}	&\textbf{0.4193}	&\textbf{0.4492}	\\
    ItemKNN CFCBF cosine    &\textbf{0.1858}	&\textbf{0.2816}	&\textbf{0.3445}	&\textbf{0.3930}	&\textbf{0.4335}	&\textbf{0.4642}	\\
    \bottomrule
    CVAE	&0.0805	&0.1569	&0.2232	&0.2760	&0.3250	&0.3687	\\
    CDL	    &0.0580	&0.1108	&0.1546	&0.1946	&0.2314	&0.2640	\\
    \bottomrule
  	\end{tabular}
%   	}
%   	\end{minipage}
\end{table}

\footnotetext{\pbetanotyetpublished}

The baseline results shown in Table \ref{tab:CVAE-results-a10} are identical to those of Table \ref{tab:CDL-results-a10}---as they were done on the same dataset and with the same evaluation protocol---except that Table \ref{tab:CVAE-results-a10} has an additional row for the results for the CVAE method.
Again, the simple hybrid baselines outperform the more complex CVAE method on all measures on this dataset.
We can, however, observe that CVAE is indeed consistently better than the CDL method, which is the main baseline method in \cite{li2017collaborativevariationalautoencoder}.
For the other dataset and sparsity configurations, our results are similar to what was reported in the previous section on CDL. %Except for the configuration CiteULike-a with \emph{P=1}, our hybrids consistently outperform CVAE on all measurements. \mf{There was an error in the previous section, the results of CDL and CVAE are identical wrt the baselines in all cases}

Overall, the authors of CVAE could show an advance with respect to CDL, but our results indicate that CDL did not represent a strong baseline method.
In the remainder of this paper, we will observe the following phenomenon several times: a neural method is introduced as improving the state-of-the-art, and subsequent works only focus on outperforming this new neural method, without considering alternative baselines.

% ---------------------------------------------------------------------------------
\subsection{Neural Collaborative Filtering (NCF)}
\label{sec:ncf}
% TODO: make sure that we use NeuMF and NCF consistently
% ---------------------------------------------------------------------------------

\paragraph{Datasets} Two datasets were used for evaluating the method, one rating dataset from MovieLens (\emph{MovieLens1M})\footnote{\url{https://grouplens.org/datasets/movielens/}} and one dataset with implicit feedback from \emph{Pinterest}\footnote{\url{https://sites.google.com/site/xueatalphabeta/academic-projects}}.
The Pinterest dataset was pre-processed by removing all users with less than 20 interactions.
After pre-processing, the dataset contained 1.5 million interactions.
For the MovieLens rating dataset, all 1 million ratings were transformed to 1 to mimic an implicit-feedback dataset, with missing entries transformed to 0.
%No justification is provided on why a different threshold was not used to binarize the dataset.
%Generally, such a transformation to binary data, while common in the literature, can be questionable from a methodological perspective. Most papers on collaborative filtering techniques explicitly or implicitly aim at providing better \emph{top-n} recommendations for users. However, what is actually evaluated with such a transformed dataset is the performance of algorithms to predict who will rate which item, independent of the rating value. As a result, an algorithm might be considered well-performing even if it ranks items highly, which the user actually did not like.

\paragraph{Evaluation} The authors use a leave-last-out procedure to evaluate their method.
For each user, the last interaction (based on its timestamp) is put into the test set. %\footnote{\textcolor{blue}{The timestamps in the MovieLens dataset actually relate to the point in time when a rating was provided, and not necessarily related to the time when the movie was watched. It therefore remains to question if a time-based splitting approach is the most suitable one when the goal is to make recommendations of movies to watch. The Pinterest dataset does not suffer from this problem as the timestamp represents the occurrence of the real interaction. \djx{Not sure we have to say this; rephrased a bit; would prefer to remove the footnote}\mf{no objections. It is just a somewhat improper use of the data. Could either remove or try to find a reference}}}.
%does not represent the occurrence of the interaction with the item, which happens outside the platform, but rather when the user provided the feedback. There can be therefore no guarantee on the ordering and time separation of the real interactions based on the available timestamp. Due to this, time-based splitting on Movielens is, to some extent, misleading or inappropriate. } }.
The resulting data splits used in the experiments are shared by the authors. To avoid to compute scores for all recommendable items, which is considered too time-consuming by the authors even for datasets of modest size, the performance of the algorithms is measured by determining how the last hidden item is ranked within other 100 randomly sampled items.
%No particular justification for the choice of this evaluation protocol is provided, even though cross-validation procedures with interactions being placed randomly in the test set are more common in the literature.
Hit Rate (HR) and NDCG at list length 10 are used as performance metrics \cite{Zhang2018CoupledCF}.
%The Hit Rate counts the average number of correct recommendations each user received, NDCG is also sensitive to their position in the recommendation list.

As personalized baselines, the authors include Matrix Factorization with Bayesian Personalized Ranking (BPR matrix factorization),
%\cite{rendle2009bprMF}
the eALS method from 2016
%\cite{He:2016:FMF:2911451.2911489}
and the ItemKNN method. The original hyperparameter optimization is done on a validation set obtained by randomly selecting one interaction per user. %\djx{please rephrase last part of previous sentence. not sure what is meant.}
For the ItemKNN method, the number of neighbors was varied, but no other configurations were tested by the authors (e.g., shrink term or normalization).
According to the reported experiments, the NCF method, and in particular the NeuMF variant, outperforms all baselines on all dataset on all performance measures.

\paragraph{Results and Discussion} We could reproduce the reported results. However, the analysis of the provided source code shows that the number of training epochs was chosen by maximising the Hit Rate on the test data. Since the number of epochs is a parameter like any other, it must be fixed before testing, e.g., through early stopping on a validation set. In our experiments, for each algorithm (including NCF) we therefore report the performance measure for the number of epochs that was considered optimal based on the validation set.

Table \ref{tab:NCF-results-movielens} and Table \ref{tab:NCF-results-pinterest} report our results for both the MovieLens and Pinterest datasets.
We report the results for list length 10, as in the original paper.
Since the authors in \cite{he2017neural} also plot the results at different list lengths from 1 to 10, we also include measurements at list lengths 1 and 5 for comparison purposes.

% Counter to allow multiple references to the same footnote. Having footnotes in tables is a mess
\newcounter{fncounteralgpredatespbeta}

\begin{table}[h!t]
    \caption{Experimental results for NCF (MovieLens 1M). We highlight in bold those entries where a baseline outperforms NCF.}
    \label{tab:NCF-results-movielens}
    \footnotesize
    % \begin{minipage}{\textwidth}
    % \resizebox{\linewidth}{!}{%
    \begin{tabular}{lcccccc}
    \toprule
	& HR@1 	& NDCG@1 &	 HR@5 	& NDCG@5 &	 HR@10 	& NDCG@10 	\\
	\midrule
    TopPopular	&0.1051	&0.1051	&0.3048	&0.2064	&0.4533	&0.2542	\\
    \midrule
    UserKNN CF asymmetric	&0.1921	&0.1921	&0.5070	&0.3546	&0.6768	&0.4100	\\
    ItemKNN CF asymmetric	&0.1843	&0.1843	&0.4906	&0.3400	&0.6627	&0.3956	\\
    \midrule
    \palpha	&0.1791	&0.1791	&0.4846	&0.3352	&0.6460	&0.3876	\\
    \pbeta \footnotemark &0.1836	&0.1836	&0.4935	&0.3419	&0.6758	&0.4011	\\
    \midrule
    \EASER  &  \textbf{0.2225} &  \textbf{0.2225} &  \textbf{0.5629} &  \textbf{0.3986} &  \textbf{0.7192} &  \textbf{0.4494} \\
    SLIM	&\textbf{0.2207}	&\textbf{0.2207}	&\textbf{0.5576}	&\textbf{0.3953}	&\textbf{0.7162}	&\textbf{0.4468}	\\
    PureSVD	&\textbf{0.2132}	&\textbf{0.2132}	&0.5339	&0.3783	&0.6937	&0.4303	\\
    \iALS	&\textbf{0.2106}	&\textbf{0.2106}	&\textbf{0.5505}	&\textbf{0.3862}	&\textbf{0.7109}	&\textbf{0.4382}	\\
    \midrule
    NCF (NeuMF variant)	    &0.2088	&0.2088	&0.5411	&0.3803	&0.7093	&0.4349	\\
    \bottomrule
  	\end{tabular}
%   	}
%   	\end{minipage}
\end{table}

\setcounter{fncounteralgpredatespbeta}{\thefootnote}

\footnotetext{\pbetanotyetpublished\label{fn:alg_predates_pbeta}}

\begin{table}[h!t]
    \caption{Experimental results for NCF (Pinterest). We highlight in bold those entries where a baseline outperforms NCF.}
    \label{tab:NCF-results-pinterest}
    \footnotesize
    % \begin{minipage}{\textwidth}
    % \resizebox{\linewidth}{!}{%
    \begin{tabular}{lcccccc}
    \toprule
	& HR@1 	& NDCG@1 	& HR@5 	& NDCG@5 &	 HR@10 	 &NDCG@10 	\\
	\midrule
    TopPopular	&0.0467	&0.0467	&0.1665	&0.1064	&0.2740	&0.1409	\\
    \midrule
    UserKNN CF jaccard	&\textbf{0.2898}	&\textbf{0.2898}	&0.7038	&\textbf{0.5056}	&0.8655	&\textbf{0.5583}	\\
    ItemKNN CF asymmetric	&\textbf{0.2903}	&\textbf{0.2903}	&\textbf{0.7117}	&\textbf{0.5096}	&0.8766	&\textbf{0.5633}	\\
    \midrule
    \palpha	&\textbf{0.2853}	&\textbf{0.2853}	&0.7022	&0.5024	&0.8700	&0.5571	\\
    \pbeta \footnotemark[\thefncounteralgpredatespbeta] &\textbf{0.2966}	&\textbf{0.2966}	&\textbf{0.7151}	&\textbf{0.5149}	&\textbf{0.8796}	&\textbf{0.5685}	\\
    \midrule
    \EASER &  \textbf{0.2909} &  \textbf{0.2909} &           0.7070 &  \textbf{0.5077} &           0.8684 &  \textbf{0.5604} \\
    SLIM    &\textbf{0.2913}	&\textbf{0.2913}	&0.7059	&\textbf{0.5072}	&0.8679	&\textbf{0.5601}	\\
    PureSVD	&0.2630	&0.2630	&0.6628	&0.4706	&0.8268	&0.5241	\\
    \iALS		&\textbf{0.2811}	&\textbf{0.2811}	&\textbf{0.7144}	&\textbf{0.5061}	&0.8761	&\textbf{0.5590}	\\
    \midrule
    NCF (NeuMF variant)	&0.2801	&0.2801	&0.7101	&0.5029	&0.8777	&0.5576	\\
    \bottomrule
  	\end{tabular}
%   	}
%   	\end{minipage}
\end{table}

On the well-known MovieLens dataset (Table \ref{tab:NCF-results-movielens}), NeuMF was competitive against the simple baselines, however was outperformed by all but one non-neural machine learning methods.
On the Pinterest dataset (Table \ref{tab:NCF-results-pinterest}), NeuMF could only outperform PureSVD, which is not optimized for implicit feedback datasets.
Most non-neural machine learning techniques, were often either similar or better than NeuMF.\footnote{It shall be noted here that after the first publication of our results \cite{Ferraridacremaetal2019}, the authors of NeuMF provided us with an alternative configuration of their method, which included new hyperparameter values taken from alternative hyperparameter ranges, and requiring other slight changes in the training procedure. While this new configuration led to slightly improved results for their method, the results of our analysis were confirmed. In this context we would like to clarify that for all neural methods investigated here better configurations than those reported in the original papers may indeed exist. Finding such configurations, e.g., in the form of better hyperparameter ranges or alternative network structures, is however not the goal of our work. Instead, our goals are to assess the reproducibility of existing works and to compare the best reported results against existing baseline techniques.}
These findings have been confirmed in a recent article by Rendle et al. \cite{ncfvsmf2020}, which discussed the problem of learning a dot product with multilayer perceptrons. The article shows that this is not a trivial task and that the computation cost of using a neural model can be unpractical for production environments where efficient algorithms based on the dot product are available.

As a side observation, we can see that machine learning methods were clearly favorable over simple techniques for the MovieLens dataset.
For the Pinterest dataset, however, it turns out that this advantage diminishes---at least in this experiment---and that a well-tuned ItemKNN method led to similar and sometimes better performance than machine learning techniques.

%\djx{Remark about our communication with NeuMF authors. Could go to footnote as well after the sentence that ends with ``similar or better than NeuMF''} \mf{It seems to me an important clarification of the purpose of this article, I would leave it here. Paolo?} \djx{Also leaning to have it here and not in the footnote.} \mf{on second thought putting it in the main text may give too much visibility to the fact that they complained to us and I would like to avoid the effect that who complains the most gets more visibility than others. On the other hand, not reporting any interaction looks to me unfair on the other end of the spectrum.} \djx{Fine with me to have it in the footnote.}

%\pc{In the discussions section we can insert a paragraph where we question the statistical significance of many of results reported in the papers we analyzed, providing as an example the comparison of Table 6 and 8 for cutoff 10, where baselines results should be identical, but are not (truly speaking, results in Table 8 should be slightly better than results in Table 6, as the negative items are 99 in Table 8 and 100 in Table 6)} \mf{Results could not be identical as we do not have the negative item set, since there is a stochastic component in their sampling. When negative items are involved, even if the "positive" test interactions are provided they still account for only half of the test data.}

% ---------------------------------------------------------------------------------
\subsection{Deep Matrix Factorization (DMF)}
\label{sec:dmf}
% ---------------------------------------------------------------------------------

\paragraph{Datasets}
Experiments were made on four public datasets: the two smallest MovieLens datasets (\emph{100k} and \emph{1M})\footnote{\url{https://grouplens.org/datasets/movielens/}}, and two publicly available datasets collected from Amazon.com (for the \emph{Movie} and \emph{Music} domains)\footnote{\url{http://jmcauley.ucsd.edu/data/amazon/}}.
All datasets contain ratings on a 1 to 5 scale.
The datasets were pre-processed (if needed) so that there were at least 20 ratings for each user.
Furthermore, for the Amazon datasets only, items were considered for which more than 5 ratings existed.

%After pre-processing, the MovieLens1M dataset was the largest and the Amazon Movie dataset the one with the highest sparsity.
%In particular for the Amazon datasets, this form of pre-processing drastically changed their characteristics, in particular in terms of sparsity.

The Amazon Music data set that resulted from the pre-processing step was shared by the authors.
It however contains users with less than 20 interactions and items with less than 5 ratings.
Therefore, it remains unclear how exactly the filtering was done.
In order to keep the results presented in this paper consistent across datasets, we have pre-processed all datasets---including Amazon Music---as described in the original paper and we have not used the Amazon Music dataset shared by the authors.
We also run our experiments on the Amazon Music dataset shared by the authors: the relative performance between baselines and DMF (not reported here) do not change with respect to the results reported here.

\paragraph{Evaluation} The evaluation procedure is \emph{leave-last-out} similar to the one used for the NCF method.
For each user, the last interaction (based on its timestamp) is held out and ranked together with 99 negative (non-interacted) random items.\footnote{The paper reports that 100 negative items are used, as described for NCF. However, the source code provided by the authors uses 99 negative items. In our experiments we have used 99 negative items.}
The Hit Rate and NDCG at list length 10 are used as metrics.
The data splits that were used in the experiments were not shared by the authors, therefore we created data splits based on the information in the paper.

As personalized baselines, the authors consider NCF \cite{he2017neural} as well as the baselines reported in that article, i.e., eALS and ItemKNN. However, the authors used NCF with binarized feedback, while DMF used explicit feedback.
Hyperparameters for the machine learning methods were tuned on a validation set built from the training set by randomly sampling one interaction per user, and the authors report that eALS and NCF were tuned as in the original papers. For the ItemKNN method, no details about neighborhood sizes or the used similarity function are provided.

\paragraph{Results and Discussion}
We reproduced the experiments reported by the authors based on the code that was provided to us upon request.
We ran DMF with both loss functions that were also evaluated in the paper but could not confirm that the normalized version \emph{nce} leads to accuracy improvements over the binary version \emph{bc}.
Our results however revealed that for three out of four datasets one of the simple baselines outperformed DMF on both measures.

Table \ref{tab:DMF-results-movielens} shows the results for the MovieLens datasets, which are among the most often used ones in the literature.
The results obtained by DMF are better than traditional nearest-neighbor baselines on the MovieLens1M dataset, but slightly worse than those obtained with the \iALS and SLIM.
For the smaller MovieLens100k dataset, the observed ranking is generally similar and again \iALS and SLIM outperform DMF on both measures.

\begin{table}
\caption{Experimental results for DMF for the MovieLens1M (left) and MovieLens100k (right) datasets. We highlight in bold those entries where a baseline outperforms DMF.}
\label{tab:DMF-results-movielens}
\resizebox{\linewidth}{!}{
\begin{minipage}{0.6\textwidth}
%\begin{table}[h!t]
%    \caption{Experimental results for DMF for the MovieLens1M (left) and  datasets.}
%    \label{tab:DMF-results-movielens}
    \begin{tabular}{lcccccc}
    \toprule
    & HR@10 	& NDCG@10 	\\
    \midrule
    TopPopular	&0.4418	&0.2475	\\
    \midrule
    UserKNN CF asymmetric	&0.6324	&0.3779	\\
    ItemKNN CF cosine	&0.6347	&0.3808	\\ \midrule
    \palpha	&0.6097	&0.3639	\\
    \pbeta \footnotemark	&0.6304	&0.3726	\\
    \midrule
    \EASER & 0.6693 &  \textbf{0.4100} \\
    SLIM 	&\textbf{0.6825}	&\textbf{0.4209}	\\
    PureSVD	&0.6570	&0.4015	\\
    \iALS	&\textbf{0.6947}	&\textbf{0.4257}	\\
    \midrule
    DMF \emph{nce}	&0.6266	&0.3768	\\
    DMF \emph{bc}	&0.6731	&0.4033	\\
    \bottomrule
  	\end{tabular}
%    \captionof{table}{MovieLens 1M}
%\end{table}

% Get the footnote number in the counter to reuse it later on
\setcounter{fncounteralgpredatespbeta}{\thefootnote}

\end{minipage}
\begin{minipage}{0.6\textwidth}
%\begin{table}[h!t]
%    \caption{DMF ML 100k}
%    \label{tab:DMF-results-ml100k}
    \begin{tabular}{lcccccc}
    \toprule
    & HR@10 	& NDCG@10 	\\
    \midrule
    TopPopular	&0.4145	&0.2342	\\
    \midrule
    UserKNN CF asymmetric	&0.5994	&0.3492	\\
    ItemKNN CF tversky	&0.6026	&0.3506	\\ \midrule
    \palpha	&0.5717	&0.3421	\\
    \pbeta \footnotemark[\thefncounteralgpredatespbeta]	&0.5685	&0.3270	\\
    \midrule
    \EASER &  \textbf{0.6089} & 0.3571 \\
    SLIM 	&\textbf{0.6238}	&\textbf{0.3765}	\\
    PureSVD	&0.5877	&0.3555	\\
    \iALS	&\textbf{0.6142}	&\textbf{0.3691}	\\
    \midrule
    DMF \emph{nce}	&0.5930	&0.3410	\\
    DMF \emph{bc}	&0.6026	&0.3623	\\
    \bottomrule
  	\end{tabular}
%     \captionof{table}{MovieLens 100k}

\end{minipage}
}
\end{table}

\footnotetext{\pbetanotyetpublished\label{fn:alg_predates_pbeta}}

The detailed results for the Amazon datasets are shown in Table \ref{tab:DMF-results-amazon}.
The results for \EASER are missing for the Amazon Movies dataset, as the author-provided Python implementation of the method needed too much memory.
For the Amazon Music dataset it is interesting to observe that the simple UserKNN and \pbeta methods work better here than other machine learning models, and also better than DMF.
For the Amazon Movies dataset, the DMF method was actually much better than all other methods on both measures.
In particular the gains in terms of the Hit Rate are substantial and much higher than the second best method \iALS.

Like for the case of CDL and the CVAE methods described in Sections \ref{sec:cdl} and \ref{sec:cvae}, a better performance could only be observed for one of the datasets.
In the case of CDL and CVAE better results were obtained for very sparse datasets with only one training interaction per user.
Looking at the Amazon Movies dataset characteristics in the context of the DMF method, we can see that it is extremely sparse after the pre-processing step, in which 80\% of the interactions were removed. In the end, there are 878k remaining interactions for over 80k movies. The Amazon Music dataset is even sparser. After the pre-processing step, which removes more than 94\% of the interactions, it has only 46k interactions for 18k items.
Further investigations are necessary to better understand why DMF works so well in this case, which could help us design algorithms that also work well on other datasets with similar characteristics.
%\pc{should we put this in the discussion: the hypothesis here is that same of the DL methods work well for extremely spars datasets (e.g., when we do have one interaction per user in the training)}
%\djx{Would rather not further elaborate on this. It is mostly speculation and reviewers might ask us to do some experiments where we artificially reduce the density level.}

% TODO: Any hypothesis why this works so well here?

\begin{table}
\caption{Experimental results for DMF for the Amazon Music (left) and Amazon Movies (right) datasets. \EASER results for Amazon Movies are missing because the code required too much memory. We highlight in bold those entries where a baseline outperforms DMF.}
\label{tab:DMF-results-amazon}
\resizebox{\linewidth}{!}{
\begin{minipage}{0.6\textwidth}
    \begin{tabular}{lcccccc}
        \toprule
    & HR@10 	& NDCG@10 	\\
    \midrule
    TopPopular	&0.5308	&0.3037	\\
    \midrule
    UserKNN CF cosine	&\textbf{0.6694}	&0.4798	\\
    ItemKNN CF cosine	&0.6647	&\textbf{0.4880}	\\
    \palpha	&0.6588	&\textbf{0.4823}	\\
    \pbeta \footnotemark[\thefncounteralgpredatespbeta]	&\textbf{0.6754}	&\textbf{0.4912}	\\
    \midrule
    \EASER & 0.6600 &  \textbf{0.4836} \\
    SLIM 	&0.6469	&0.4744	\\
    PureSVD	&0.5912	&0.4190	\\
    \iALS	&0.6600	&\textbf{0.4880}	\\
    \midrule
    DMF nce	&0.4799 & 0.3371 \\
    DMF bc	&0.6659 & 0.4815 \\
    \bottomrule
  	\end{tabular}
\end{minipage}
\begin{minipage}{0.6\textwidth}
    \begin{tabular}{l|cccccc|}
    \toprule
    & HR@10 	& NDCG@10 	\\
    \midrule
    TopPopular	&0.5794	&0.3489	\\
    \midrule
    UserKNN CF cosine	&0.7327	&0.5132	\\
    ItemKNN CF asymmetric	&0.6986	&0.4914	\\ \midrule
    \palpha	&0.6972	&0.5028	\\
    \pbeta	\footnotemark[\thefncounteralgpredatespbeta] &0.7107	&0.5078	\\
    \midrule
    \EASER 	& - 	& - 	\\
    SLIM 	&0.6981	&0.5005	\\
    PureSVD	&0.6021	&0.4156	\\
    \iALS	&0.7352	&0.5230	\\
    \midrule
    DMF \emph{nce}	& 0.6832 & 0.4677 \\
    DMF \emph{bc}	& \textbf{0.7818} & \textbf{0.5417} \\
    \midrule
\end{tabular}
\end{minipage}
}
\end{table}

Regarding methodological aspects, we found that the authors reported the best Hit Rate and NDCG results across different epochs.
%, as was done for the NCF method. \mf{no it was not done for NCF}
We therefore report the numbers here that were obtained after determining a suitable number of epochs on the validation set. In that context, the provided code shows that the authors sample different negative items to be used for testing in each training epoch. This seems questionable as well, in particular when considered in combination of the practice of reporting the best value for each metric across epochs. In our experiments we use the same negative item set for all evaluations.

From a conceptual perspective, the authors argue that they combine implicit feedback and explicit feedback in their approach. While this might be true in some interpretation, the authors mainly rely on the explicit ratings and add zeros to the empty matrix cells. Furthermore, when comparing their method with NCF, they only fed the binarized data to NCF, even though this method could deal with explicit rating data as well.
% TODO Check if this is true.

% ---------------------------------------------------------
\subsection{Variational Autoencoders for Collaborative Filtering (Mult-VAE)}
\label{sec:vae}
% ---------------------------------------------------------

\paragraph{Datasets}
The authors use three datasets for evaluation.
The first two datasets contain explicit feedback in the form of movie ratings (\emph{MovieLens20M}\footnote{\url{https://grouplens.org/datasets/movielens/}} and \emph{Netflix}\footnote{\url{https://www.kaggle.com/netflix-inc/netflix-prize-data}}).
The third dataset contains play counts for musical tracks.
All datasets are binarized. For the movie datasets, ratings higher than three are considered positive signals and only users with more than five interactions are retained. For the music dataset, users with more than 20 interactions are retained; tracks that were listed less than 200 times are filtered out. After pre-processing, the datasets are still relatively large, having between 10 and almost 57 millions interactions.\footnote{We did not run experiments for the music dataset as the original paper did not contain sufficient information to guarantee we used the dataset in the exact same way as the authors.}

\paragraph{Evaluation}
Four machine learning models are used as baselines, \iALS, %\cite{hu2008IALS}
SLIM,
%\cite{ning2011SLIM}
NCF \cite{he2017neural} (see Section \ref{sec:ncf}) and the Collaborative Denoising Autoencoder (CDAE) method proposed in \cite{wu2016collaborativeCDAE} in 2016.
For evaluation purposes, the datasets are split into training, validation and test splits by holding out users.
For instance, for the MovieLens20M dataset (136k users overall), 10k users are removed for validation and 10k users are removed for testing.
For each hold-out user, 80\% of the interactions are used as user profile, and the remaining 20\% are used as ground truth to measure the performance metrics. The models are optimized for NDCG@100 on the validation set.
Performance results for Recall@20 and Recall@50 are reported as well.
%Then, a special protocol is used in this research to evaluate the method under ``strong generalization''. In the end, after training, 80\% of the interactions of each test users is used to learn user-level representations, and the remaining 20\% are used to determine the performance metrics. The models are optimized for NDCG@100 on the validation set. Performance results for Recall@20 and Recall@50 are reported as well.

%\pc{Check the consistency of eALS -> iALS}
%\djx{Moved this part to the evaluation section, as it is not about results. I also found it confusing to discuss two versions and then say we used one of them. I have slightly revised the text, removing the part about the mathematically equivalence please check.}
Note that in order to be able to use matrix factorization baselines on cold users we built the cold users' latent factors based upon both their user profile and the latent factors of the warm items. In particular, we added a hyperparameter to the matrix factorization models to select how those cold user's latent factors are estimated, either via an item based similarity or an item embeddings average, see \cite{Cremonesi:2010:PRA:1864708.1864721}. The first case is inspired by the \emph{folding-in} technique \cite{sarwar2002incremental} for the PureSVD model. It can be mathematically derived that PureSVD has an equivalent ItemKNN formulation, in which the similarity matrix is the product of the items' latent factor matrix by its transpose. In this case a further parameter is introduced to control the number of neighbors, as in the other neighborhood-based methods. In the second case the latent factors of a user are the product of the user profile and the items' latent factors, resulting in the average of the embeddings of the items the user interacted with. Both methods proved effective.

%Although the two approaches are mathematically equivalent, their numerical stability is sometimes different, leading to different results.

%In order to be able to use matrix factorization baselines on cold users we built the cold users' latent factors based upon both their user profile and on the latent factors of the warm items. In particular, we added a hyperparameter to the matrix factorization models to select how those cold user's latent factors are estimated, either via an item based similarity or an item embeddings average, see \cite{Cremonesi:2010:PRA:1864708.1864721}. In the first case an ItemKNN model is created by defining the similarity matrix as the product of the items' latent factor matrix by its \textcolor{blue}{transpose}. In the second case the latent factors of a user are the product of the user profile and the items' latent factors, resulting in the average of the embeddings of the items the user interacted with. \mf{Not sure about this}

%Although the two approaches are mathematically equivalent, their numerical stability is sometimes different, leading to different results. Generating an ItemKNN similarity matrix proved to be the most effective solution.

\paragraph{Results and Discussion}
Using the code, the data splits and information about the seed for the random number generator that the authors provided, we could reproduce the results from the paper.
In the original paper NDCG@100 is used as an optimization goal, however no reason is provided for this, as well as why Recall@20 and Recall@50 are used as additional measures, but not, e.g., Recall@100.
To obtain a more comprehensive picture, we made additional measurements at the corresponding but missing cut-off values: Recall@100, NDCG@20 and NDCG@50. Table \ref{tab:Mult-VAE-results-movielens} shows our results for the MovieLens20M dataset and Table \ref{tab:Mult-VAE-results-netflix} those for the Netflix data.

\begin{table}[h!t]
\caption{Results for Mult-VAE for the MovieLens20M dataset. UserKNN could not be applied because of the evaluation protocol (hold-out of users). We highlight in bold those entries where a baseline outperforms Mult-VAE.}
\label{tab:Mult-VAE-results-movielens}
\footnotesize
% \resizebox{.9\linewidth}{!}{%
    \begin{tabular}{lcccccc}
    \toprule
	& REC@20 	& NDCG@20 	& REC@50 	& NDCG@50 	& REC@100 	& NDCG@100 	\\
	\midrule	
    TopPopular	&0.1441	&0.1201	&0.2320	&0.1569	&0.3296	&0.1901	\\
    \midrule
    UserKNN 	& - 	& - 	& - 	& - 	& - 	& - 	\\
    ItemKNN CF asymmetric	&0.2937	&0.2444	&0.4486	&0.3087	&0.5709	&0.3527	\\
    \midrule
    \palpha	&0.2620	&0.2168	&0.4047	&0.2742	&0.5287	&0.3182	\\
    \pbeta	&0.3006	&0.2501	&0.4540	&0.3133	&0.5797	&0.3583	\\
    \midrule
    \EASER & 0.3530 &  \textbf{0.3074} & 0.5147 &  \textbf{0.3755} &           0.6353 &  \textbf{0.4196} \\
    SLIM 	&0.3356	&0.2920	&0.4893	&0.3576	&0.6110	&0.4017	\\
    PureSVD   &  0.2935 &  0.2514 &  0.4371 &  0.3117 &  0.5544 &  0.3538 \\
    \iALS	&  0.2968 &  0.2496 &  0.4406 &  0.3090 &  0.5631 &  0.3521 \\
    \midrule
    Mult-VAE              &  \textbf{0.3541} &           0.2988 &  \textbf{0.5222} &           0.3690 &  \textbf{0.6517} &           0.4158 \\

    \bottomrule
    \end{tabular}
    % }
\end{table}

\begin{table}[h!t]
\caption{Results for Mult-VAE for the Netflix dataset. UserKNN could not be applied because of the evaluation protocol (hold-out of users). We highlight in bold those entries where a baseline outperforms Mult-VAE.}
\label{tab:Mult-VAE-results-netflix}
\footnotesize
% \resizebox{.9\linewidth}{!}{%
    \begin{tabular}{lcccccc}
    \toprule
	& REC@20 	& NDCG@20 	& REC@50 	& NDCG@50 	& REC@100 	& NDCG@100 	\\
	\midrule	
    TopPopular	&0.0786	&0.0762	&0.1643	&0.1159	&0.2717	&0.1570	\\
    \midrule
    UserKNN 	& - 	& - 	& - 	& - 	& - 	& - 	\\
    ItemKNN CF cosine	&0.2091	&0.1970	&0.3387	&0.2592	&0.4598	&0.3092	\\
    \midrule
    \palpha	&0.1960	&0.1759	&0.3325	&0.2412	&0.4633	&0.2962	\\
    \pbeta	&0.2210	&0.2053	&0.3633	&0.2739	&0.4932	&0.3281	\\
    \midrule
    \EASER &  \textbf{0.2681} &  \textbf{0.2591} &  \textbf{0.4170} &  \textbf{0.3334} &  \textbf{0.5471} &  \textbf{0.3890} \\
    SLIM 	&0.2555	&\textbf{0.2479}	&0.4002	&\textbf{0.3203}	&0.5299	&\textbf{0.3752}	\\
    PureSVD	&  0.2271 &  0.2184 &  0.3593 &  0.2840 &  0.4784 &  0.3342 \\
   \iALS	&  0.1956 &  0.1839 &  0.3138 &  0.2410 &  0.4216 &  0.2862 \\
    \midrule
    Mult-VAE & 0.2615 & 0.2423 & 0.4127 & 0.3167 & 0.5456 & 0.3730 \\
    \bottomrule
    \end{tabular}
    % }
\end{table}

%\pc{Some results from the tables are missing} \mf{Added missing results and draft description on how the MF models have been applied to cold users}
%\djx{I would be interesting to see the results when the hold out is not done per user but randomly. But this is probably not in the scope of our work.} \mf{I tried that for the recsys paper and Mult-VAE remained competitive}

For the MovieLens dataset, we observed a positive result and could confirm the claims made by the authors of Mult-VAE.
On all measurements, both the original and the additional ones, Mult-VAE leads to better performance results than all baseline methods which were available at the time the algorithm was proposed.
SLIM is the second best method in this evaluation, with performance results that are around 1\% to 2\% lower in terms of the NDCG.

For the Netflix dataset, the claims of the authors could not be confirmed to the full extent.
In terms of NDCG, which is the optimization criterion, SLIM outperforms Mult-VAE at all list lengths.
Mult-VAE is however better in terms of Recall.

Overall, with Mult-VAE a method was found which was easy to reproduce, thanks to all needed material being made by the authors publicly available.
Furthermore, as our results indicated, the method consistently outperformed previous methods at least on one well-known and comparably large dataset. We could also confirm that \EASER leads to improvements over Mult-VAE in most cases, supporting the claim that shallow models are a competitive solution.

% ------------------------------------------------------------------------
\subsection{NeuRec: On Nonlinear Transformation for Personalized Ranking}
\label{sec:neurec}
% ------------------------------------------------------------------------
% \paragraph{Method} \emph{NeuRec} \cite{Zhang2018Neurec} was presented at IJCAI '18. The work aims at learning user-item relationships from implicit feedback and combines latent factor models with neural networks in order to capture both linear and non-linear dependencies in the data. Technically, the user-item interaction matrix is first mapped into a low-dimensional space with multi-layered networks. Recommendations are then generated by computing the inner product of item and user latent factors. A user-based and an item-based variant are proposed.

\paragraph{Datasets} The authors use four public datasets for their evaluations (\emph{MovieLens1M}\footnote{\url{https://grouplens.org/datasets/movielens/}}, \emph{HetRec}\footnote{\url{https://grouplens.org/datasets/hetrec-2011/}}, \emph{FilmTrust}\footnote{\url{https://www.librec.net/datasets/filmtrust.zip}}, \emph{Frappe}\footnote{The original download link was not active anymore at the time or our analysis. We provide an alternative download link \url{https://github.com/hexiangnan/neural_factorization_machine/archive/master.zip}}).
Three of them contain movie ratings, which are binarized by converting all ratings to 1 and the missing entries to 0.
The largest dataset (MovieLens1M) comprises 1 million interactions.
The fourth dataset (Frappe) is from the domain of mobile app recommendation and contains about 100k interactions, reduced to 20k after pre-processing, which consisted of the removal of multiple interactions between the same user-item pairs.
Data splits were not provided online but could be reproduced based on the information in the paper.

\paragraph{Evaluation} The authors use five random training-test splits (80\%/20\%) for evaluation and report the average results. As performance metrics, the authors use Precision and Recall at list lengths 5 and 10, as well as Mean Average Precision, MRR, and the NDCG, at list length 10.
% TODO: Could not find this information in the paper

As non-trivial baselines, the authors consider SLIM, BPR matrix factorization, NeuMF and the GMF model, which is part of NCF \cite{he2017neural}.
%\pc{BPR: I assume is Matrix Factorization with BPR loss function. If so, we should name it BPR matrix factorization. This appears in several other places.}
Information about hyperparameter tuning for the baselines is not provided except for GMF and NeuMF, which are said to use the ``default'' configuration as described in the original article.
Hyperparameters for NeuRec were determined through grid search and the finally used values are reported in the paper in detail.
The number of training epochs is not reported in the paper.\footnote{According to an exchange of emails with the authors, the training was done for a large number of epochs and the best performance values on the test set were reported.}
As usual we selected the number of epochs via early-stopping on a validation split.

\paragraph{Results and Discussion}
%We reproduced the work based on the public code and datasets.
%\mf{Correction, we could NOT reproduce their results in any way.}
Even though the authors published a runnable implementation of their method and provided detailed information on the hyperparameters, we could \emph{not} obtain the results reported in the original paper. We contacted the authors but we were not able to reconstruct an experimental pipeline (from pre-processing to hyperparameter optimization) that led to results that were comparable to the ones reported in the original paper. In the end, the reason for this discrepancy could not be clarified.
The outcome of our evaluation is that NeuRec is outperformed on any data set and almost on any measure by at least one, but usually several, of the baselines in our comparison.

Since the detailed results are comprehensive, given the number of datasets and evaluation measures, we only provide in Table \ref{tab:neurec-results} the results for the most commonly used MovieLens1M dataset and for list lengths 5 and 10. All other results can again be found in the online appendix (Section \ref{sec:methodology}).

Looking at the results, we observe that on the MovieLens dataset even the simplest baselines are better than NeuRec and that the performance of the best baselines is better by a large margin. For the HetRec and FilmTrust datasets (not shown in detail here), the picture is mostly the same. Finally, for the small and rarely used Frappe dataset, NeuRec actually leads to the best results for Precision@5, but is outperformed, e.g., by \pbeta on all other measures.\footnote{We point out that the Frappe dataset is very small and exhibits very unstable results depending on the random split. In particular, compared to the results reported in the original paper, our TopPopular algorithm exhibits results that are four times higher; also NeuRec's values are two times higher.} %\mf{I believe we should remove the results on Frappe. The dataset is microscopic and exhibits incredibly unstable results, our TopPopular is 4x the one reported in the original paper and NeuRec 2x.}  \djx{Would add this information in a footnote.}

\begin{table}[h!t]
    \caption{Experimental results for NeuRec for the MovieLens1M dataset. We highlight in bold those entries where a baseline outperforms NeuRec.}
    \label{tab:neurec-results}
    \footnotesize
    \resizebox{\linewidth}{!}{
    \begin{tabular}{lccccc|ccccc}
    \toprule
    	& \multicolumn{5}{c}{@5}  \vline & \multicolumn{5}{c}{@10} \\
		& PREC 	& REC 	& MAP 	& NDCG 	& MRR 	& PREC 	& REC 	& MAP 	& NDCG 	& MRR 		\\
	\midrule
    TopPopular	&0.2105	&0.0402	&0.1531	&0.0689	&0.3621	&0.1832	&0.0685	&0.1168	&0.0939	&0.3793	\\
    \midrule
    UserKNN CF asymmetric	&\textbf{0.4212}	&\textbf{0.1065}	&\textbf{0.3441}	&\textbf{0.1674}	&\textbf{0.6399}	&\textbf{0.3617}	&\textbf{0.1726}	&\textbf{0.2774}	&\textbf{0.2230}	&\textbf{0.6509}	\\
    ItemKNN CF asymmetric	&\textbf{0.3995}	&\textbf{0.0984}	&\textbf{0.3244}	&\textbf{0.1563}	&\textbf{0.6179}	&\textbf{0.3452}	&\textbf{0.1590}	&\textbf{0.2618}	&\textbf{0.2084}	&\textbf{0.6293}	\\ \midrule
    \palpha	&\textbf{0.4041}	&\textbf{0.1007}	&\textbf{0.3286}	&\textbf{0.1596}	&\textbf{0.6250}	&\textbf{0.3456}	&\textbf{0.1627}	&\textbf{0.2627}	&\textbf{0.2121}	&\textbf{0.6362}		\\
    \pbeta	&\textbf{0.4080}	&\textbf{0.1007}	&\textbf{0.3325}	&\textbf{0.1602}	&\textbf{0.6260}	&\textbf{0.3508}	&\textbf{0.1639}	&\textbf{0.2676}	&\textbf{0.2137}	&\textbf{0.6374}\\
    \midrule
    \EASER &  \textbf{0.4488} &  \textbf{0.1134} &  \textbf{0.3717} &  \textbf{0.1779} &  \textbf{0.6620} &  \textbf{0.3857} &  \textbf{0.1820} &  \textbf{0.3035} &  \textbf{0.2364} &  \textbf{0.6717}		\\
    SLIM 	&\textbf{0.4437}	&\textbf{0.1106}	&\textbf{0.3692}	&\textbf{0.1749}	&\textbf{0.6578}	&\textbf{0.3813}	&\textbf{0.1770}	&\textbf{0.3003}	&\textbf{0.2321}	&\textbf{0.6679}	\\
    PureSVD	&\textbf{0.4123}	&\textbf{0.0987}	&\textbf{0.3371}	&\textbf{0.1586}	&\textbf{0.6266}	&\textbf{0.3575}	&\textbf{0.1624}	&\textbf{0.2722}	&\textbf{0.2132}	&\textbf{0.6380}		\\
    \iALS	&\textbf{0.4164}	&\textbf{0.1036}	&\textbf{0.3373}	&\textbf{0.1635}	&\textbf{0.6327}	&\textbf{0.3628}	&\textbf{0.1702}	&\textbf{0.2743}	&\textbf{0.2200}	&\textbf{0.6443}	\\
    \midrule
    INeuRec	&0.3280	&0.0663	&0.2554	&0.1110	&0.5003	&0.2839	&0.1094	&0.2027	&0.1500	&0.5129	\\
    UNeuRec	&0.2098	&0.0395	&0.1560	&0.0684	&0.3663	&0.1856	&0.0688	&0.1199	&0.0944	&0.3852	\\
    \bottomrule
  	\end{tabular}
 	}
\end{table}

Regarding methodological aspects, we found again that researchers optimized the number of epochs on the test set and apparently reported the best results of NeuRec for different measures at, potentially, different training epochs.
Furthermore, it is unclear from the paper if hyperparameter optimization was done for the baselines.
For the NCF method, the authors state that they used the ``configuration'' proposed in the original paper, but it is unclear if this refers to the network structure, the hyperparameters, or both.

% ------------------------------------------------------------------------
\subsection{CoupledCF: Learning Explicit and Implicit User-item Couplings}
\label{sec:coupledCF}
% ------------------------------------------------------------------------
% \paragraph{Method}
% \emph{CoupledCF} \cite{Zhang2018CoupledCF} was also presented at IJCAI '18. The approach is based on the observation that users and items in real-world datasets are not independent and identically distributed. The proposed method therefore aims to learn implicit and explicit couplings between users and items and to thereby leverage available side information (e.g., user demographics, item features) more effectively. Technically, a complex architecture is used, involving a CNN for learning the couplings based on the side information, and a deep CF model that considers explicit and implicit interactions between users and items.
% TODO: Check if the descriptions are ok.
%The authors' claim is that the method \enquote{significantly outperforms two latest neural recommenders}.

\paragraph{Datasets}
Experiments were made on two public datasets, the \emph{MovieLens1M}\footnote{\url{https://grouplens.org/datasets/movielens/}} rating dataset and a dataset called \emph{Tafeng}\footnote{The original download link was not active anymore at the time or our analysis. We provide an alternative download link \url{https://www.kaggle.com/chiranjivdas09/ta-feng-grocery-dataset}} containing grocery store transactions.
The Tafeng dataset has about 750k transactions (i.e., less than the MovieLens1M dataset), but is much more sparse as it contains many more users and items.
The explicit ratings in the MovieLens1M dataset are transformed into binary ratings, where each rating is considered as a positive interaction.
Both datasets contain side information about users and items that is used by the CoupledCF algorithm.
Therefore, we have included item-based and user-based content techniques among the baselines.
The authors of the paper provided us with the train-test splits, including the sampled test negative items they had used during the evaluation.
% In their training, the authors sampled four negative items per positive interaction.
% These samples were provided to us as well.
% TODO: why for negative for one positive, why not 50? \mf{Those negatives are used for training and are not provided to us, only the test negatives are.}

\paragraph{Evaluation}
A leave-one-out procedure is used, where for each user one random interaction is put in the test set. For evaluating the performance, 99 items are sampled for which there was no interaction for the given user. The 100 items are then ranked by the algorithm and the Hit Rate and the NDCG are used to evaluate the performance. Cut-off list lengths between 1 and 10 were considered.

The hyperparameters of the proposed model were systematically fine-tuned by the authors. Information about hyperparameter tuning for the baselines is not provided. The considered baselines include NCF \cite{he2017neural}, and Google's Wide\&Deep method.% \cite{wideanddeep2016}.
From the original source code we observed that the authors reported the best results for each metric on the test data across different epochs.
Since this is inappropriate, we report the values of the metrics after the optimal number of epochs is chosen with early stopping on the validation set.

\paragraph{Results and Discussion}
We could not fully reproduce the results by the authors based on the provided code and data splits.
%\mf{The CoupledCF part has a convolution, the authors state they use 8,8 filters while in the code there is a 3,3 filter. If I change it to 8,8 the code does not work anymore. DeepCF is reproducible while Coupled CF is not}
Different variations of the proposed model were tested in the original paper. We used the best-performing one (CoupledCF) in our experiments as well as their simplest variant (DeepCF), where the former is the only one for which we could fully reproduce the results. Since there were some apparent issues regarding the way the authors did the sampling, which we discuss below, we recreated the train-test splits according to the descriptions in the paper. We also used a validation set for hyperparameter tuning. While the numerical results obtained using the split provided by the original authors and the one generated by us are different, the relative ordering of the algorithms in terms of their recommendation quality remains the same. We report the results obtained on the data split generated by us.
As previously mentioned, we observe there were some apparent issues regarding the way the authors sampled the negative items. For both Movielens and Tafeng the negative item data contains duplicates leading to many users (72\% for Movielens, 28\% for Tafeng) having less of the desired 99 negative items, e.g., some have only 93. For the Tafeng dataset, cumulatively, almost 3,000 negative items (0.1\% of the total number of negative items) also appeared as train items or test items for that same user. If the number of unique negative items is not constant, different users will be evaluated under \emph{slightly} different conditions, in particular, the recommendation problem becomes easier as the number of negatives decreases due to the reduction in the pool of alternatives the recommender has to choose from.
We nonetheless assume that the impact of these inconsistencies will generally be small, since both the original split and the one generated by us exhibit the same ranking of the algorithms in terms of their performance.
More strikingly, however, for 8\% of the users in the Tafeng dataset we observed problems regarding the usefulness of the test data. Some of these users have a test item but no negative items, meaning that there is only a single item that can be ranked and even a random recommender will exhibit perfect recommendations. Other users have negative items but no test item, meaning that no correct recommendations are possible.

Our experimental results are shown in Table \ref{tab:coupledcf-results-movielens} (MovieLens) and Table \ref{tab:coupledcf-results-tafeng} (Tafeng).

\begin{table}[h!t]
    \caption{Experimental results for CoupledCF for the MovieLens1M dataset. We highlight in bold those entries where a baseline outperforms CoupledCF.}
    \label{tab:coupledcf-results-movielens}
    \footnotesize
% \resizebox{.8\linewidth}{!}{%
    \begin{tabular}{lcccccc}
    \toprule
		& HR@1 	& NDCG@1 	& HR@5 	& NDCG@5 	& HR@10 	& NDCG@10 	\\
	\midrule
    TopPopular   &   0.1593 &   0.1593 &   0.4217 &   0.2936 &   0.5813 &   0.3451 \\
    \midrule
    UserKNN CF asymmetric	&   0.3546 &   0.3546 &   0.6914 &   0.5343 &   0.8114 &   0.5735 \\
    ItemKNN CF cosine	 &   0.3305 &   0.3305 &   0.6682 &   0.5080 &   0.7940 &   0.5488 \\
    \midrule
    \palpha	 &   0.3316 &   0.3316 &   0.6543 &   0.5031 &   0.7687 &   0.5402 \\
    \pbeta	 &   0.3464 &   0.3464 &   0.6743 &   0.5198 &   0.7959 &   0.5591 \\
    \midrule
    \EASER &  \textbf{0.4003} &  \textbf{0.4003} &  \textbf{0.7258} &  \textbf{0.5738} &  \textbf{0.8343} &  \textbf{0.6093} \\
    SLIM 	  &  \textbf{0.3906} &  \textbf{0.3906} &  \textbf{0.7116} &  \textbf{0.5625} &  \textbf{0.8315} &  \textbf{0.6014} \\
    PureSVD	 &  \textbf{0.3735} &  \textbf{0.3735} &  \textbf{0.7088} &  \textbf{0.5522} &   0.8132 &  \textbf{0.5861} \\
    \iALS	&  \textbf{0.3816} &  \textbf{0.3816} &  \textbf{0.7121} &  \textbf{0.5581} &   0.8200 &  \textbf{0.5933} \\
    \midrule
    ItemKNN CBF asymmetric	 &   0.0884 &   0.0884 &   0.2586 &   0.1752 &   0.3780 &   0.2137 \\
    UserKNN CBF tversky	&   0.1714 &   0.1714 &   0.4427 &   0.3108 &   0.6065 &   0.3636 \\
    \midrule
    ItemKNN CFCBF cosine  &   0.3328 &   0.3328 &   0.6694 &   0.5107 &   0.7985 &   0.5526 \\
    UserKNN CFCBF  dice	&  \textbf{0.3555} &  \textbf{0.3555} &           0.6869 &           0.5328 &           0.8008 &           0.5698 \\
    \midrule
    DeepCF   &   0.3550 &   0.3550 &   0.7017 &   0.5388 &   0.8272 &   0.5794 \\
    CoupledCF         &   0.3522 &   0.3522 &   0.7018 &   0.5374 &   0.8247 &   0.5775 \\
    \bottomrule
  	\end{tabular}
%  	}
\end{table}

\begin{table}[h!t]
    \caption{Experimental results for CoupledCF for the Tafeng dataset. We highlight in bold those entries where a baseline outperforms CoupledCF.}
    \label{tab:coupledcf-results-tafeng}
    \footnotesize
% \resizebox{.8\linewidth}{!}{%
    \begin{tabular}{lcccccc}
    \toprule
		& HR@1 	& NDCG@1 	& HR@5 	& NDCG@5 	& HR@10 	& NDCG@10 	\\
	\midrule
    TopPopular     &  \textbf{0.2654} &  \textbf{0.2654} &           0.5194 &           0.3965 &           0.6549 &           0.4402 \\
    \midrule
    UserKNN CF cosine	&  \textbf{0.3215} &  \textbf{0.3215} &  \textbf{0.5412} &  \textbf{0.4369} &           0.6415 &  \textbf{0.4693} \\
    ItemKNN CF cosine	&  \textbf{0.3314} &  \textbf{0.3314} &  \textbf{0.5424} &  \textbf{0.4427} &           0.6376 &  \textbf{0.4735} \\
    \palpha    &  \textbf{0.3245} &  \textbf{0.3245} &  \textbf{0.5503} &  \textbf{0.4437} &           0.6404 &  \textbf{0.4730} \\
    \pbeta    &  \textbf{0.3202} &  \textbf{0.3202} &  \textbf{0.5525} &  \textbf{0.4424} &           0.6470 &  \textbf{0.4732} \\
    \midrule
    \EASER &  \textbf{0.3272} &  \textbf{0.3272} &  \textbf{0.5452} &  \textbf{0.4417} &           0.6435 &  \textbf{0.4736} \\
    SLIM 	&  \textbf{0.3233} &  \textbf{0.3233} &  \textbf{0.5438} &  \textbf{0.4389} &           0.6476 &  \textbf{0.4726} \\
    PureSVD	 &           0.2462 &           0.2462 &           0.4889 &           0.3714 &           0.6260 &           0.4156 \\
    \midrule
    ItemKNN CBF asymmetric	&           0.0589 &           0.0589 &           0.0958 &           0.0769 &           0.1467 &           0.0931 \\
    UserKNN CBF asymmetric	&           0.2464 &           0.2464 &           0.4654 &           0.3600 &           0.5798 &           0.3970 \\
    \midrule
    ItemKNN CFCBF asymmetric	&  \textbf{0.3331} &  \textbf{0.3331} &  \textbf{0.5434} &  \textbf{0.4442} &           0.6314 &  \textbf{0.4727} \\
    UserKNN CFCBF asymmetric &  \textbf{0.3424} &  \textbf{0.3424} &  \textbf{0.5882} &  \textbf{0.4713} &  \textbf{0.6937} &  \textbf{0.5055} \\
    \midrule
    DeepCF     &           0.2647 &           0.2647 &           0.5244 &           0.3995 &           0.6583 &           0.4428 \\
    CoupledCF  &           0.2641 &           0.2641 &           0.5175 &           0.3948 &           0.6499 &           0.4377 \\
    \bottomrule
  	\end{tabular}
%  	}
\end{table}

For the MovieLens datasets we can observe that CoupledCF is almost consistently able to outperform the simple neighborhood-based methods and the hybrids (except for very short list lengths).
However, relatively simple non-neural methods %\pc{is it correct to say "simple linear models"} \djx{I guess so. the term is used in both papers (on slim and easer) added: ``relatively''}
like \iALS and \EASER are consistently better than CoupledCF.
The differences between the CoupledCF and the DeepCF variant are tiny and the simpler DeepCF is sometimes even better. This stands in contrast to the results in the original article, where the differences were large.

%\mf{In their article that difference was huge. CoupledCF is not reproducible}

For the Tafeng dataset, even the nearest-neighbor methods outperform CoupledCF by far.
Only the pure content-based baselines do not reach the performance level of CoupledCF.
% \pc{In this experiments we introduce user-based content-based. Check if we did explain them in the baseline section.} \mf{Done}
Generally, on this dataset, the performance of the proposed method is at the level of the TopPopular baseline. The simpler DeepCF method also leads to better accuracy results than the CoupledCF variant.

% Methodological considerations as mentioned in the long paper.
Looking at methodological aspects, it seems that the baselines were not properly optimized and default hyperparameters were used.
Furthermore, from the provided code it seems that the number of epochs was determined on the test set, as was done in other papers examined in this work.
A specific problem in this work also lies in the creation of the train and test splits which are inconsistent with the description reported in the article and are likely the result of an erroneous splitting procedure. %\mf{explained in detail before}% We found that for some users in the test set there were no negative samples. %\pc{Maurizio, can you explain this better to me? It is not in coherent with the problem reported before: duplicate negative instances and test interaction in the training.}
% TODO Check.
%In other cases, the positive (hidden) item was part of the negative samples used for ranking.
% TODO: Could mention that with NCF a method was used as a baseline that does not use side information

Further methodological problems were discussed in a recent article by Ferrari Dacrema et al. \cite{FerrariDacrema2020convolution} which questioned the claim that embedding maps derived from the outer product of the embeddings are analogous to images. The article shows that embedding maps do not have a semantically relevant topology and do not share image properties, therefore the use of CNNs is not well justified. Moreover, contrary to what stated in the original paper, CoupledCF does not benefit from the additional parameter space provided by the embedding correlations which can be removed from the model without altering its accuracy.

\subsection{DELF: A Dual-Embedding based Deep Latent Factor Model for Recommendation}
\label{sec:delf}
% ------------------------------------------------------------------------
% \paragraph{Method}
% The \emph{DELF} model \cite{Cheng2018DELFAD}, presented at IJCAI '18, was designed for \emph{top-n} recommendation tasks given implicit feedback data.
% Inspired by previous work (NSVD) \cite{Paterek2007ImprovingRS}, the authors propose to learn \emph{dual} embeddings to capture certain interactions in the data.
% Instead of using only the common user embedding, the authors propose to learn an additional item-based user embedding and vice versa for item embeddings. The embeddings are then combined to model non-linear interactions between users and items within a deep learning architecture. Through this approach the authors  generalize ideas of NSVD and Neural Collaborative Filtering (NCF). Two variants of the approach, \emph{DELF-MLP} and \emph{DELF-EF} were investigated in the original paper.

\paragraph{Datasets} Two public rating datasets are used for the evaluation. One is the well-known \emph{MovieLens1M}\footnote{\url{https://grouplens.org/datasets/movielens/}} dataset and the other one the \emph{Amazon Music}\footnote{\url{http://jmcauley.ucsd.edu/data/amazon/}} dataset.
%, which was also used, e.g., for the evaluation of the DMF model discussed in Section \ref{sec:dmf}.
The rating datasets were binarized by transforming each non-zero rating to 1.
Pre-processing was applied so that for both datasets only users were retained for which more than 20 interactions were observed.
Through this pre-processing, the Amazon Music dataset was reduced to less than one tenth of its original size in terms of interactions (only 76k interactions for 40k items remain in the Amazon Music dataset). The Amazon Music dataset contains, on average, only 2 interactions per item. Due to this, when the train-test split is performed according to the timestamp, 52\% of the items in the test data are cold (i.e., they never appear in the train data for any user). An evaluation so strongly oriented towards cold items is particularly surprising, considering that DELF is a pure collaborative algorithm that, as such, cannot be able to learn a representation for any of those cold items.
%The sparsity of the Amazon dataset was still much higher than for the MovieLens dataset.

%\pc{Check for consitency on verbs when we describe soa algos: sometimes we use past, some times we use present. Use the same tense for all DL algos.}

\paragraph{Evaluation}
The evaluation procedure was similar to the one used for NCF %CoupledCF
as discussed in Section \ref{sec:ncf}: a leave-last-out procedure was applied using the interaction timestamp, the hidden element was ranked within 99 randomly sampled negative items, and the Hit Rate and the NDCG at a cut-off length of 10 were used as performance measures. %\footnote{As previously discussed in Section \ref{sec:ncf} the timestamp reported in the Movielens dataset should not be used in this fashion as it does not represent the occurrence of the original interaction but rather when the feedback was added to the system.}
The non-trival baselines include BPR matrix-factorization, %\cite{rendle2009bprMF},
\iALS, %\cite{He:2016:FMF:2911451.2911489},
DMF as discussed in Section \ref{sec:dmf},
and two variants of the NCF model \cite{he2017neural} described in Section \ref{sec:ncf}.

The hyperparameters for the proposed model were systematically optimized on a validation set built with the second most recent interaction. No information is provided regarding the hyperparameter optimization of the baselines.

\paragraph{Results and Discussion}
We reproduced the results using a validation set which we constructed in the same way as reported by the authors.
NDCG@10 was used as an optimization criterion.

% \mf{NEW} \djx{Small revision, and discussing ML1M first, please double-check}
The obtained results are shown in Table \ref{tab:delf-movielens} (MovieLens1M) and Table \ref{tab:delf-amazon-colditems} (Amazon Music).
For the MovieLens1M dataset, all our machine learning baselines outperformed \emph{DELF} on all measures. The traditional UserKNN method is also competitive with DELF on this dataset.
On Amazon Music, however, DELF EF leads to substantially better accuracy results than all baselines. As we previously observed, the test data of Amazon Music contains an anomalously high number of cold items, for which pure collaborative filtering approaches cannot learn a representation. %Moreover the DELF EF variant takes the popularity of the items into account when computing the recommendations.
% \djx{I could not understand what this is about the popularity, commented it out. If it is popularity-biased, it should almost never recommend items that never appeared in the training data. The discussion should be improved.} \mf{Yes it was unclear. DELF EF is based on a heuristic which takes the item popularity into account, I suppose there is a numerical effect when that is zero that has the effect of pushing the cold items up, because this effect is not present on Movielens.}

% \djx{I also found the discussion here unclear. Tried to improve it and shorten it..}
In order to assess if the performance of DELF EF is due to this uncommon test dataset, we ran additional experiments in which we removed the cold items from the test sets. This allows us to assess the true modeling capacity of the approach. The results are shown are shown in Table \ref{tab:delf-amazon} (Amazon Music) and Table \ref{tab:delf-movielens} (MovieLens1M). Compared to the situation where cold items are considered in the test set, the recommendation quality of DELF EF drops below many of our baselines for the Amazon Music dataset. For the MovieLens dataset, where there are no cold items, %\djx{not many or no?}
the performance does not change much as expected. Overall, these observations lead to the conclusion that DELF EF has a tendency to push never seen items to higher ranks of the recommendation list, i.e., higher than the previously seen negative samples. This might be a consequence of the DELF EF heuristic, which takes the the popularity of the items into account.

Overall, when considering only situations where cold items are not recommended, DELF was never the best-performing one.
%Across both datasets (MovieLens1M and Amazon Music when the cold items are not recommended) and across all measurements, the proposed model was never the best-performing one.
For the MovieLens1M data, \iALS, \EASER, and SLIM outperformed DELF on all measures. The UserKNN method was also consistently better in terms of the NDCG (the optimization criterion).

On the Amazon dataset, \iALS was better than all other methods. Furthermore, all neighborhood models and SLIM outperformed the new methods in all but one measurement (NDCG@20). A consistent ``win'' for DELF method was only observed over the PureSVD method.

% \djx{Suggest to discuss MovieLens first because it is more relevant. Also the order of the tables should be changed accordingly. Did not touch because of the footnote thing.}

\begin{table}[h!t]
    \caption{Experimental results for DELF on the MovieLens dataset. We highlight in bold those entries where a baseline outperforms DELF.}
    \label{tab:delf-movielens}
    \footnotesize
% \resizebox{.8\linewidth}{!}{%
    \begin{tabular}{lcccccc}
    \toprule
		& HR@5 	& NDCG@5 	& HR@10 	& NDCG@10 	& HR@20 	& NDCG@20 	\\
	\midrule
    TopPopular & 0.3302 & 0.2229 & 0.4696 & 0.2674 & 0.6577 & 0.3148 \\
    \midrule
    UserKNN CF asymmetric &	\textbf{0.5205} &  \textbf{0.3635} &  \textbf{0.6852} &  \textbf{0.4168} & 0.8329 &  \textbf{0.4542} \\
    ItemKNN CF cosine	& 0.4936 & 0.3426 & 0.6677 & 0.3989 & 0.8243 & 0.4387 \\ 
    \midrule
    \palpha & 0.4945 & 0.3438 & 0.6574 & 0.3965 & 0.7952 & 0.4313 \\
    \pbeta & 0.5138 & 0.3559 & 0.6809 & 0.4102 & 0.8276 & 0.4475 \\
    \midrule
    \EASER &  \textbf{0.5716} &  \textbf{0.4064} &  \textbf{0.7258} &  \textbf{0.4566} &  \textbf{0.8516} &  \textbf{0.4887} \\
    SLIM 	&  \textbf{0.5706} &  \textbf{0.4038} &  \textbf{0.7306} &  \textbf{0.4557} &  \textbf{0.8586} &  \textbf{0.4882} \\
    PureSVD	&  \textbf{0.5513} &  \textbf{0.3891} &  \textbf{0.7021} &  \textbf{0.4382} &           0.8303 &  \textbf{0.4708} \\
    \iALS	 &  \textbf{0.5643} &  \textbf{0.3975} &  \textbf{0.7228} &  \textbf{0.4489} &  \textbf{0.8354} &  \textbf{0.4776} \\
    \midrule
    DELF MLP & 0.5168 & 0.3587 & 0.6809 & 0.4119 & 0.8342 & 0.4508 \\
    DELF EF & 0.4805 & 0.3305 & 0.6504 & 0.3852 & 0.8043 & 0.4243 \\
    \bottomrule
  	\end{tabular}
%  	}
\end{table}

\begin{table}[h!t]
    \caption{Experimental results for DELF for the Amazon Music Data when the recommendation of cold items is allowed. \EASER results are missing because the code required too much memory on these datasets. We highlight in bold those entries where a baseline outperforms DELF.}
    \label{tab:delf-amazon-colditems}
    \footnotesize
    \centering
    \begin{tabular}{lcccccc}
    \toprule
		& HR@5 	& NDCG@5 	& HR@10 	& NDCG@10 	& HR@20 	& NDCG@20 	\\
	\midrule
    % Random	&0.0490	&0.0290	&0.0997	&0.0452	&0.2000	&0.0702	\\
    TopPopular	&0.2452	&0.1726	&0.3057	&0.1921	&0.3744	&0.2094	\\
    \midrule
    UserKNN CF cosine	&0.3248	&0.2544	&0.3760	&0.2708	&0.4376	&0.2864	\\
    % UserKNN CF dice	&0.3210	&0.2522	&0.3760	&0.2700	&0.4371	&0.2854	\\
    % UserKNN CF jaccard	&0.3210	&0.2526	&0.3760	&0.2704	&0.4371	&0.2858	\\
    % UserKNN CF asymmetric	&0.3188	&0.2516	&0.3749	&0.2698	&0.4365	&0.2853	\\
    % UserKNN CF tversky	&0.3221	&0.2527	&0.3760	&0.2701	&0.4371	&0.2855	\\
    % ItemKNN CF cosine	&0.3204	&0.2528	&0.3733	&0.2698	&0.4371	&0.2858	\\
    % ItemKNN CF dice	&0.3117	&0.2441	&0.3717	&0.2632	&0.4338	&0.2789	\\
    % ItemKNN CF jaccard	&0.3090	&0.2439	&0.3602	&0.2604	&0.4256	&0.2767	\\
    ItemKNN CF asymmetric	&0.3204	&0.2566	&0.3711	&0.2731	&0.4327	&0.2886	\\
    % ItemKNN CF tversky	&0.3046	&0.2431	&0.3619	&0.2615	&0.4278	&0.2780	\\
    \midrule
    \palpha	&0.3188	&0.2524	&0.3684	&0.2684	&0.4300	&0.2839	\\
    \pbeta	&0.3155	&0.2494	&0.3684	&0.2663	&0.4272	&0.2811	\\
    \midrule
    \EASER 	& - 	& - 	& - 	& - 	& - 	& - 	\\
    % SLIM BPR 	&0.3139	&0.2446	&0.3717	&0.2632	&0.4392	&0.2801	\\
    SLIM &0.3199	&0.2577	&0.3678	&0.2730	&0.4354	&0.2900	\\
    % MF  AsySVD 	& - 	& - 	& - 	& - 	& - 	& - 	\\
    % MF BPR 	&0.2376	&0.1896	&0.2768	&0.2023	&0.3520	&0.2213	\\
    % MF FunkSVD 	&0.2545	&0.2035	&0.2916	&0.2155	&0.3417	&0.2280	\\
    PureSVD	&0.2627	&0.2141	&0.3084	&0.2290	&0.3537	&0.2405	\\
    % NMF	&0.2921	&0.2306	&0.3510	&0.2498	&0.4087	&0.2644	\\
    \iALS	&0.3319	&0.2604	&0.3717	&0.2732	&0.4229	&0.2860	\\
    \midrule
    DELF MLP & 0.2986 & 0.2339 & 0.3619 & 0.2542 &  \textbf{0.4561} & 0.2778 \\
    DELF EF &  \textbf{0.5422} &  \textbf{0.3632} &  \textbf{0.7439} &  \textbf{0.4290} &  \textbf{0.8578} &  \textbf{0.4583} \\
    \bottomrule
  	\end{tabular}
\end{table}

%%%%%%% RESULTS
\begin{table}[h!t]
    \caption{Experimental results for DELF for the Amazon Music Data when the recommendation of cold items is \emph{not} allowed.  \EASER results are missing because the code required too much memory on this dataset.  We highlight in bold those entries where a baseline outperforms DELF.}
    \label{tab:delf-amazon}
    \footnotesize
% \resizebox{.8\linewidth}{!}{%
    \begin{tabular}{lcccccc}
    \toprule
		& HR@5 	& NDCG@5 	& HR@10 	& NDCG@10 	& HR@20 	& NDCG@20 	\\
	\midrule
    TopPopular	&0.2474	&0.1730	&0.3041	&0.1913	&0.3738	&0.2090	\\
    \midrule
    UserKNN CF cosine	&\textbf{0.3150}	&\textbf{0.2495}	&\textbf{0.3471}	&\textbf{0.2600}	&0.3738	&\textbf{0.2668}	\\
    ItemKNN CF asymmetric	&\textbf{0.3090}	&\textbf{0.2506}	&\textbf{0.3401}	&\textbf{0.2609}	&0.3717	&\textbf{0.2689}	\\ \midrule
    \palpha	&\textbf{0.3074}	&\textbf{0.2465}	&\textbf{0.3373}	&\textbf{0.2564}	&0.3689	&\textbf{0.2644}	\\
    \pbeta	&\textbf{0.3046}	&\textbf{0.2434}	&\textbf{0.3379}	&\textbf{0.2543}	&0.3651	&\textbf{0.2611}	\\
    \midrule
    \EASER	& - 	& - 	& - 	& - 	& - 	& - 	\\
    SLIM 	&\textbf{0.3101}	&\textbf{0.2526}	&\textbf{0.3411}	&\textbf{0.2625}	&0.3711	&\textbf{0.2701}	\\
    PureSVD	&0.2627	&0.2141	&0.3084	&0.2290	&0.3542	&0.2406	\\
    \iALS	&\textbf{0.3319}	&\textbf{0.2604}	&\textbf{0.3706}	&\textbf{0.2729}	&\textbf{0.4109}	&\textbf{0.2831}	\\
    \midrule
    DELF MLP & 0.2905 & 0.2239 & 0.3275 & 0.2361 & 0.3787 & 0.2489 \\
    DELF EF & 0.2883 & 0.2224 & 0.3313 & 0.2364 & 0.3831 & 0.2496 \\
    \bottomrule
  	\end{tabular}
%  	}
\end{table}

Looking at methodological aspects,
%data pre-processing for the Amazon Music dataset changed the characteristics of the dataset significantly. After pre-processing, the dataset was not only significantly smaller, on average there were also less than two interactions per item left.
like in other works discussed here, the authors did not optimize the number of epochs on the validation set but took the best values using the test.
This is a methodological issue leading to information leakage from the test data.
%can lead to overfitting on the test data. %\mf{Our results are substantially different from the original paper -0.12 HR and -0.05 NDCG. Further checks are in progress...}
%\djx{For other papers, we were more negative about such a procedure than we are here, where we only indicate a potential overfitting and not a methodological issue.}

% ------------------------------------------------------------------------

% ------------------------------------------------------------------------
\subsection{Outer Product-based Neural Collaborative Filtering (ConvNCF)}
\label{sec:oncf}
% ------------------------------------------------------------------------
% \paragraph{Method} The \emph{ConvNCF} method \cite{He:2018:OPN:3304889.3304969} was presented at IJCAI '18. Its main idea is to explicitly model the pairwise correlations between the dimensions of the embedding using an outer product. With this technique, the authors aim to create an \emph{interaction map}, which is more expressive than existing methods that use simple concatenations of embeddings or element-wise products.

\paragraph{Datasets} The proposed method is evaluated on two public implicit-feedback datasets, \emph{Gowalla}\footnote{\url{5http://dawenl.github.io/data/gowalla_pro.zip}} and \emph{Yelp}\footnote{\url{https://github.com/hexiangnan/sigir16-eals}}.
Both datasets contain multiple implicit interactions at different timestamps for the same user-item pair.
These interactions are merged by keeping only the earliest for each user-item pair.\footnote{For the Gowalla dataset the authors use the first interaction appearing in the file as the \emph{earlier} interaction.
}
Both dataset have been filtered by removing items with less than 10 interactions and users with less than 2 (Gowalla) or 10 (Yelp) interactions.
% Both datasets and train/test splits are provided by the authors.
The Yelp dataset has about 69k interactions after processing.
The Gowalla dataset is sparser and with more interactions (1.2M).
Both filtered datasets with their train/test splits have been provided by the authors.
%The used Gowalla dataset is one that was already binarized and does not contain timestamps.
%The authors use the first interaction appearing in the file as the first interaction, but it is unclear if this corresponds to the true order.

\paragraph{Evaluation} Each dataset is split into training and test data according to a leave-last-out protocol. Evaluation is done by randomly selecting 999 negative items, and the algorithm has to rank these items together with the hidden item. Different methodological issues were observed, as will be discussed below. The Hit Rate and the NDCG at different list lengths are used as evaluation measures. Hyperparameter optimization is done via a validation set, except for the embedding size, which is kept constant for all methods.

\paragraph{Results and Discussion}
We could reproduce the results using the code and the data provided by the authors. The results for the Yelp dataset are shown in Table \ref{tab:convncf-yelp}. Those for the larger Gowalla dataset are given in Table \ref{tab:convncf-gowalla}.
% TODO: Could just report the Yelp dataset

\begin{table}[h!t]
    \caption{Experimental results for ConvNCF for the Yelp dataset. We highlight in bold those entries where a baseline outperforms ConvCF.}
    \label{tab:convncf-yelp}
    \footnotesize
% \resizebox{.8\linewidth}{!}{%
    \begin{tabular}{lcccccc}
    \toprule
    & HR@5 	& NDCG@5 	& HR@10 	& NDCG@10 	& HR@20 	& NDCG@20 	\\
    \midrule
    TopPopular&0.0817	&0.0538	&0.1200	&0.0661	&0.1751	&0.0799	\\
    \midrule
    UserKNN CF asymmetric	&\textbf{0.2131}	&\textbf{0.1400}	&\textbf{0.3209}	&\textbf{0.1747}	&\textbf{0.4482}	&\textbf{0.2068}	\\
    ItemKNN CF cosine	&\textbf{0.2521}	&\textbf{0.1686}	&\textbf{0.3669}	&\textbf{0.2056}	&\textbf{0.4974}	&\textbf{0.2385}	\\
    \midrule
    \palpha &\textbf{0.2146}	&\textbf{0.1395}	&\textbf{0.3211}	&\textbf{0.1737}	&0.4442	&\textbf{0.2049}	\\
    \pbeta	&\textbf{0.2202}	&\textbf{0.1431}	&\textbf{0.3323}	&\textbf{0.1793}	&\textbf{0.4667}	&\textbf{0.2132}	\\
    \midrule
    \EASER &  \textbf{0.2349} &  \textbf{0.1557} &  \textbf{0.3419} &  \textbf{0.1902} &  \textbf{0.4617} &  \textbf{0.2205} \\
    SLIM 	&\textbf{0.2330}	&\textbf{0.1535}	&\textbf{0.3475}	&\textbf{0.1904}	&\textbf{0.4799}	&\textbf{0.2238}	\\
    PureSVD	&\textbf{0.2011}	&\textbf{0.1307}	&0.3002	&\textbf{0.1626}	&0.4238	&0.1938	\\
    \iALS	&\textbf{0.2048}	&\textbf{0.1348}	&\textbf{0.3080}	&\textbf{0.1680}	&0.4319	&\textbf{0.1993}	\\
    \midrule
    ConvNCF	&0.1947	&0.1250	&0.3059	&0.1608	&0.4446	&0.1957	\\
    \bottomrule
  	\end{tabular}
%  	}
\end{table}

\begin{table}[h!t]
    \caption{Experimental results for ConvNCF for the Gowalla dataset. \EASER results are missing because the code required too much memory on this dataset. We highlight in bold those entries where a baseline outperforms ConvCF.}
    \label{tab:convncf-gowalla}
    \footnotesize
% \resizebox{.8\linewidth}{!}{%
    \begin{tabular}{lcccccc}
    \toprule
	& HR@5 	& NDCG@5 	& HR@10 	& NDCG@10 	& HR@20 	& NDCG@20 	\\
	\midrule
    TopPopular		&0.2188	&0.1652	&0.2910	&0.1884	&0.3803	&0.2110	\\
    \midrule
    UserKNN CF cosine	&\textbf{0.7131}	&\textbf{0.5879}	&\textbf{0.7939}	&\textbf{0.6142}	&0.8532	&\textbf{0.6293}	\\
    ItemKNN CF tversky	&\textbf{0.7047}	&\textbf{0.5864}	&0.7790	&\textbf{0.6105}	&0.8331	&\textbf{0.6244}	\\
    \midrule
    \palpha	&\textbf{0.6926}	&\textbf{0.5703}	&0.7674	&\textbf{0.5948}	&0.8158	&\textbf{0.6071}	\\
    \pbeta	&\textbf{0.6836}	&\textbf{0.5525}	&0.7723	&\textbf{0.5814}	&0.8361	&\textbf{0.5976}	\\
    \midrule
    \EASER & - 	& - 	& - 	& - 	& - 	& - 	\\
    SLIM 	&0.6365	&\textbf{0.5284}	&0.7083	&0.5517	&0.7608	&0.5651	\\
    PureSVD	&0.5653	&0.4482	&0.6627	&0.4798	&0.7393	&0.4993	\\
    \iALS	&0.6460	&0.5081	&0.7554	&0.5436	&0.8356	&0.5641	\\
    \midrule
    ConvNCF	&0.6702	&0.5233	&0.7799	&0.5590	&\textbf{0.8623}	&0.5799	\\
    \bottomrule
  	\end{tabular}
%  	}
\end{table}

For the Yelp dataset, ConvNCF is consistently outperformed by the traditional nearest-neighbor methods, \pbeta,  and SLIM. The other baselines outperform ConvNCF as well in most cases.
For the Gowalla case, ConvNCF is slightly more competitive.
In all but one measurement, however, it is outperformed by the UserKNN method. Interestingly, the simple machine learning methods do not work better on this dataset than the simple baselines. % TODO: would be interesting to know why.

A number of methodological issues were observed with this paper.
First, based on the provided source code the number of epochs, as in other papers, was determined on the test data. Furthermore, the authors decided to set the embedding size to the constant value of 64 for all baselines.
However, the embedding size is a hyperparameter to be tuned for each dataset and for each embedding-based algorithm, as different models with different objective functions or training procedures will likely require different values for it.
There were also issues with the provided test splits.
Negative test samples contained duplicates and partially overlapped with the train data. This results in virtually no user having the correct number of 999 unique negative items. Furthermore, the users having less than 980, are only 1\% for Gowalla but a remarkable 60\% for Yelp, with a minimum of 938 negative items.
%\textcolor{blue}{(Based on the code provided by the authors we could also observe that test data was indirectly used for training, as the function responsible for the sampling of training negative items guaranteed them to neither be in the train data nor in the test data. We report that the error was later fixed by the authors on the public Github repository. In our experiments we never used test data during training and TODO: This needs a proper explanation here; or we leave it out. \mf{Is it clear now? Should we keep it? The funny thing is that the error was fixed after we published the preprint of the recsys article and after we had sent emails to the author. My guess is that they knew and fixed it.})} \mf{Removed}
\footnote{We report that at the time of our experiments the published version of the algorithm contained a bug that caused information leakage from the test data. Items not present in the train data were divided in two categories, those \emph{not} present in the test data were sampled as negative items during the training of the model, while those present in the test data were not. %In particular it was guaranteed that test items could never be sampled as negative items during training.
% This may result in test items to be advantaged over other items.
Items unobserved in the train data should all be potential negative items to be sampled during training, regardless of the test data, otherwise test items will be advantaged over other items. This bug was later fixed on the public Github repository. In our experiments, during the training we only relied upon train data and never used any information from the test data.%\djx{I think we can have this here But it should be made clearer. It is a bit confusing because we say there is a test data leakage, then we say it was ensured that test items were \emph{not} considered (which might be an indirect leakage). and then we say we do not use test item information.} \mf{I'll try to rephrase it.}
} %\mf{leave as a footnote? To guarantee us against possible complaints, since the github repo reports quite clearly the commit "fix the bug that gtItem is excluded when sampling the negative item"}}

In addition, a recent article by Ferrari Dacrema et al. \cite{FerrariDacrema2020convolution} questioned the claim that embedding maps derived from the outer product of the embeddings are analogous to images. The article shows that embedding maps do not have a semantically relevant topology and do not share image properties, therefore the use of CNNs is not well justified. Moreover, contrary to what stated in the original paper, ConvNCF does not allow to model the embedding correlations which can be removed from the model without altering its accuracy.

% Could add thoughts about the use of CNNs here. There is no order.
% epochs based on the test data.
% set embedding size same value for all models and datasets. this is inappropriate, number of epochs on the test set.
% ------------------------------------------------------------------------
\subsection{Leveraging Meta-path based Context (MCRec)}
\label{sec:mcrec}
% ------------------------------------------------------------------------
% \paragraph{Method}
% The \emph{MCRec} \cite{hu2018leveragingmetapathcontext} method was published at KDD '18. It is a hybrid method that uses side information about the recommendable items in the recommendation process. The side information is represented as a network structure, and meta-paths are relation sequences that connect objects in this graph. Technically, the authors use a priority-based sampling technique to select more informative paths instances and a novel co-attention mechanism to improve the representations of meta-path based context, users and items. % TODO: Check that the text is not the same as in the original paper. (also for other papers).

\paragraph{Datasets} Three datasets are used for the evaluation. The historical \emph{MovieLens100k}\footnote{\url{http://jmcauley.ucsd.edu/data/amazon/}} dataset, another small data set containing listening logs from \emph{Last.fm}\footnote{\url{https://www.last.fm}}, and a dataset containing user feedback from \emph{Yelp}\footnote{\url{http://www.yelp.com/dataset-challenge}}.
With almost 200k ratings, the Yelp dataset is the largest one.
The available meta-data includes genres for movies, artists for tracks, and city and category for businesses. Only for the MovieLens100k dataset the used data splits are publicly available.
The rating information in the datasets is transformed to binary data.

\paragraph{Evaluation} The datasets are split into 80\% training and 20\% test splits. Ten percent of the training data are used for validation.
% For each positive item in the test set, 50 negative test interactions are randomly selected in such a way that the negative interactions in train and test data are disjoint.
The algorithm is evaluated by ranking the positive items of a user against a set of negative items randomly sampled among the items the user did not interact with. For each user, a set of exactly 50 negative items is sampled for each of the user's test items. The negative interactions in train and test data are disjoint.\footnote{This evaluation protocol is different from more traditional protocols. Usually, both the number of positive items and negative items associated to each user is constant (e.g., 1 positive item ranked with 100 negative items), in MCRec instead different users are tested against a different number of both positive and negative samples (e.g., if the positive items are 10 the negative samples will be 500, if the positive items are 30 the negative samples will be 1500).} For the MovieLens100k dataset, 23\% of the users have long user profiles to the point that they cannot have the stated number of 50 negative items per each positive because the number of items which do not appear in the user profile is too low.\footnote{In particular, the longest user profile (including all available data, therefore train, validation and test interactions) that allows to sample 50 negative items for each positive (test) item is 151. However, the longest user profile in the dataset contains 737 items, which would require to sample 7370 negative items while the dataset contains 1662 and the number of unseen items to choose from is only 925.}
%Like in other approaches discussed here, the evaluation procedure is also based on sampling negative items for each positive item in the test set.
%However, this time it is not a leave-one-out procedure, but the evaluation is done by user, which leads to a varying number of positive items (and negative samples) for ranking.
Precision, Recall, and the NDCG are then used as evaluation measures at a cut-off length of 10.
% TODO: if there is only one positive item, how do they compute precision and recall, or: shouldn't this be the same?
The averaged results of ten randomized runs are reported in the paper. Since the NDCG measure was implemented in an unusual way by computing the ``ideal'' NDCG only based on the successfully recommended items, the values reported in the paper are much higher than the ones obtained by us using a standard implementation of the NDCG. Looking at the evaluation code provided by the authors, we furthermore observed that the authors reported the best results for each metric across different epochs.
Since this is inappropriate, we report the values of the metrics after the optimal number of epochs is chosen with early stopping on the validation set.

As baselines, the authors use ItemKNN, two collaborative MF methods (one based on BPR loss, the other based on cross entropy loss), two hybrid MF methods, two methods designed for metadata networks, and a number of variants of their own method.
For their own method, the final hyperparameters are reported, which are the same for both datasets.
For the MF and the NeuMF method, hyperparameters and configuration are taken from the original papers. Other baselines are  reported to be systematically optimized on a validation set.

\paragraph{Results and Discussion}
The meta-path information is hard-coded for the MovieLens dataset in the provided code and the preprocessed version of the data is publicly available. No code is available for the other two datasets. Furthermore, since the code for pre-processing the Yelp and Last.fm is not available, we only did an evaluation for the MovieLens dataset.\footnote{The pre-processing code for Movielens was made available later by the authors on their GitHub repository.} The results are reported in Table \ref{tab:mcrec-movielens}. They show that MCRec is outperformed both by the traditional neighborhood-based methods, the more complex learning-based methods and one of our simple hybrids.
Our pure content-based baseline led to generally weak results (worse than TopPopular) and was also outperformed by MCRec.

\begin{table}[h!t]
    \caption{Experimental results for MCRec for the MovieLens dataset. We highlight in bold those entries where a baseline outperforms MCRec.}
    \label{tab:mcrec-movielens}
    \footnotesize
% \resizebox{.6\linewidth}{!}{%
    \begin{tabular}{lcccccc}
    \toprule
		& PREC@10 	& REC@10 	& NDCG@10 	\\
	\midrule
    TopPopular	&0.1907	&0.1180	&0.1361	\\
    \midrule
    UserKNN CF dice		&\textbf{0.3442}	&\textbf{0.2237}	&\textbf{0.2692}	\\
    ItemKNN CF asymmetric	&\textbf{0.3320}	&\textbf{0.2171}	&\textbf{0.2601}	\\ \midrule
    \palpha	&\textbf{0.3305}	&0.2081	&\textbf{0.2554}	\\
    \pbeta	&\textbf{0.3435}	&\textbf{0.2191}	&\textbf{0.2588}	\\
    \midrule
    \EASER &  \textbf{0.3739} &  \textbf{0.2430} &  \textbf{0.2905} \\
    SLIM 	&\textbf{0.3770}	&\textbf{0.2441}	&\textbf{0.2957}	\\
    PureSVD	&\textbf{0.3545}	&\textbf{0.2247}	&\textbf{0.2719}	\\
    \iALS	&\textbf{0.3596}	&\textbf{0.2283}	&\textbf{0.2759}	\\
    \midrule
    ItemKNN CBF cosine	&0.0455	&0.0185	&0.0254	\\
    \midrule
    ItemKNN CFCBF cosine	&\textbf{0.3398}	&\textbf{0.2239}	&\textbf{0.2646}	\\
    \midrule
    MCRec	&0.3110	&0.2113	&0.2466	\\
    \bottomrule
  	\end{tabular}
%   	}
\end{table}

From a methodological perspective, it is worth mentioning that the used datasets are very small, at least compared to the usual MovieLens and Netflix datasets, where we have millions of recorded interactions. We will discuss aspects of scalability later in this paper. % TOOD: add section number and add discussion

% ------------------------------------------------------------------------
\subsection{Collaborative Memory Network for Recommendation System (CMN)}
\label{sec:cmn}

% ------------------------------------------------------------------------
% \paragraph{Method} Collaborative Memory Networks (CMN) \cite{ebesu2018collaborative} was presented at SIGIR '18, and it represents a collaborative-filtering approach based on memory networks and neural attention mechanisms. The underlying idea of the approach is to combine latent factor with neighborhood (memory-based) approaches in a non-linear fashion.
% % TODO? Say something more here?

\paragraph{Datasets} The method was evaluated on three public datasets. One containing data from \emph{Epinions}\footnote{\url{http://www.trustlet.org/downloaded_epinions.html}}, a \emph{CiteULike} dataset used in previous works (\emph{CiteULike-a})\footnote{See Section \ref{sec:cdl}.}, and a dataset from \emph{Pinterest}\footnote{\url{https://sites.google.com/site/xueatalphabeta/academic-projects}}. The last dataset is the largest one and has about 1.5M interactions. The smallest Epinions dataset contains ratings on scale from one to five, which are binarized (all non-zero ratings are set to one).
The train/test splits for CiteULike and Pinterest are provided by the authors.

\paragraph{Evaluation}
The authors use a leave-one-out methodology similar to the one used in previous works.
The test set contains one positive item per user, the train set all other user-item pairs.
If the user rated only one item, this interaction is kept in the training set.
To evaluate the algorithm, for each user 100 unobserved (negative) items are sampled and ranked together with the positive item.
The Hit Rate and the NDCG at a cut-off length of 10 are used as evaluation metrics.

As baselines, the authors include both simple, non-neural, and neural methods in there experiments: ItemKNN, BPR matrix-factorization, SVD++, %\cite{koren2008factorization}
two variants of NCF \cite{he2017neural}, % TODO Check
and the Collaborative Denoising Auto Encoder.%\cite{wu2016collaborativeCDAE}.

Hyperparameters are tuned on a validation set. Details for the hyperparameter optimization process are provided for the proposed model, but not for the baselines. The number of training epochs is not mentioned in the paper. %--it is 30 according to the provided code for the CiteULike dataset. \mf{Removed because we don't mention it for any other algorithm}
In our evaluation, we applied early stopping as described previously.

\paragraph{Results and Discussion}
We could reproduce the results of the paper using the data splits provided by the authors for two datasets, CiteULike and Pinterest. Since the split used for Epinions was not made available we had to recreate it based on the information provided in the paper, in that case, however, we could not reproduce the original results. %\mf{Rephrased to highlight that we can reproduce it using the authors split for two datasets, the third dataset does not have a public split, we created it ourselves and on that the results are NOT reproducible.}
%Both set of results are consistent.
In Table \ref{tab:cmn-results-pinterest}, we show the results for the Pinterest dataset, the largest one.
For the CMN method, we report the results for the \emph{CMN-3} variant, which led to the best results.

\begin{table}[h!t]
    \caption{Experimental results for CMN for the Pinterest dataset. We highlight in bold those entries where a baseline outperforms CMN.}
    \label{tab:cmn-results-pinterest}
    \footnotesize
% \resizebox{.6\linewidth}{!}{%
    \begin{tabular}{lcccc}
    \toprule
		& HR@5 	& NDCG@5 	& HR@10 	& NDCG@10 	\\
		\midrule
    TopPopular	&0.1665	&0.1064	&0.2740	&0.1409	\\
    \midrule
    UserKNN CF asymmetric	&0.7005	&\textbf{0.5037}	&0.8630	&\textbf{0.5567}	\\
    ItemKNN CF cosine		&\textbf{0.7132}	&\textbf{0.5116}	&\textbf{0.8781}	&\textbf{0.5653}	\\
    \midrule
    \palpha	&0.6990	&\textbf{0.5034}	&0.8596	&\textbf{0.5559}	\\
    \pbeta	&\textbf{0.7147}	&\textbf{0.5150}	&\textbf{0.8772}	&\textbf{0.5680}	\\
    \midrule
    \EASER &  \textbf{0.7072} &  \textbf{0.5129} & 0.8567 &  \textbf{0.5617} \\
    SLIM &\textbf{0.7084}	&\textbf{0.5107}	&\textbf{0.8683}	&\textbf{0.5628}	\\
    PureSVD	&0.6619	&0.4721	&0.8146	&0.5219	\\
    \iALS	&\textbf{0.7219}	&\textbf{0.5175}	&\textbf{0.8677}	&\textbf{0.5652}	\\
    \midrule
    CMN	&0.7013	&0.5005	&0.8674	&0.5547	\\
	\bottomrule
  	\end{tabular}
%   	}
\end{table}

The results show that CMN led to competitive results, but is actually slightly outperformed on all measures by algorithms from all families, including traditional nearest-neighbor methods. On the CiteULike dataset, the main observations are the same. On this smaller dataset, however, the performance of CMN is often much lower than the one achieved by nearest-neighbor methods and machine learning techniques.

A quite surprising result is found for the Epinions dataset.
Here, the trivial TopPopular method consistently led to the best values across all measurements. The difference to all other methods is often huge.
This indicates that the popularity distribution of the items in this dataset is very skewed, which makes it difficult to make personalized recommendations that are better in terms of information retrieval measures than the TopPopular method. The detailed results are available in the online material (Section \ref{sec:methodology}).
%We made additional experiments with the BPR method, which has a tendency to recommend popular
% TODO: Should we include a figure that illustrates the distribution and sparsity of this dataset?

% methodology: some issues with the splits. should be mentioned? Maybe too detailed.
% ------------------------------------------------------------------------
\subsection{Spectral Collaborative Filtering (SpectralCF)}
\label{sec:spectralcf}
% ------------------------------------------------------------------------
% \paragraph{Method} \emph{SpectralCF}, presented at RecSys '18 \cite{Zheng:2018:SCF:3240323.3240343} is a graph-based approach. Users and items are represented as a bipartite graph.
% The novelty of this method is a convolution approach which operates on the \emph{spectral domain}.
% The method considers both proximity and connectivity information in the graph, which is assumed to be particularly helpful for cold-start problems.

\paragraph{Datasets}
Three public datasets are used for the evaluation, \emph{MovieLens1M}\footnote{\url{https://grouplens.org/datasets/movielens/}}, another movie rating dataset (\emph{HetRec})\footnote{\url{https://grouplens.org/datasets/hetrec-2011/}}, and an \emph{Amazon} dataset (\emph{InstantVideo})\footnote{\url{http://jmcauley.ucsd.edu/data/amazon/}}.
Explicit ratings are binarized.
Further pre-processing is applied to, e.g., remove users associated to less than 5 interactions.
After pre-processing, the MovieLens1M dataset is shrinked to one fifth of the original size (226k interactions). The other datasets are even smaller (71k and 22k interactions, respectively).
The data split for the MovieLens1M dataset is provided online by the authors.
Due to an apparent problem on how the splits were generated---see our discussions below---we created our own splits based on the information provided in the paper.

\paragraph{Evaluation}
Two evaluation scenarios are tested, a regular one and a cold-start setup. We evaluated both scenarios. For the main scenario, 80\% of the interactions of each user are randomly put into the training set and the rest is used for evaluation.
The random process is repeated five times and averaged results are reported.
Recall and MAP at different list lengths are used as metrics for this scenario.

In the cold-start scenario, the training set is built with different degrees of sparsity by varying the number $P$ of interactions associated with each user, where $P$ is varied from one to five.
The remaining items associated with each users are used as test set.
Recall@20 and MAP@20 are used in this scenario for evaluation.

As baselines, the authors consider ItemKNN, BPR matrix factorization, iALS, NCF \cite{he2017neural} and two graph-based methods, GNMF
%\cite{GNMF}
and GCMC,
%\cite{Berg2017GraphCM}
originally designed for explicit feedback datasets.
The hyperparameters are tuned on a validation set and the parameter ranges are reported in the paper.
For NCF the authors stated that they used the configuration reported in the original paper.

\paragraph{Results and Discussion}
We reproduced the results obtained by the authors based on the provided code and following the information in the paper.
Our first set of experiments showed that on HetRec and Amazon Video datasets the performance of SpectralCF was relatively weak, whereas it performed very well and better than all baselines when using the provided data splits for the MovieLens dataset.
We therefore investigated the provided data splits and found that these splits were unlikely the result of the splitting procedure described in the paper, see the discussions below.
For this reason, we created new data splits ourselves following the described procedure.

The results obtained for the MovieLens1M dataset are shown in Table \ref{tab:spectralcf-movielens}. The results show that SpectralCF is outperformed by all baselines in our comparison often exhibiting a recommendation quality equal to the TopPopular method. The observations for the other datasets are similar, the full results are available in the online material (Section \ref{sec:methodology}).

% TODO: Reformat the table like in the RecSys paper.
\begin{table}[h!t]
    \caption{Experimental results for SpectralCF for the MovieLens dataset.  We highlight in bold those entries where a baseline outperforms SpectralCF.}
    \label{tab:spectralcf-movielens}
    \footnotesize
\resizebox{\linewidth}{!}{%
    \begin{tabular}{l|cccccccccc|}
    \toprule
		& REC@20 	& MAP@20 	& REC@40 	& MAP@40 	& REC@60 	& MAP@60 	& REC@80 	& MAP@80 	& REC@100 	& MAP@100 	\\
		\midrule
    TopPopular	&\textbf{0.1892}	&\textbf{0.0584}	&\textbf{0.2788}	&\textbf{0.0636}	&\textbf{0.3356}	&\textbf{0.0666}	&\textbf{0.3834}	&\textbf{0.0687}	&0.4226	&\textbf{0.0702}	\\
    \midrule
    UserKNN CF jaccard	&\textbf{0.3001}	&\textbf{0.1201}	&\textbf{0.4134}	&\textbf{0.1285}	&\textbf{0.4901}	&\textbf{0.1335}	&\textbf{0.5457}	&\textbf{0.1367}	&\textbf{0.5884}	&\textbf{0.1388}	\\
    ItemKNN CF asymmetric	&\textbf{0.2876}	&\textbf{0.1134}	&\textbf{0.4000}	&\textbf{0.1213}	&\textbf{0.4768}	&\textbf{0.1263}	&\textbf{0.5367}	&\textbf{0.1295}	&\textbf{0.5820}	&\textbf{0.1317}	\\
    \midrule
    P3alpha	&\textbf{0.2939}	&\textbf{0.1141}	&\textbf{0.4150}	&\textbf{0.1233}	&\textbf{0.4900}	&\textbf{0.1285}	&\textbf{0.5463}	&\textbf{0.1318}	&\textbf{0.5903}	&\textbf{0.1342}	\\
    RP3beta	&\textbf{0.2737}	&\textbf{0.1044}	&\textbf{0.3879}	&\textbf{0.1124}	&\textbf{0.4664}	&\textbf{0.1173}	&\textbf{0.5234}	&\textbf{0.1206}	&\textbf{0.5726}	&\textbf{0.1230}	\\
    \midrule
    \EASER &  \textbf{0.3085} &  \textbf{0.1249} &  \textbf{0.4255} &  \textbf{0.1340} &  \textbf{0.4986} &  \textbf{0.1391} &  \textbf{0.5559} &  \textbf{0.1425} &  \textbf{0.6010} &  \textbf{0.1448} \\
    SLIM 	&\textbf{0.3069}	&\textbf{0.1265}	&\textbf{0.4246}	&\textbf{0.1356}	&\textbf{0.5010}	&\textbf{0.1410}	&\textbf{0.5564}	&\textbf{0.1443}	&\textbf{0.6001}	&\textbf{0.1466}	\\
    PureSVD	&\textbf{0.2595}	&\textbf{0.1008}	&\textbf{0.3638}	&\textbf{0.1083}	&\textbf{0.4378}	&\textbf{0.1131}	&\textbf{0.4913}	&\textbf{0.1161}	&\textbf{0.5347}	&\textbf{0.1182}	\\
    \iALS	&\textbf{0.3033}	&\textbf{0.1183}	&\textbf{0.4201}	&\textbf{0.1273}	&\textbf{0.4933}	&\textbf{0.1326}	&\textbf{0.5493}	&\textbf{0.1360}	&\textbf{0.5925}	&\textbf{0.1383}	\\
    \midrule
    SpectralCF		&0.1813	&0.0533	&0.2643	&0.0581	&0.3274	&0.0613	&0.3823	&0.0635	&0.4261	&0.0651	\\
    % SpectralCF article default	&0.1785	&0.0540	&0.2590	&0.0586	&0.3232	&0.0614	&0.3689	&0.0632	&0.4101	&0.0646	\\
	\bottomrule
  	\end{tabular}
 	}
\end{table}

\begin{figure}
\begin{subfigure}{.48\textwidth}
  \centering
    \includegraphics[width=\linewidth]{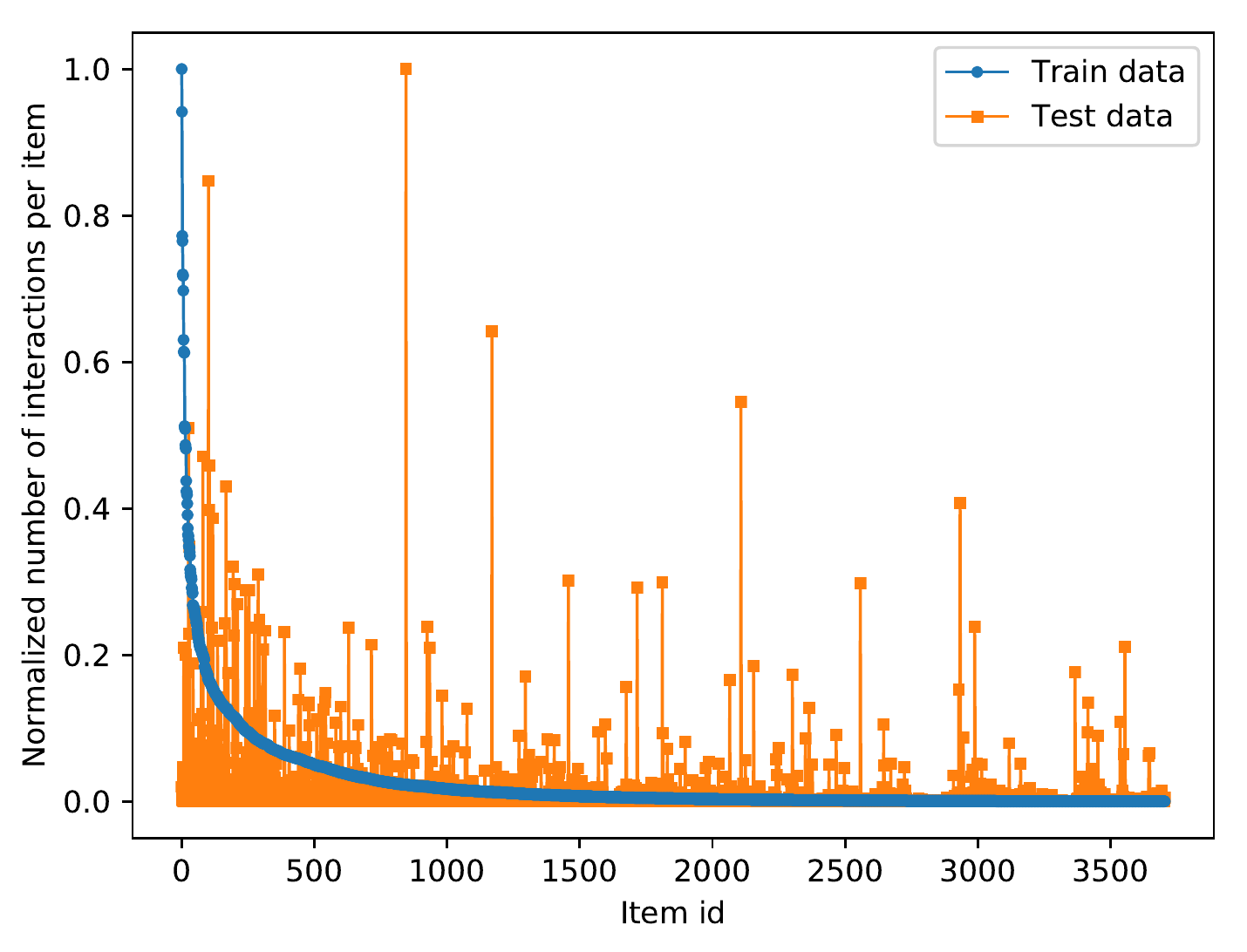}
    \caption{Normalized popularity distributions of the train and test splits provided by the original authors.}
    \label{fig:spectralcf-appendix-observed-distribution}
\end{subfigure}%
\hfill
\begin{subfigure}{.48\textwidth}
  \centering
\includegraphics[width=\linewidth]{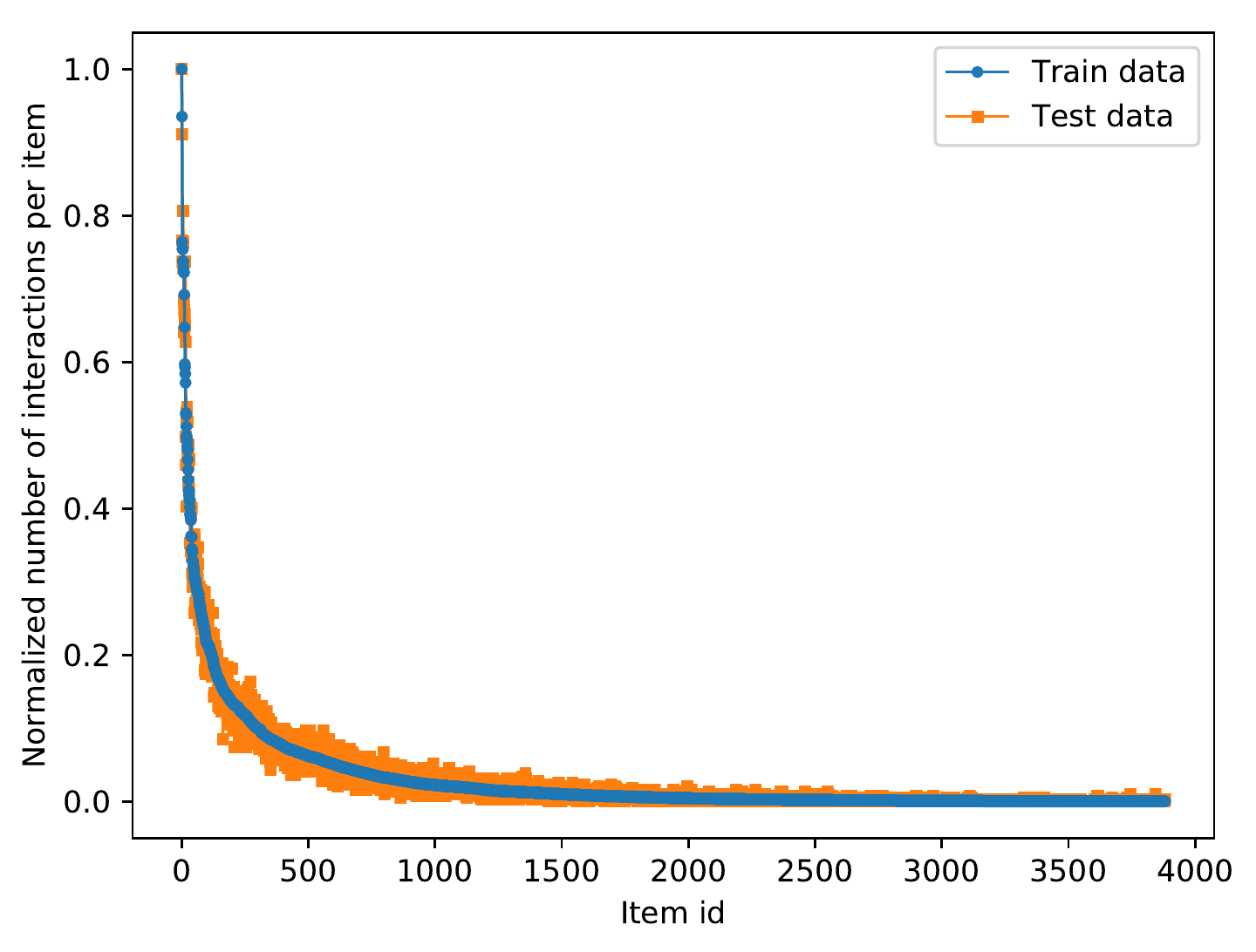}
    \caption{Normalized popularity distributions of the train and test splits generated by us.}
    \label{fig:spectralcf-appendix-expected-distribution}
\end{subfigure}
\label{fig:spectralcf-appendix-comparison-distribution}
\caption{Normalized popularity distributions of the train and test splits for SpectralCF, the value 1 corresponds to the most popular item in that split. For a random split, as can be seen in Figure \ref{fig:spectralcf-appendix-expected-distribution}, the normalized values of both splits are, on average, similar. In the split provided by the original authors, however, as can be seen in Figure \ref{fig:spectralcf-appendix-observed-distribution}, train and test data have quite different distributions.}
\end{figure}

To illustrate the data splitting problem, we compared the popularity distribution of the items in the training and the test split in the provided data, see Figure \ref{fig:spectralcf-appendix-expected-distribution}. The figure plots the normalized popularity (i.e., the popularity of an item over the popularity of the most popular item) for both training and test items.
Items are ranked based on their popularity in the training set, with the most popular training items being on the left.
In case of a true random split, the normalized popularity values of the items in the training and the test split should be relatively close.
However, the figure shows the split provided by the authors has some major deviations.
Figure \ref{fig:spectralcf-appendix-expected-distribution} shows the popularity distributions of our random split, which are almost identical between training and test sets.

Besides the visual analysis, we also computed numerical statistics like the Gini index and Shannon entropy. %, reported in Table XXX.
In a true random split, the Gini index should be close to the value obtained on the unsplitted data (in this case 0.78) for both the training and test splits.
%The Gini index for the provided training split is similar to what we obtained,
However, the Gini index of the provided test split is much higher (0.92).
This indicates that the provided test split has a much higher popularity bias than we would expect.
A similar consideration applies for the Shannon entropy: the entropy of the original dataset is close to 10 and is similar to the entropy of our random train/test split.
However, the provided test split has a lower entropy (8.5), i.e., it is easier to predict.

%\pc{change the axis in this way - vertical: Normalized number of interactions per item - horizontal: Item id}

Regarding other methodological aspects, we found that the authors only report one set of hyperparameters, whereas one would expect hyperparameter settings for each dataset. In our evaluation, we therefore optimized the hyperparameters for all baselines and all datasets individually.
% usual problems. original paper configuration
% only one set of parameters reported. used for all experiments?

%\pc{check for consitency top-n vs top-n} \mf{DONE}

\clearpage
\section{Hyperparameter Ranges}
\label{sec:hyperparameter_range}

%This allows for multiple references to the same footnote when the footnote is not a number
\makeatletter
\newcommand\footnoteref[1]{\protected@xdef\@thefnmark{\ref{#1}}\@footnotemark}
\makeatother

\begin{table}[h]
\caption{Hyperparameter list, value ranges and distributions for the baselines reported in this paper.}
\label{tab:ML_hyperparameters_our_baselines}
\resizebox{.90\linewidth}{!}{%
    \begin{minipage}{\textwidth}
    \begin{tabular}{cl|cccc}
    \toprule
    Algorithm	& Hyperparameter	&  Range	 & Type     & Distribution	\\
    \midrule
    \multirow{4}{*}{\begin{tabular}{c}UserKNN, ItemKNN \\ cosine\end{tabular}}  	
    				&topK	        & 5 - 1000 	& Integer   & uniform 	\\
    				&shrink	        & 0 - 1000 	& Integer   & uniform 	\\
    				%&similarity	    & cosine 	& Categorical 	& 	\\
    				&normalize\footnote{\label{foot:knn_normalize}The \emph{normalize} hyperparameter in KNNs refers to the use of the denominator when computing the similarity.} 	    & True, False 	& Categorical 	& 	\\
    				&feature weighting	& none, TF-IDF, BM25 	& Categorical 	& 	\\
    \midrule
    \multirow{3}{*}{\begin{tabular}{c}UserKNN, ItemKNN \\ dice\end{tabular}}  	
    				&topK	        & 5 - 1000 	& Integer   & uniform 	\\
    				&shrink	        & 0 - 1000 	& Integer   & uniform 	\\
    				% &similarity	    & dice 	& Categorical 	& 	\\
    				&normalize\footnoteref{foot:knn_normalize}	    & True, False 	& Categorical 	& 	\\
    \midrule
    \multirow{3}{*}{\begin{tabular}{c}UserKNN, ItemKNN \\ jaccard\end{tabular}}  	
    				&topK	        & 5 - 1000 	& Integer   & uniform 	\\
    				&shrink	        & 0 - 1000 	& Integer   & uniform 	\\
    				% &similarity	    & jaccard 	& Categorical 	& 	\\
    				&normalize\footnoteref{foot:knn_normalize}	    & True, False 	& Categorical 	& 	\\
    \midrule
    \multirow{5}{*}{\begin{tabular}{c}UserKNN, ItemKNN \\ asymmetric\end{tabular}}  	
    				&topK	        & 5 - 1000 	& Integer   & uniform 	\\
    				&shrink	        & 0 - 1000 	& Integer   & uniform 	\\
    				% &similarity	    & asymmetric 	& Categorical 	& 	\\
    				&normalize\footnoteref{foot:knn_normalize}	    & True   	& Categorical 	& 	\\
    				&asymmetric alpha	& 0 - 2	& Real   & uniform 	\\
    				&feature weighting	& none, TF-IDF, BM25 	& Categorical 	& 	\\
    \midrule
    \multirow{5}{*}{\begin{tabular}{c}UserKNN, ItemKNN \\ tversky\end{tabular}}  	
    				&topK	        & 5 - 1000 	& Integer   & uniform 	\\
    				&shrink	        & 0 - 1000 	& Integer   & uniform 	\\
    				% &similarity	    & tversky 	& Categorical 	& 	\\
    				&normalize\footnoteref{foot:knn_normalize}	    & True   	& Categorical 	& 	\\
    				&tversky alpha		& 0 - 2	& Real   & uniform 	\\
    				&tversky beta		& 0 - 2	& Real   & uniform 	\\
    \midrule
    \multirow{3}{*}{P3alpha}  	
    				&topK	        & 5 - 1000 	& Integer   & uniform 	\\
    				&alpha	        & 0 - 2	& Real   & uniform 	\\
    				&normalize similarity\footnote{\label{foot:normalize_similarity}The \emph{normalize similarity} hyperparameter refers to applying L1 regularisation on the rows of the similarity matrix}	& True, False 	& Categorical 	& 	\\
    \midrule
    \multirow{4}{*}{RP3beta}  	
    				&topK	        & 5 - 1000 	& Integer   & uniform 	\\
    				&alpha	        & 0 - 2	& Real   & uniform 	\\
    				&beta	        & 0 - 2	& Real   & uniform 	\\
    				&normalize similarity\footnoteref{foot:normalize_similarity}	& True, False 	& Categorical 	& 	\\
% 	\bottomrule
%   	\end{tabular}
%   	\end{minipage}
%   	}
%     \caption{Hyperparameter values for our KNN baselines.}
%     \label{tab:KNN_hyperparameters_our_baselines}
% \end{table}

% \begin{table}[h]
% \resizebox{\linewidth}{!}{%
%     \begin{minipage}{\textwidth}
%     \begin{tabular}{ll|ccl}
%     \toprule
%     Algorithm	& Hyperparameter	&  Range	 & Type     & Distribution	\\
    \midrule
    % \multirow{7}{*}{SLIM BPR }  	
    % 				&topK	        & 5 - 1000 	& Integer   & uniform 	\\
    % 				&epochs	        & 1 - 1500 	& Integer 	& early-stopping 	\\
    % 				&symmetric	    & True, False 	& Categorical 	& 	\\
    % 				&sgd mode	    & sgd, adam, adagrad    & Categorical 	& 	\\
    % 				&lambda i   	& $10^{-5}$ - $10^{-2}$     & Real   & log-uniform 	\\
    % 				&lambda j   	& $10^{-5}$ - $10^{-2}$     & Real   & log-uniform 	\\
    % 				&learning rate 	& $10^{-4}$ - $10^{-1}$     & Real   & log-uniform 	\\
    % \midrule
    \multirow{3}{*}{SLIMElasticNet}  	
    				&topK	        & 5 - 1000 	& Integer   & uniform 	\\
    				&l1 ratio	    & $10^{-5}$ - $10^{0}$     & Real   & log-uniform 	\\
    				&alpha	        & $10^{-3}$ - $10^{0}$     & Real   & uniform 	\\
    \midrule
    \multirow{1}{*}{PureSVD}  	
    				&num factors	& 1 - 350 	& Integer   & uniform 	\\
    % \midrule
    % \multirow{4}{*}{NMF}  	
    % 				&num factors	& 1 - 350 	& Integer   & uniform 	\\
    % 				&solver	        & mult. update, coord. descent 	 & Categorical 	& 	\\
    % 				&init type	    & nndsvda, random                & Categorical 	& 	\\
    % 				&beta loss  	& kullback-leibler, frobenius 	 & Categorical 	& 	\\
    \midrule
    \multirow{6}{*}{\iALS}  	
    				&num factors    & 1 - 200 \footnote{\label{foot:num_factor_lower_slow}The number of factors is lower than PureSVD due to \iALS being slower.} 	    & Integer       & uniform 	\\
    				&epochs	        & 1 - 500  	& Integer 	& early-stopping 	\\
    				&confidence scaling	&linear, log    & Categorical 	& 	\\
    				&alpha	    & $10^{-3}$ - $5\cdot10^{+1}$ \footnote{\label{foot:hyperparameter_value_original_article}The maximum value of this hyperparameter had been suggested in the article proposing the algorithm.}      & Real   & log-uniform 	\\
    				&epsilon	& $10^{-3}$ - $10^{+1}$ \footnoteref{foot:hyperparameter_value_original_article}     & Real   & log-uniform 	\\
    				&reg	    & $10^{-5}$ - $10^{-2}$     & Real   & log-uniform 	\\
    \midrule
    \multirow{1}{*}{\EASER}  	
    				&l2 norm	& $10^{0}$ - $10^{+7}$     & Real   & log-uniform 	\\
	\bottomrule
   	\end{tabular}
   	\end{minipage}
  	}
\end{table}

% Ensure nothing is mixed with the references
% \clearpage

% \bibliographystyle{ACM-Reference-Format}
% \bibliographystyle{plain}
% \bibliography{acmart}

\clearpage
\bibliographystyle{plain}
\bibliography{acmart}

\end{document}